\documentclass[rmp,twocolumn,showkeys,superscriptaddress]{revtex4-2}
\usepackage{graphicx}
\begin{document}
\title{Transport in electron-photon systems}

\author{Jian-Sheng Wang}
\email[]{phywjs@nus.edu.sg}
\affiliation{Department of Physics, National University of Singapore, Singapore 117551, Republic of Singapore\smallskip\looseness=-2}
\author{Jiebin Peng}
\affiliation{Center for Phononics and Thermal Energy
Science, China-EU Joint Center for Nanophononics, Shanghai
Key Laboratory of Special Artificial Microstructure Materials
and Technology, School of Physics Science and Engineering,
Tongji University, 200092 Shanghai, People's Republic of China\smallskip\looseness=-2}
\author{Zu-Quan Zhang}
\affiliation{Department of Physics, Zhejiang Normal University, Jinhua 321004, Zhejiang, People's Republic of China\smallskip\looseness=-2}
\author{Yong-Mei Zhang}
\affiliation{College of Physics, Nanjing University of Aeronautics and Astronautics, Jiangsu 210016, People's Republic of China\smallskip\looseness=-2}
\author{Tao Zhu\medskip}
\affiliation{School of Electronic and Information Engineering, Tiangong University, Tianjin 300387, People's Republic of China, 
and Beijing National Laboratory for Condensed Matter Physics, Institute of Physics, Chinese Academy of Sciences, Beijing 100190, People's Republic of China\smallskip\looseness=-2}


\date{6 August 2022}

\begin{abstract}
We review the description and modeling of transport phenomena among the electron systems coupled via scalar or vector photons.  It consists of 
three parts.  The first part is about scalar photons, i.e., Coulomb interactions.   The second part is with transverse photons described by vector potentials.   The third part is on
$\phi=0$ or temporal gauge, which is a full theory of the electrodynamics.   We use
the nonequilibrium Green's function (NEGF) formalism as a basic tool to study steady-state 
transport.   Although with local equilibrium it is equivalent to the fluctuational 
electrodynamics (FE), the advantage of NEGF is that it can go
beyond FE due to its generality.  We have given a few examples in the review, such as transfer of heat between
graphene sheets driven by potential bias, emission of light by a double quantum dot,
and emission of energy, momentum, and angular momentum from a graphene nanoribbon.
All of these calculations are based on a generalization of the Meir-Wingreen formula
commonly used in electronic transport in mesoscopic systems, with materials
properties represented by photon self-energy, coupled with the Keldysh equation 
and the solution to the Dyson equation.   
\end{abstract}


\maketitle

\section{Introduction}

Electron and photon interaction plays an important role in our understanding of basic physics in condensed matter, atomic physics, and nano-optics \cite{Tannoudji89,Bloch22}.  
It is also the principle governing photosynthesis and thermal radiation, and underpins many functioning devices such as lasers and solar cells.   This review focuses on one 
aspect of the electron-photon interaction, that is, the transfer of energy and other conserved quantities mediated by photons between metals in local 
thermal equilibrium and nonequilibrium states.   

Radiation of energy happens for any object, due to the electrically charged nature of 
ions and electrons that make up materials microscopically.
The fluctuation of charge and thus current is inevitable, leading to thermal radiation.  At the turn of the twentieth century, Planck, in attempting to understand
blackbody radiation, discovered the quantum nature of radiation and proposed his famous formula for the intensity of the radiation \cite{Planck14}.  He is aware of the
limitation of his theory, i.e., the cavity of the blackbody or the length scale of the object should be much larger than the wavelength
of the photons.   But it took 70 years to see the effect of geometry from so-called near-field radiative heat transfer.  Heat transfer increases
when the distance between objects decreases. Motivated by the earlier 
experimental results \cite{Hargreaves69,Domoto70}, Polder and van Hove worked out a theory \cite{Polder71}, now known as fluctuational electrodynamics (FE), which in turn is  based on
the idea of Rytov of Maxwell's equations with stochastic random currents \cite{Rytov53,Rytov89}.   Together with the fluctuation dissipation theorem as a corner stone \cite{Callen51}, 
the theory is complete. The theory has been applied to a variety of situations and geometries \cite{Krueger11prl,Otey14,GTang21} and has been reviewed extensively \cite{Joulain05,Basu09,Song15,Volokitin07,Biehs21,Bimonte17,Henkel17}.  Recent experiments 
\cite{Kittel05,Shen09,Ottens11,Kim15} mostly
confirm the theory, but there remain some controversies \cite{Cui17,Kloppstech17,Tokunaga21}.  An interesting
recent development is the super-Planckian heat transport \cite{Fernandez-Hurtado18,Cuevas19}, e.g., two sheets side by side instead of face to face, where it is found that 
the enhancement of transport occurs even for the far field \cite{Thompson18}. 

Another line of research, developed in parallel to the radiation problem, is the Casimir force.  Casimir predicted, for ideal metal surfaces with a vacuum gap,
there is an attractive force between them, due to vacuum fluctuations of the field \cite{Casimir48}.  A more complete theory is developed by Lifshitz from
fluctuational electrodynamics for nonzero temperatures \cite{Lifshitz56}. The effect is small, and experimental verification was only attained much later in
the 1980s and 90s \cite{Blokland78,Lamoreaux97,Mohideen98,Obrecht07,Klimchitskaya09,Garrett18}.  These forces can be utilized for ultra-sensitive measuring devices \cite{Stange21}.   The recent development has been in the dynamic Casimir effect \cite{Wilson11,Vezzoli19,XZhang19},
where the objects can move and couple to the vacuum fluctuations.   There is a close relationship between the two problems above.   The radiation is due to
thermal fluctuations, thus, a nonzero temperature is necessary.   But the force can exist even at zero temperature.   Both problems share some common features.
Their solutions all require knowledge of the response functions of the materials and solutions to Maxwell's equations.   They differ in that 
energy radiation is related to the difference of the Bose functions $N$ at different temperatures, while the force is related to $N+1/2$, the Bose function 
with the zero-point motion contribution.   It is important that we see these connections and have a unified theory.

Another transport quantity is the angular momentum related to the rotational symmetry of the problem.  The radiation carries, in addition to energy and
momentum, angular momentum \cite{Maghrebi19,Katoh17,XGao21}.   The radiation field patterns are related to the orbital angular momentum \cite{M-Chen18}.   This can be used for transfer of information
as extra communication channels \cite{Nagali09}.  It turns out that emission of angular momentum requires a broken time reversal symmetry.  
More generally, nonreciprocity has been an emerging area of active research \cite{Asadchy20,Khandekar21}.  Both the momentum and angular momentum radiation of a single
object require nonreciprocity.   

In the theoretical and computational approaches for radiative heat and momentum transfers, earlier theories have been developed for the special geometry
of parallel plates with solutions to the scattering problems.   For arbitrary geometries, 
Messina {\it et al.} \cite{Messina11EPL,Messina11PRA}, and 
Kr\"uger {\it et al.} \cite{Krueger11prl,Krueger12prb} developed a formalism based on the $T$ operators
and Lippmann-Schwinger equation \cite{Lippmann50}.   The fluctuation of the electromagnetic field can also be viewed as coming from dipoles \cite{ben-Abdallah11}.   A general
theory for many objects is developed and recently reviewed \cite{Biehs21}.   Other computational techniques are developed from the continuum engineering perspectives \cite{Rodriguez11,Rodriguez12}.  

In this review, we systematically develop a nonequilibrium Green's function approach with the aim of going beyond fluctuational electrodynamics (FE). 
The nonequilibrium Green's function has been the standard approach to study mesoscopic electron transport \cite{Datta95,DiVentra08}, and also, to some extent, phononic
transport in the ballistic regime \cite{Wang-review-1,Yu-lifa21}.   For some reason, the use of it for photon transport is rather scarce \cite{Janowicz03,Aeberhard11}.    The nonequilibrium Green's function method is 
a powerful tool to study transport as transport, by its very nature, must deal with nonequilibrium situations.   Since Maxwell's equations
are linear, the associated equations for the Green's functions are also linear.  This implies a formally exactly solvable problem, just like in the free electron
or ballistic phonon case.   The solutions are encapsulated as a pair of equations for two Green's functions, namely, retarded and lesser \cite{Haug08}.   The retarded Green's function satisfies
a Dyson equation, $D^r = v + v \Pi^r D^r$, where $v$ is the free Green's function of photons when the materials are absent.  
This equation describes the ``dynamics''.  A related equation, $D^< = D^r \Pi^< D^a$, known
as the Keldysh equation, describes the thermal distribution.  Finally, the transported physical observables can be expressed in terms of 
$D$ and materials properties $\Pi$ known as the Meir-Wingreen formulas.
Here $\Pi$ is the charge-charge or current-current correlation under a
random phase approximation.     
The advantage of this language and methodology is that it is fairly general:  1) it is fully quantum-mechanical with quantum electrodynamics; 2) local equilibrium 
or not, where
the fluctuation-dissipation theorem may or may not hold,  the theory remains the same; 3) reciprocal or not, the formalism applies to both; 
4) the nonequilibrium system is set up explicitly by modelling the electrons connected to fermionic baths; 5) change of materials properties $\Pi$ due to extreme proximity can be
handled self-consistently via many-body formalism.   These are not in the usual spirit
of FE, which requires explicit local equilibrium assumptions and sometimes explicit reciprocity.  However, if local equilibrium is assumed, we recover
the standard fluctuational electrodynamics results.   The global or local equilibrium implies the fluctuation-dissipation theorem \cite{Eckhardt84}.   In
the language of NEGF, this is,  $D^< = N(D^r - D^a)$ and $\Pi^< = N(\Pi^r - \Pi^a)$,  
in a concise form for the field correlation and materials properties. 

Unlike FE formulated in a gauge independent manner in the electric field $\bf E$ and magnetic induction $\bf B$, the NEGF theory is formulated
in terms of the gauge dependent scalar and vector potentials.   This is necessary for a more fundamental theory as quantization requires a 
specification or choice of gauge.   Our review consists of three parts.   Part I is a theory based purely on Coulomb interaction for the
electrons, which we can also formulate as a scalar field theory.   In part II, the quantization for the vector field in transverse gauge
is presented.  This is a standard choice in condensed matter physics and quantum optics.   In the last part III, we develop a full theory in
the temporal, or also called axial gauge, which sets the scalar field to zero.   The last choice is more economical, and is very closely related to the  
gauge independent FE.  For example, the Green's function here is the same as the dyadic Green's function in FE up to a constant. 
In order not to make the paper too long, we do not attempt to review the basics of NEGF.   Reader unfamiliar with the general NEGF
method should consult the relevant literature cited \cite{Keldysh65,Haug08,Wang-review-1,Wang-review-2}.   We believe that the NEGF approach is not
reinventing the wheel, but rather, opens up new avenues to study transport in more general settings.

\newpage

\part{Scalar photons}

The electromagnetic field, in the limit that the speed of light $c$ goes to infinity, reduces to a rather simple theory of
the Poisson equation, $-\nabla^2 \phi  = \rho/\epsilon_0$, relating charge density and scalar potential $\phi$ instantaneously.   We shall
call such an approximation to electrodynamics the non-retardation limit.   This is not a bad approximation; in fact the whole of electronic
structure theories, such as the density functional theory, and the bulk of condensed matter physics, are based on such Coulomb interactions.   When the typical distance is much smaller compared to $c\tau$, where $\tau$ is some characteristic time scale, 
the non-retardation limit is a good approximation.   For heat transport mediated by electromagnetic fields, this distance scale
is the thermal de Broglie wavelength for the massless photon, $\lambda \approx c\hbar/(k_B T)$.
Here the time scale is identified with the thermal energy scale.  At 300\,K it is about a few micrometers.  In this first part, 
we discuss theories for the pure Coulomb problems which is the dominant mechanism of near-field heat transport \cite{Mahan17} at  sub-micron distances.  

\section{Heat transfer from capacitor physics\label{sec-capacity-physics}}
Consider a collection of good conductors, each of which is characterized by a constant potential on the conductor, $\phi_i$,  and the charge
on the conductor, $q_i$.   Given the potentials on the conductors and at infinity, as well as the geometry of the setup, the Poisson equation determines the potential distribution in space everywhere, thus also determines the charges on the conductors by Gauss's law.    
For this electrostatic equilibrium and time-independent situation, the charges are related to the potentials linearly by \cite{Jackson3rd,Smolic21}
\begin{equation}
\label{eq-eq1-qv}
q_i = \sum_j C_{ij} \phi_j,
\end{equation}
where $C_{ij}$ is defined as the capacitance of conductor $i$ induced by $j$.   Symbolically, we will write the above equation as
$q = v^{-1} \phi$, where $q$ and $\phi$ denote column vectors and $v^{-1}$ is a matrix formed by the capacitance $C_{ij}$.  
Formally, the potential produced by the charges is just $\phi = v q$.  The total electrostatic energy is 
$\frac{1}{2} \sum_{i,j} C_{ij} \phi_i \phi_j$.   It is clear that the matrix must be symmetric,  $C_{ij} = C_{ji}$.   In a simple parallel plate 
capacitor, we have two conductors 1 and 2. The overall charge neutrality requires $q_1 + q_2 = 0$, and the charges should 
only depend on the potential difference due to gauge invariance  ($\phi_i \to \phi_i + {\rm const}$).   If the total charge is not zero, it 
would mean that the total electrostatic energy is infinite due to a nonzero electric field outside the capacitor, which is unreasonable.    
These conditions imply that the four coefficients of capacitance are not independent, and should
satisfy  $C_{11} = C_{22} = - C_{12} = - C_{21}$, given by the usual formula of $C=C_{11}=\epsilon_0 A/d$.  Here $\epsilon_0$ is the vacuum
permittivity, $A$ is the area of the plate, and $d$ is the distance between the plates.   This recovers the textbook definition of
capacitance by $C = q/ \Delta \phi$.  

Can the capacitor system be considered as a heat transfer system mediated by the Coulomb force?   In principle, yes; the charges
on the conductors can fluctuate and thus transfer energy.    If each conductor is maintained at a different temperature, there will be 
a net transfer of energy.   However, if the conductors are macroscopic in size, the fluctuation will be very small.   We expect
substantial fluctuations only when the system is at the nanoscale.   There is one constraint it must fulfill;  this is the total charge conservation.   
We imagine that each conductor is connected to a battery maintained at a certain temperature and chemical potential.   If we focus only
on the conductor, the charge on the conductor is not conserved as it can go into or out of the battery from time to time.   So, we need
an equation to describe this process.   We look towards the fluctuational electrodynamics \cite{Rytov53}, in the non-retardation limit.

On a phenomenological ground, we put down the following Langevin-like stochastic equation for the fluctuating charge capacitor 
system \cite{Wang18pre},
\begin{equation}
\label{eq-discrete-poisson}
v^{-1}\phi = \delta q + \Pi \phi.
\end{equation}
Here $v^{-1}$ is the capacitor matrix, $\delta q$ is a vector of the fluctuational charges, $\Pi \phi$ gives the induced charge response when
the potential is changed by $\phi$.   In the above equation, we use the symbol $\phi$ to mean the change relative to the 
static value due to the time-dependent change $\delta q$ in
charge.    The charge response is retarded, in the sense that $\Pi \phi$ depends on the history of the potential,
\begin{equation}
\Pi \phi \to  \int^t \Pi(t-t') \phi(t') dt'.
\end{equation}
Equation (\ref{eq-discrete-poisson}) generalizes the static case, Eq.~(\ref{eq-eq1-qv}), for the fluctuational charge and field in time. 
It is best to think of the stochastic equation in the frequency domain, defined by
\begin{equation}
\phi(t) = \int_{-\infty}^{+\infty} \frac{d\omega}{2\pi} \phi(\omega) e^{-i \omega t},  
\end{equation}
and similarly for $\delta q(t)$ and $\Pi(t)$, 
then the convolution in time becomes multiplication in the frequency domain.    Note that we can think of 
$\bigl(D^{r}\bigr)^{-1} = v^{-1} - \Pi$ as a frequency-dependent capacitance matrix.   It is this frequency dependence 
of the material response $\Pi$ to the field fluctuations that gives rise to energy transfer. The Green's function
$D^r = v + v \Pi D^r$ will be needed to express the energy transport.  The solution to Eq.~(\ref{eq-discrete-poisson}) is $\phi = D^r \delta q$.  

No matter what the detailed mechanisms are for the charge fluctuation, in thermal equilibrium, the charge fluctuations must be related to the 
linear response by the fluctuation-dissipation theorem \cite{Callen51,Kubo91},
\begin{equation}
\label{eq-FDT}
\frac{1}{i\hbar} \left\langle \delta q \,\delta q^T\right\rangle_\omega = \left( N(\omega)+\frac{1}{2}\right)\bigl(\Pi(\omega) - \Pi^\dagger(\omega)\bigr) = \bar{\Pi}.
\end{equation}
Here, the superscript $T$ is matrix transpose, and the dagger $\dagger$  is for Hermitian conjugate.  The left-hand side is the 
fluctuational charge correlation, 
$\bigl\langle \delta q(t) \delta q(t')^T\bigr\rangle/(i\hbar)$, which is a function of the 
time difference in steady state, Fourier transformed into $\omega$ space.   $N(\omega) = 1/\bigl( \exp(\beta \hbar\omega) - 1\bigr)$ is the Bose (or Planck) function
at temperature $T = 1/(k_B \beta)$.  
N.B. In many-body theory, the charge-charge correlation is the susceptibility $\chi = \Pi + \Pi v \Pi  + \cdots$  \cite{Giuliani-Vignale05}, but here it is the
polarizability $\Pi$ in the fluctuational electrostatics.  The reason is that $\phi$ here is the total actual field on the matter.  
As we will see, this is consistent with the Dyson equation and the Keldysh equation
in a more rigorous treatment later.    For the moment, we consider
$\Pi$ as a phenomenological function that is given.  Subsequently, when we build a microscopic (quantum) model, its functional
form can be worked out.   Here with this minimum information, how can we describe the transfer of energy among the conductors 
when they are out of equilibrium?

In order not to deviate too much from thermal equilibrium, we assume local equilibrium.   This is possible if the charge-charge 
correlation can be localized so that different regions are not correlated, and each region has its separate temperature and chemical
potential.   To be definite, let us partition the conductors as belonging to the left ($L$) or right ($R$) such that $\Pi$ is block-diagonal,
\begin{equation}
\Pi = \left( \begin{array}{cc} 
\Pi_L & 0 \\
0 & \Pi_R
\end{array}
\right).
\end{equation}
The fluctuation-dissipation theorem, Eq.~(\ref{eq-FDT}), remains the same, except that it is applied to the left sites or right sites separately with $T_L$ or $T_R$,  
and the charge correlation is zero between the left side and right side. 

Next, we consider the average energy transfer from left to right, based on Joule heating, and the local fluctuation-dissipation theorem.   
 On a continuum, ${\bf j} \cdot {\bf E}$ is the work density done by the field to the charge. Using the relation between the field and 
potential ${\bf E} = - {\nabla} \phi$,  the continuity equation $\dot{\rho} + \nabla \cdot {\bf j} = 0$ for charge conservation,  and an integration by parts, we also have the work as $- \dot{\rho} \phi$ per unit volume \cite{Yu17}.    For discrete charge, this is $-\dot{q}^T \phi$.  Since Eq.~(\ref{eq-discrete-poisson}) is linear in $\phi$, we can consider the effect of random noises of two regions separately.  Turning off $\delta q_R$, the energy transfer  to the right side per unit time 
due to the fluctuation of charge $\delta q_L$ of the left side is
\begin{equation}
\label{eqqdotphi}
I_{L\to R} = -\langle \dot{q}_R^T \phi_R\rangle,
\end{equation}
where $q_R = \Pi_R \phi_R$ and $\phi_R = D^{r}_{RL} \delta q_L$.  
Here we collect all the values of discrete potentials on the right side as a column vector $\phi_R$.   The charge $q_R$ in Eq.~(\ref{eqqdotphi}) is the induced one due to the field.  
These are time domain quantities,
for example, 
\begin{equation}
\phi_R(t) = \int D_{RL}^r(t-t') \delta q_L(t') dt'.
\end{equation}
This is the solution to Eq.~(\ref{eq-discrete-poisson}) restricted to the right side for the potential. 
We assume that the system is in steady state and $I_{L\to R}$ is in fact independent of time.
Representing all the time domain quantities by their Fourier transforms in frequency domain, after
some lengthy but straightforward algebra, we find
\begin{equation}
\label{eq-I1to2}
I_{L\to R} =  \int_{-\infty}^{+\infty} \frac{d\omega}{2\pi} \hbar \omega
{\rm Tr}\bigl(D^a_{LR} \Pi^a_R D^r_{RL} \bar{\Pi}_L  \bigr).
\end{equation}
Here ${\rm Tr}$ stands for matrix trace, $\Pi^a_R = (\Pi_R)^\dagger$, and $D^a_{LR} = (D^r_{RL})^\dagger$. 
The last factor is due to the noise correlation by the local fluctuation-dissipation theorem.  

The energy pumped from $R$ to $L$ by $\delta q_R$ can be obtained similarly by swapping the
index $L \leftrightarrow R$.   The overall net heat current from left to right is given by the difference,
$I_L = I_{L\to R} - I_{R\to L}$.  The expression can be simplified using the fact that
(1) $I_{L\to R}$ and $I_{R\to L}$ are real, so that we can take the Hermitian conjugate of
the factors inside the matrix trace and add them, then divide by 2; (2) we can perform cyclic permutation
under trace; (3)  both $D^r$ and $\Pi$ are symmetric matrices, thus, e.g.,
$D^a = (D^r)^\dagger = (D^r)^{*}$.  [This last condition of reciprocity is not really necessary;
we can replace it by the fact that when all the baths are at the same temperature, the heat
transfer must be zero, to be consistent with the second law of thermodynamics.]
(4) The integrand is even in $\omega$.    
With these manipulations, the expression can be
simplified to a standard Caroli form, 
\begin{equation}
\label{eq-landauer-IL}
I_{L} =  \int_{0}^{\infty} \frac{d\omega}{2\pi} \hbar \omega
 {\rm Tr}\bigl(  D^r_{LR} \Gamma_R  D^a_{RL} \Gamma_L  \bigr) (N_L - N_R).
\end{equation}
Here we define $\Gamma_L = i (\Pi_L - \Pi_L^\dagger)$ and similarly for $\Gamma_R$, and
$N_{L}$ is the Bose function at temperature $T_L$, and $N_R$ is at $T_R$.  We shall call this
result the Landauer/Caroli formula, although the original formula was for mesoscopic electron 
transport \cite{Landauer57,Caroli71,Datta95}. 

Let us consider the simplest possible case of a parallel plate capacitor represented as two dots \cite{capacitors} with
the Dyson equation in the form
\begin{equation}
\left( D^r \right)^{-1} = 
\left( \begin{array}{rr} 
C & -C \\
-C & C
\end{array}
\right) 
-\left( \begin{array}{cc} 
\Pi_L & 0 \\
0 & \Pi_R
\end{array}
\right),
\end{equation}
where $C = \epsilon_0 A/d$ is the capacitance of the parallel plate capacitor.   The retarded Green's
function can be solved explicitly, given 
\begin{equation}
D^r = \frac{1}{\Pi_L \Pi_R - (\Pi_L + \Pi_R)C}  
\left( \begin{array}{cc} 
C-\Pi_R & C \\
C & C-\Pi_L
\end{array}
\right)
\end{equation}
as a $2 \times 2$ matrix.   With this, the heat transfer takes the Landauer form, Eq.~(\ref{eq-landauer-IL}), 
with the transmission function given by 
\begin{equation}
\label{eq-transmission}
{\rm Tr}\bigl(  D^r \Gamma_R D^a \Gamma_L  \bigr)  
= { 4 C^2 {\rm Im} \Pi_L\, {\rm Im} \Pi_R \over
\bigl| \Pi_L \Pi_R - (\Pi_L + \Pi_R) C \bigr|^2 }.
\end{equation}
This formula tells us, at long distances, the heat transfer is proportional to $1/d^2$ and is constant for short 
distances, simply because the capacitance $C \propto 1/d$.  The crossover distance of the two types of behaviors is controlled
by the energy scale $E_\alpha \sim e^2/C$ (see Eq.~(\ref{eq-Pi-alpha}) below). 
The distance dependence behavior also appears in many more realistic systems, such as between two graphene 
sheets \cite{jiang17,zhu1}.

\begin{figure}
  \includegraphics[width=\columnwidth]{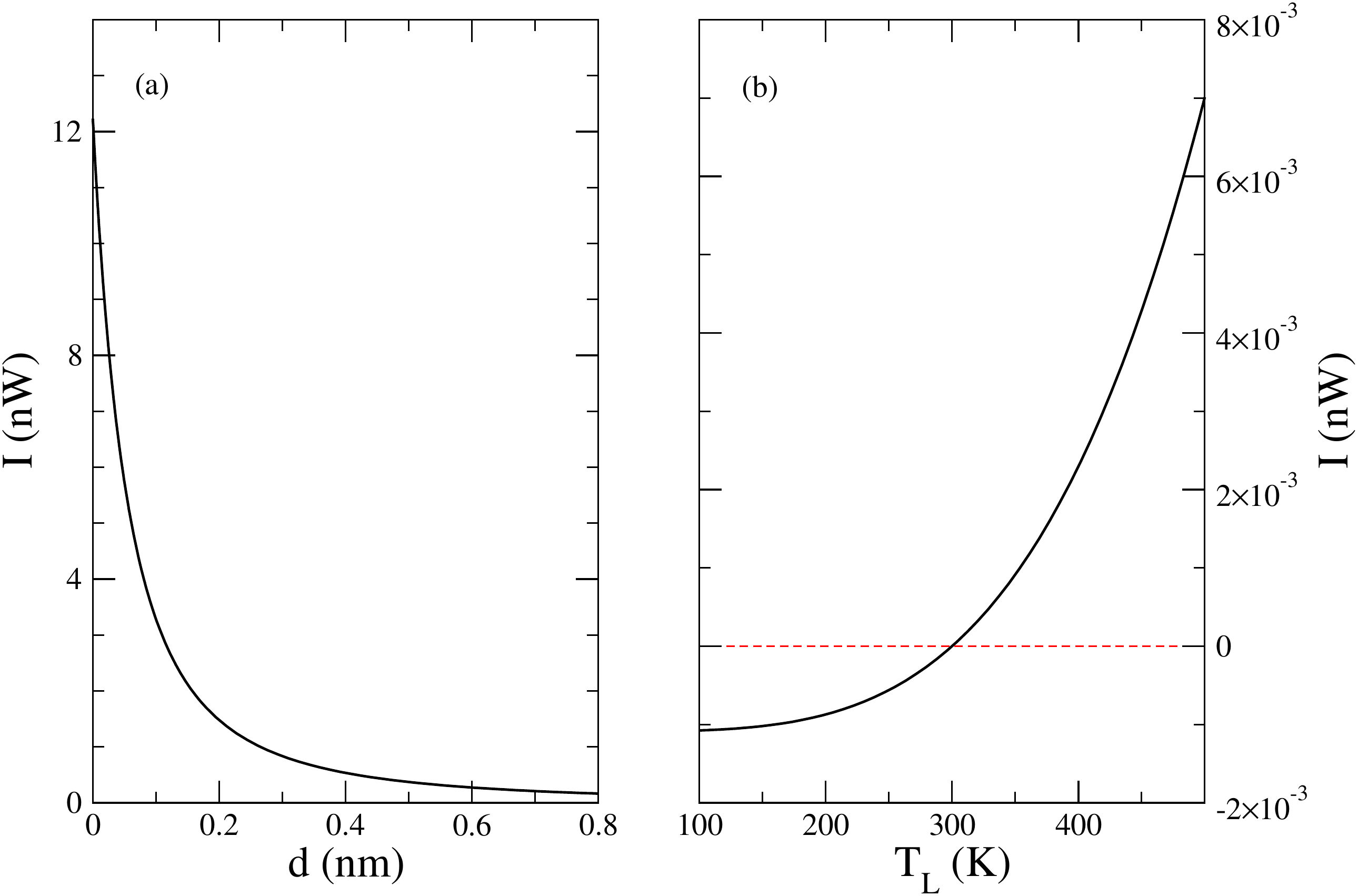}
  \caption{Heat current of the simple capacitor model, from Eq.~(\ref{eq-landauer-IL}), (\ref{eq-transmission}), and (\ref{eq-Pi-alpha}).  (a) Current  $I_L$ vs.\ 
  distance $d$ at $T_L = 1000\,$K and $T_R = 300\,$K.  (b)  The current $I_L$ vs.\ $T_L$ while fixing $d=1\,$nm, $T_R = 300\,$K.  Parameters are area 
  $A=1\,$nm$^2$,  energy scale $E_L = E_R = U_L = U_R = 1\,$eV. }
 \label{fig-capacitor}
\end{figure} 

A simple model for $\Pi_\alpha$ with $\alpha = L,R$ can be given based on the analytic properties over frequency $\omega$.  
The polarizability $\Pi_\alpha$ is from a correlation of real quantities, thus the real part must be symmetric in $\omega$ and 
the imaginary part must be antisymmetric in $\omega$, so for low enough thermal energy $k_B T$, we can take
\begin{equation}
\label{eq-Pi-alpha}
\Pi_\alpha = - \frac{e^2}{E_\alpha} - i \frac{ e^2 \hbar \omega}{U_\alpha^2}, 
\end{equation}
where $e$ is the magnitude of electron charge,  $E_\alpha$ and $U_\alpha$ are some constants of order eV.   This assumption for $\Pi_\alpha$ leads to a robust Stefan-Boltzmann law of $T^4$ with 
a coefficient that depends on the details.    Fig.~\ref{fig-capacitor} illustrates the results of heat current as a function of distance and
temperature for the simple capacitor model. 

Is the assumption that the left dot and right dot are uncorrelated realistic?   If the capacitor is connected to the same battery, the individual 
fluctuations of the charges on the two plates seem to violate charge conservation.   However, if the systems are more complex than two dots, the charge conservation can be fulfilled by moving the charge on the same side from one place to another.   In fact, the
derived Landauer/Caroli formula is generally valid for a tight-binding model of a system with $v_{ij} = 1/(4\pi\epsilon_0 r_{ij})$ as
the bare Coulomb interaction between two electrons at site $i$ and $j$ with a distance $r_{ij}$.   Here a conductor becomes a hopping site.  Although the validity to a two-dot capacitor can be doubtful,
the validity of the theory for general electron systems with two sides not directly connected can be proved rigorously.   This will be the subject of the 
next section.

\section{Coulomb interaction model\label{sec-coulomb-interaction}}
In this section, we consider the following model of $N$ pieces of metal objects described by a free
electron Hamiltonian $c^\dagger H c = \sum_\alpha c_\alpha^\dagger H_\alpha c_\alpha$, plus a Coulomb interaction.   
We assume that each object labeled by $\alpha$ is not directly connected to other objects such that electric currents are absent so 
that we can focus on the energy (or heat) transport among the objects.   This means that the matrix $H$ is block-diagonal.  
An example of such a system is two graphene sheets with a vacuum gap. 
Although the electrons cannot jump from
one object to another, they can still exchange energy through a Coulomb interaction.  We take the electron to be
spinless, and the Coulomb term is
\begin{equation}
\label{eqqq-Coulomb}
\frac{e^2}{2} \sum_{i,j} \frac{1}{4 \pi \epsilon_0 |{\bf r}_i -{\bf r}_j|} c^\dagger_i c^\dagger_j c_j c_i. 
\end{equation}
Here $i$ and $j$ run over all the sites of all the objects.  This  is just the standard sum of Coulomb interaction between two charges
at the position ${\bf r}_i$ and ${\bf r}_j$ in the second quantization notation, excluding
the self-interaction terms. 

If our system of interest is finite, it is impossible to maintain a local thermal equilibrium and a steady-state heat transport.
To realize heat transport, we connect each object to a bath so that an infinite amount of energy can be
supplied.   This is done in the nonequilibrium Green's function approach (NEGF) \cite{Haug08} by giving a self-energy of the bath to the degrees of the particular object
while the bath is maintained in thermal equilibrium at temperature $T_\alpha$ and chemical potential 
$\mu_\alpha$.   Thus, the object $\alpha$ before turning on the Coulomb interaction and in local thermal equilibrium is described by
two Green's functions, the retarded one, according to
\begin{equation}
G^r_\alpha(E)= \bigl( (E + i \eta)I - H_\alpha - \Sigma^r_\alpha\bigr)^{-1},
\end{equation} 
where $I$ is the identity matrix, $\eta >0$ is an infinitesimal quantity describing the electron relaxation,  and the lesser Green's function 
describing the electron occupation,
\begin{equation}
G^{<}_\alpha(E) = - f_\alpha(E) \bigl( G^r_\alpha(E) - G^a_\alpha(E) \bigr),
\end{equation}
where $G^a_\alpha = \left( G^r_\alpha\right)^\dagger$ is the advanced Green's function, and 
$f_\alpha(E) = 1/\bigl( e^{\beta_\alpha (E-\mu_\alpha)} + 1 \bigr)$ is the Fermi distribution function of the 
bath connected to object $\alpha$,  $\mu_\alpha$ is the chemical potential.   If the object is not in thermal equilibrium, e.g., one object connected to two baths, we need to use 
the Keldysh equation $G^< = G^r \Sigma^< G^a$ instead \cite{Keldysh65}.  The electron Green's functions are our starting point to characterize 
the materials properties, such as the polarizability $\Pi$.

We now derive the Meir-Wingreen formula \cite{Meir92,Jauho94} for the energy current which is an exact formula
if the Coulomb problem can be solved exactly.   To derive the equation, we first make a notational
simplification and consider only one bath and one object with the one-particle Hamiltonian
\begin{equation}
H  = H_d + V = \left( \begin{array}{cc} 
H^O &  0 \\
0 & H^B
\end{array}
\right)  + 
 \left( \begin{array}{cc} 
 0 & V^{OB} \\
V^{BO} & 0
\end{array}
\right),
\end{equation}
here $H^O$ is the Hamiltonian of the object, and $H^B$ is the isolated bath, and $V$ is the coupling
between the object and the bath.   The energy transfer per unit time out of the bath is obtained by the average decrease of the bath
Hamiltonian, 
\begin{eqnarray}
I_B &=& - \left\langle \frac{d \hat{H}^B }{dt}\right\rangle =  \frac{1}{i \hbar} \left\langle [\hat V, \hat H^B ]\right\rangle \nonumber\\
&=& \left\langle c^\dagger V^{OB} \dot{d} + {\dot{d}}^\dagger V^{BO} c \right\rangle.
\end{eqnarray}
Here we use a convention that the hatted operators are in second quantization notations with the creation operators multiplied 
from the left and the annihilation operators multiplied from the right to the one-particle matrix elements of the same letter without
the hat, e.g., $\hat{H}^B = d^\dagger H^B d$.   We use $c,c^\dagger$ for the object and $d,d^\dagger$ for the bath, all of these are column or row vectors.
We have used the Heisenberg equation for the rate of changes.   Here the dot means evolution by $\hat{H}^B$ only, holding the system fixed.  We can replace it
by the full evolution of the full Hamiltonian, including a possible Coulomb interaction at the dot or center (but not at the bath or 
between dot and bath), we obtain
\begin{equation}
\dot{d} = \frac{d\,d}{dt} - \frac{1}{i\hbar} V^{BO} c.
\end{equation}
Here $d/dt$ represents the full evolution.  Similarly, we obtain $\dot{d}^\dagger$ by Hermitian conjugation.  Putting this result
into the energy current expression, we see that  $c^\dagger V^{OB} V^{BO} c$ terms cancel, we obtain
\begin{equation}
I_B = \left\langle   c^\dagger V^{OB} \frac{d\,d}{dt} + \frac{d\,d^\dagger}{dt} V^{BO} c\right\rangle. 
\end{equation}
This expression can be further expressed in terms of the lesser Green's function connecting the bath with
the object,  $G^{OB,<}_{kj}(t,t') = -  \bigl\langle  d^\dagger_j(t') c_k(t)\bigr\rangle/(i \hbar)$, or the reversed
version $G^{BO}$.  A key step is to remove the reference to the bath variables by using the Green's function
of the object only.   It will have taken us too long here to repeat this argument, thus we refer to the standard
NEGF textbooks \cite{Haug08,Stefanucci13} on this, given then by the so-called Langreth rule \cite{Langreth76,Wang-review-1}
\begin{equation}
G^{BO,<} = g_B^r V^{BO} G^< + g_B^< V^{BO} G^a,
\end{equation}
here the small $g_B$ is the Green's function of the bath when it is isolated from the object, and capital $G$ is
the Green's function of the object when it is interacting with the bath (as well as internal Coulomb
interaction and other baths not in our focus).   The product of two Green's functions means a convolution
in time domain and a multiplication in the energy domain.   Then we have \cite{lv07}, using the Fourier representation of the
Green's functions at time $t=0$,
\begin{eqnarray}
I_B &=&  - i \hbar \frac{\partial\ }{\partial t} 
{\rm Tr} \left(V^{OB} G^{BO,<}(t) - G^{OB,<}(t) V^{BO}  \right)\Big|_{t=0} \nonumber \\  
&=&  \int_{-\infty}^{+\infty} \frac{dE}{2\pi \hbar} E\, {\rm Tr} \left(
 (G^r - G^a) \Sigma_B^<  -G^< (\Sigma^r_B - \Sigma^a_B) \right) \nonumber \\
&=& \int_{-\infty}^{+\infty} \frac{dE}{2\pi \hbar} E\, {\rm Tr} \Bigl( G^> \Sigma_B^< 
- G^< \Sigma_B^> \Bigr). \label{MWeq1}
\end{eqnarray}
We have used the fact that the bath self-energy is related to the isolated bath Green's function
by $\Sigma_B = V^{OB}g_B V^{BO}$, and used the relations among Green's functions that
$G^r  - G^a = G^> - G^<$, also valid for the self-energies.   For a multiple-lead (or bath)
setup, we simply replace $B$ by a more general index $\alpha$ for the bath or object $\alpha$.
The formalism is also valid if we connect several baths to the same object, in order to establish
a situation that is not in local equilibrium for the object.  

As we can see from the derivation, we have not used specific properties of the objects other than
the fact that the couplings between the object and bath are bilinear in the creation and annihilation 
operators.  As a result, the Meir-Wingreen formula is valid, having the Coulomb interaction or not. 
However, the Coulomb interaction is one of the
hardest problems in condensed matter physics, and it cannot be solved exactly.   Here we use the 
simplest and also standard approximation for the Coulomb interaction self-energy
$\Sigma_n$ by the Fock term and for the electron screening $\Pi$ by the random phase approximation 
(RPA) \cite{Bohm53}.  
The main point below is to get rid of the electron Green's functions, and to relate the energy current to the 
(scalar) photon Green's functions and polarizability.   From this, we obtain a similar Meir-Wingreen formula in terms of
the photon Green's function $D$ and polarizability $\Pi_\alpha$ of object $\alpha$.   Further approximation
with the local thermal equilibrium gives the Landauer/Caroli formula.  
The Meir-Wingreen form is more general as no local equilibrium is assumed (although
each object still needs to be connected to one or more baths, the bath is in thermal equilibrium).

To make progress with the Coulomb problem in energy transport,  one further approximation is needed, 
which is the lowest order expansion approximation \cite{Paulsson05} in
terms of the Coulomb interaction.  Such approximation preserves energy conservation exactly.  An alternative
to this is the self-consistent Born approximation through iterative solutions; we will make a comment on this latter approach 
at the end of this section.  
For notational simplicity, we will here consider two-bath situation with two objects, call them
$L$ and $R$. The Green's functions for the electrons and self-energies 
are block diagonal since two sides are not directly connected, and the Meir-Wingreen formula needs only the block $\alpha$.  
We focus on the object $L$, and Green's function $G_{L}^>$ can be expressed by the Keldysh equation as
\begin{equation}
G_L^> = \bigl[ G^r( \Sigma_L^> +\Sigma_R^> + \Sigma_n^> ) G^a \bigr]_{LL}.
\end{equation}
Here $\Sigma_{L,R}^>$ are the lead self-energies, and $\Sigma_n^>$ is the Fock term of Coulomb
interaction.
For the $LL$-subblock, $\Sigma_R$ is 0 since bath $R$ is not directly connected to object $L$.  
This expression requires the knowledge of the full retarded Green's function $G^r$ with Coulomb interaction.
We prefer to work with the free electron Green's functions.
An approximation we can use is the lowest order expansion,
\begin{equation}
\label{LOEeq}
G^> \approx G_0^> + G_0^r \Sigma_n^r G_0^> + G_0^r \Sigma_n^> G_0^a + G_0^> \Sigma_n^a G_0^a.
\end{equation}
We obtain such terms if we expand the contour ordered Dyson equation, $G = G_0 + G_0 \Sigma_n G
\approx G_0 + G_0 \Sigma_n G_0 + \cdots$, and then take the greater component using the 
Langreth rule.  The subscript $0$ denotes the left system that is free of Coulomb interaction, i.e., a perfect 
ballistic system with quadratic Hamiltonian $c^\dagger H c$ (with baths).   We drop the subscript 0 from now on.

\begin{figure}[htp] 
  \centering
  \includegraphics[totalheight=55mm]{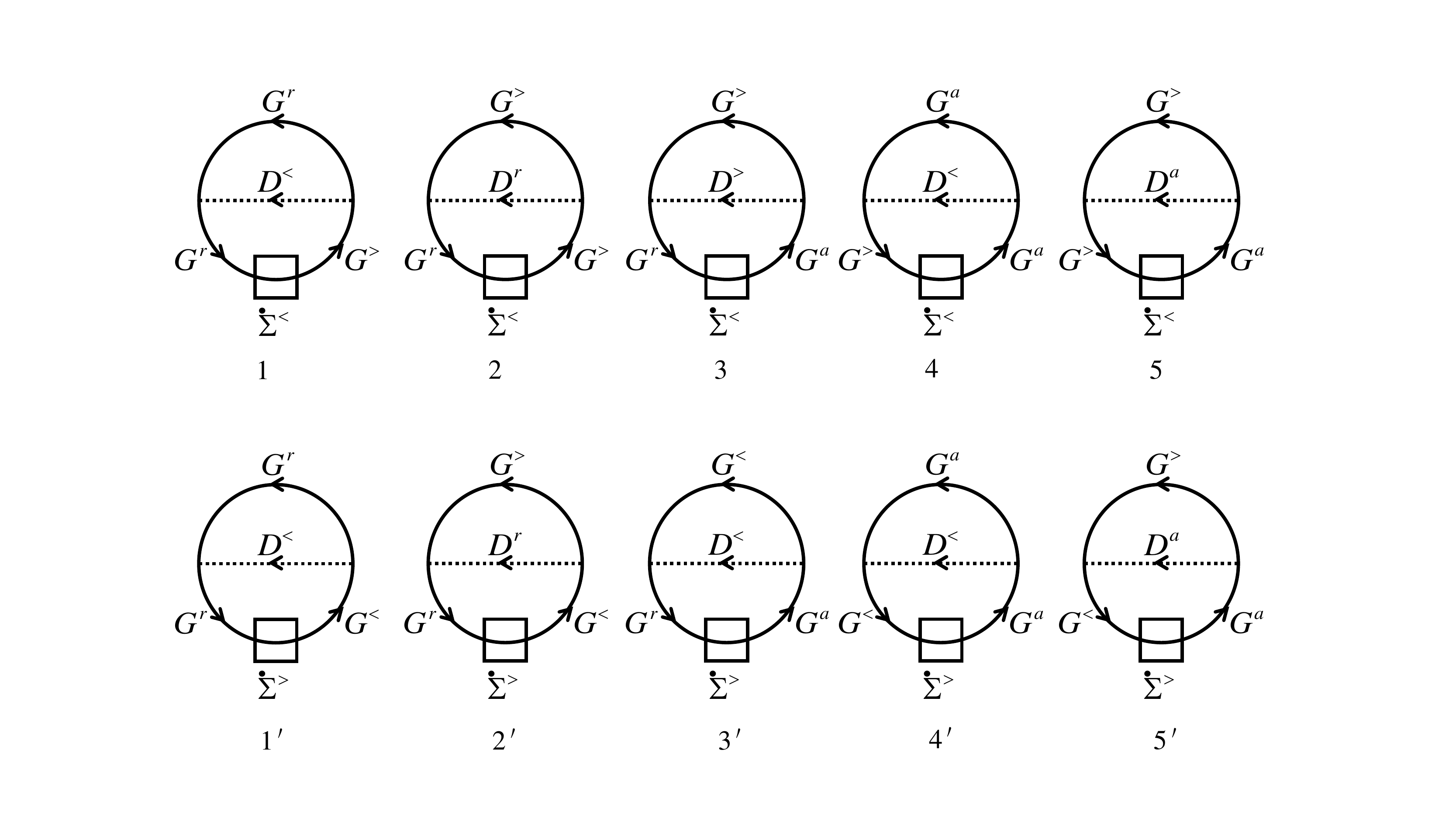}
  \caption{Diagrams for the heat current in the lowest order expansion.  The solid lines denote the electron Green's functions,
  and the dash lines for the photon Green's functions.  The boxes denote derivative of the bath self energy.  The end of the arrow 
  is associated with the first argument and beginning of the arrow with the second argument of the Green's functions or self energies.  
  The electron-photon vertex is associated with the matrix $M^l$.}
 \label{feynman-Diagrams}
\end{figure} 

It is useful for symmetry reason we express the current by vacuum diagrams in time domain.  
We use the inverse Fourier transform to change the integral in energy to time, and also the Fock diagram result \cite{lv07,Dash10}, 
\begin{equation}
\Sigma_n^>(t,t') = i \hbar \sum_{l,l'} M^l G^>(t,t') M^{l'} D_{ll'}^>(t,t').   
\end{equation}
A matrix multiplication, $M G M$, is implied in the electron index space.  Similar expressions are given for the retarded and advanced
self-energies as $\Sigma_n^r \propto G^r D^<  + G^> D^r $, and $\Sigma_n^a \propto 
G^a D^<  + G^> D^a $.
Here for generality, we assume that the interaction bare vertex
takes the form $\sum_{l} c^\dagger M^l c \phi_l$, where $c$ is a column vector of electron annihilation operators 
and $c^\dagger$ is a row vector of Hermitian conjugate, and $\phi_l$ is a scalar field at site $l$, and $M^l$ is a Hermitian matrix.  
We will explain the scalar field $\phi$ in more detail in the next section.   Here it is sufficient to know that the electron
Fock diagram is given by $GD$ (in electron many-body theory $D = v + v \Pi D$ is called 
screened Coulomb $W$ where $v$ is the bare Coulomb interaction).   We have also ignored the Hartree diagram
as the Hartree term only shifts the energy levels of single-particle Hamiltonian and does not contribute to transport
in our order of approximation.
By plugging Eq.~(\ref{LOEeq}) into (\ref{MWeq1}), 
the expansion leads to 10 terms, represented
by the 10 diagrams in Fig.~\ref{feynman-Diagrams}.  We will label these diagrams as 1 to $5$, and $1'$ to $5'$, as shown. The diagrammatic rule follows the usual convention with all the (real) times
as dummy integration variables and space indices summed.   The current is obtained by
$(i\hbar)^2/T$ times the value of the diagram.  Since all the times are integration variables on an equal footing, 
the integral diverges due to time translational invariance, the $1/T$ factor cancels the last integral interpreted as $\int_{-T/2}^{T/2} dt \cdots$.   As 
an example, the graph 3 represents the contribution to current as
\begin{eqnarray}
3) &=& \frac{(i\hbar)^2}{T} \int dt dt' dt_1 dt_2 \sum_{l,l'} D_{ll'}^>(t,t') \times \\
 &&{\rm Tr}\left[ M^l G^{>}(t,t') M^{l'} G^a(t',t_1) 
\frac{\partial \Sigma^<(t_1, t_2)}{\partial t_1}  G^r(t_2, t) \right]. \nonumber
\end{eqnarray}
Note the partial derivative on the first argument of $\Sigma^<$, which is represented by a dot in the diagram.
The partial derivative can  be moved around with repeated integration by parts.  The trouble with this 
expression is that $G^a \Sigma^< G^r$ is in the wrong order to be a $G^<$.  

A crucial identity \cite{Datta95},
\begin{eqnarray}
G^r( \Sigma^> - \Sigma^< ) G^a &=&  
G^a( \Sigma^> - \Sigma^< ) G^r \nonumber \\
\label{eq-Datta-identity} &=& G^r - G^a = G^>  - G^<, 
\end{eqnarray}
is needed to show that the 10 diagrams cancel among themselves and reduce to only two with the
correct order.   Here the self-energies are total 
 lead self-energy (for object $L$ only).  This identity is a simple consequence of the Dyson equation
 $(G^r)^{-1} = (g_c^r)^{-1} - \Sigma^r$, where $g_c^r$ is the Green's function of the isolated center.
 From the above equation we can show that
 \begin{equation}
 G^a \Sigma^{<} G^r = G^{<}  + C,
 \end{equation}
here we define $C = G^a \Sigma^{>,<} G^r - G^r \Sigma^{>,<} G^a$, and is the same for greater and lesser
components.   $C$ is anti-Hermitian, $C^\dagger = - C$.  $C=0$ if matrices are actually 1 by 1, or
if the system is reciprocal (i.e., $(G^r)^T = G^r$, $(G^<)^T=G^<$), but not so in general.
From this, ignoring the proportionality constant, integration variables, and $M$ factors, we can
write, symbolically, 
\begin{equation}
\label{eq-delta3}
\Delta 3 + \Delta 3') = {\rm Tr} \bigl[ (D^> G^> - D^< G^< ) C\bigr].
\end{equation}
Here the notation $\Delta$ means that the term when $G^a$ and $G^r$ are swapped to form $G^>$ or 
$G^<$ has been subtracted off.   We show that Eq.~(\ref{eq-delta3})  cancels all the other 8 diagrams. 
To this end, we define
\begin{equation}
B = G^> \Sigma^< - G^< \Sigma^>.
\end{equation}
Using the same identities, we have $B G^r = - C$, thus $BG^r - G^a B^\dagger = -2C$, and
$B G^r + G^a B^\dagger = 0$. 

We can factor out common factors in the remaining diagrams.  Using $B$, we can write
\begin{eqnarray}
1\!+\!1') + 2 \!+\!2') &=& D^< {\rm Tr}(G^r B G^r) + D^r {\rm Tr}(G^> B G^r), \nonumber \\
4\!+\!4') + 5\! +\!5') &=& D^< {\rm Tr}(G^a G^a B^\dagger) + D^a {\rm Tr}(G^> G^a B^\dagger).\nonumber
\end{eqnarray}
Further simplification is possible because
\begin{equation}
D^< G^r + D^r G^> = D^> G^> - D^< G^< + D^< G^a + D^a G^>. \nonumber 
\end{equation}
Now, putting all the terms together, and using the identities obtained, we see
$\Delta 3 + \Delta 3')$ cancels all the rest as claimed.

The remaining two terms in the correct $G^r \Sigma^{>,<} G^a$ order can be transformed into the desired form.  
First, we need to move the derivative to other places, for example, from graph $3  - \Delta 3)$, we can write
\begin{equation}
- D^>(t,t') {\rm Tr}\bigl[ G^>(t,t') {\partial \over \partial t} G^<(t',t) \bigr].
\end{equation}
The extra minus sign is due to an integration by parts.   We can combine a similar term
from $3' - \Delta 3')$ so that it becomes $\partial \Pi^<(t',t)/\partial t$, using integration by parts
and cyclic permutation of trace whenever needed.  We define the polarization or charge-charge correlation as 
\begin{eqnarray}
\Pi^{<}_{l'l}(t',t) &=& \frac{1}{i\hbar} \bigl\langle q_{l}(t) q_{l'}(t')  \bigr\rangle \nonumber \\
\label{eq-RPA-pi}
&=& - i\hbar\, {\rm Tr}\bigl[  M^{l'} G^<(t',t) M^{l} G^{>}(t,t') \bigr], 
\end{eqnarray}
where the charge operator is $q_l = c^\dagger M^l c$, and the second line is obtained 
assuming that Wick's theorem \cite{Fetter71,Ford65} is valid.   This is called RPA.  $\Pi^>$ is obtained by swapping the positions of the charge
operators and by swapping $G^< \leftrightarrow G^>$.  
Fourier transforming the final expression to frequency domain, we obtain \cite{Peng1805arxiv}
\begin{equation}
\label{eq-photon-IL-WM}
I_L =- \frac{1}{4\pi} \int_{-\infty}^{+\infty}\!\!
d\omega\, \hbar \omega\, {\rm Tr} \bigl(D^> \Pi^<_L - D^< \Pi^>_L \bigr).
\end{equation}

The derivation under the lowest order of expansion for the electron Green's function is rather long and complicated. 
In fact, if we take a self-consistent Born approximation \cite{lv07} (also known as the self-consistent $GW$ method), we obtain
the same result, with $\Pi$ being computed by a self-consistent $G$.   To this end, we note that since we do not
allow the electrons to jump from one object to another, the Green's functions and electron self-energies are block
diagonal.  We can treat each as a separate system, although they are connected through $D$.  As a result, 
we have the following identity \cite{Haug08},
\begin{eqnarray}
{\rm Tr}\left(G^>_\alpha \Sigma^{<,\rm tot}_\alpha - G^<_\alpha\Sigma^{>,\rm tot}_\alpha   \right) = 0, \\
\Sigma^{>,<\rm tot}_\alpha = \Sigma^{>,<}_{\alpha} + \Sigma^{>,<}_{n,\alpha}.
\end{eqnarray} 
Since the isolated center is non-dissipative, the equation reflects the 
conservation laws.   The self-energy is the total, including baths and nonlinear term, for object $\alpha$.  
Using this identity, starting with the Meir-Wingreen formula in $G$ and bath self-energy $\Sigma_L$, Eq.~(\ref{MWeq1}),
we can replace the bath self-energy with the nonlinear Coulomb self-energy, keeping $G^>$ or $G^<$
as it is.   The next step is to use the self-consistent Fock term to replace the nonlinear self-energy.  This results
in a double integral in energy $E$ and frequency $\omega$.   With a change of variable,  such as $E \to 
E - \hbar \omega$, and the symmetry of the Green's function $\bigl( D^>(\omega) \bigr)^T = 
D^<(-\omega)$, we can rewrite the result in the form of Eq.~(\ref{eq-photon-IL-WM}).

Finally, if we can assume local thermal equilibrium for each object, i.e., $\Pi_\alpha^< = -i N_\alpha \Gamma_\alpha$,
$\Pi_\alpha^> = -i (N_\alpha+1) \Gamma_\alpha$,  $\alpha = L,R$,  together with the Keldysh equation 
$D^{>,<} = D^r \sum_{\alpha} \Pi_\alpha^{>,<} D^a$, the equation can be put into the
Landauer/Caroli form, as \cite{Zhang18prb}
\begin{equation}
\label{eq-second-Laudauer}
I_L = \int_0^\infty \frac{d\omega}{2\pi} \hbar \omega {\rm Tr} \bigl( 
D^r \Gamma_R D^a \Gamma_L \bigr) (N_L - N_R),
\end{equation}
here we have used a fact that the integrand is an even function of the frequency $\omega$, 
thus, we can perform integration over the positive frequency, and times 2.  A multiple lead formula of 
B\"uttiker type \cite{Buttiker88} is obtained if we replace  $R$ by a summation index $\alpha'$, i.e., 
for the current out of lead $\alpha$, it is 
$I_\alpha = \int_0^\infty {d\omega}/{(2\pi)} \, \hbar \omega \sum_{\alpha'} {\rm Tr} \bigl( 
D^r \Gamma_{\alpha'} D^a \Gamma_\alpha \bigr) (N_\alpha - N_{\alpha'})$.

If we apply the formula to a parallel plate geometry, the result is identical to the usual fluctuational electrodynamics result 
\cite{Volokitin07,Song15,Biehs21}, 
taking the speed of light to infinity.  Unlike the original Meir-Wingreen formula in terms of the electron Green's function $G$ and
bath self-energies $\Sigma_B$, this final expression cannot be an exact result (even in the non-retardation limit) mainly due to
approximations in $\Pi$.   Let us recapitulate 
the approximations going into the derivations.  (1) From the Hedin equations \cite{Hedin65} point of view, which give a formally exact solution
to the Coulomb problem, we have disregarded vertex corrections, so that the electron interaction self-energy is of the Fock form
$G D$, and the polarization is the RPA form $GG$.   We also ignored the Hartree term.   (2) To compute the transport quantities, 
we have further used the lowest order of expansion in terms of the interaction strength $e$.  Alternatively, we can also use the more accurate self-consistent 
$GW$ \cite{Martin16}.  (3)  To obtain the Landauer form, we have assumed local equilibrium for each object.  
(4) The vertex correction as well as the Hartree potential can be incorporated with Eq.~(\ref{eq-photon-IL-WM}) remaining valid.  The most severe limitation appears
to be the assumption of additivity of $\Pi$.  At the RPA level of approximation, $\Pi$ is block-diagonal; but this is not true at higher order 
in charge $e$.  It will be helpful to think of these approximations used here, when we assess the validity of the usual fluctuational
electrodynamics approach.

\section{Scalar field model}
The usual approach to study the Coulomb interaction \cite{Mahan00} in condensed matter physics is to eliminate the scalar field $\phi$ and to focus on
the instantaneous charge-charge interaction as given by expression (\ref{eqqq-Coulomb}).    The quantization of the electromagnetic field
is only for the transverse vector field $\bf A$.  One disadvantage in this approach is that the (scalar) photon Green's function $D$ appears 
to be somewhat {\it ad hoc} as a convoluted construction out of the electron screening perspective, unrelated to the scalar field.  
Here in this section, we take the scalar $\phi$ as fundamental \cite{capacitors,Peng1703arxiv} and define $D$ in terms of $\phi$ in the usual way of NEGF.  
As a result, the electron Green's function $G$ and photon Green's function $D$ stand on an equal footing.   This also 
makes the symmetry properties of the Green's function $D$ transparent. 
The only technical problem with this is that $\phi$, because it is instantaneous, does not have a free field dynamics.  

To overcome the difficulty of no conjugate momentum for $\phi$, we introduce a fictitious speed of light $\tilde{c}$.  Then
the usual machinery of quantum field theory applies \cite{Tannoudji89,Weinberg05}.   We take the limit that $\tilde{c}$ goes to infinity at the end of the calculation.   In this limit, the theory becomes
equivalent to the instantaneous Coulomb problem.   But it turns out that the free field energy must be negative definite, in order
to be consistent with the Poisson equation, and removing the zero-point motion contribution to the Poynting vector (in the
$c\to \infty$ limit, it becomes the ``Poynting'' scalar), we must use an anti-normal order, instead of the usual normal order.   A 
photon bath at infinity must have a negative temperature.   All these exotic features actually disappear and have no physical
consequences if, at the end of the calculation, we take the limit $\tilde{c} \to \infty$ \cite{Peng1703arxiv}.

The electron and scalar photon coupled system can be described by the following Lagrangian,
\begin{eqnarray}
L &=& \int dV \frac{\epsilon_0}{2}\left( - \frac{\dot{\phi}^2}{\tilde{c}^2} + (\nabla \phi)^2 \right) +  \nonumber \\
&& \, c^\dagger \left(i \hbar \frac{dc}{dt} - H c\right) +
e \sum_j c_j^\dagger c_j \phi({\bf r}_j).
\label{eq-scalar-L}
\end{eqnarray}
Here the first term is for the free photons, the second term is for the free electrons, and the last term is the electron-photon
interaction (electron charge is $-e$).    This last term is expressed as  $-\sum_l c^\dagger M^l c \phi_l$ in the previous section.  
This Lagrangian is equivalent to a Lorenz gauge choice with the vector field setting to zero. 
While the electrons are allowed to sit on a set of discrete lattice sites with a tight-binding Hamiltonian, the field exists in
the whole space and the contribution to the Lagrangian is an integration over all $\bf r$. 
We notice that the kinetic energy of the scalar photons is negative.   With this Lagrangian, 
the variational principle, $ \delta \int L dt =0$, reproduces the Poisson equation for $\phi$ and the Schr\"odinger equation in a field
of $\phi$ for the electrons.  Here we treat $c^\dagger$, $c$, $\phi$ as independent variables. 

The reason we start with the Lagrangian $L$ instead of Hamiltonian is that we need to identify the proper conjugate momenta for 
the fundamental dynamic variables, and to aid us for the canonical quantization.  It is clear that the electrons are already in the quantum form
with the anti-commutation relations.  Here we mainly focus on the scalar field $\phi$.     The conjugate momenta are given by
\begin{eqnarray}
P_{c_j} &=& \frac{\partial L }{\partial \dot{c}_j } = i \hbar c_j^\dagger, \\
P_{c_j^\dagger} &=& \frac{\partial L }{\partial \dot{c}_j^\dagger } = 0, \\
\Pi_\phi({\bf r}) &=&  \frac{ \delta L }{\delta \dot{\phi}({\bf r})} = - \epsilon_0 \frac{\dot{\phi}}{\tilde{c}^2}.
\end{eqnarray}
The $\delta$ in the last equation means a functional derivative, as $L$ depends on $\phi({\bf r})$ functionally. 
The ominous minus sign, which says that the momentum $\Pi_\phi$ is opposite to the velocity  $\dot{\phi}$, already is an indication of something unusual.
From the conjugate momenta we obtain the Hamiltonian as $H = \sum p \dot{q} - L$, or
\begin{eqnarray}
\hat H &=& - \int dV \frac{\epsilon_0}{2}\left(  \frac{\dot{\phi}^2}{\tilde{c}^2} + (\nabla \phi)^2 \right) +  \nonumber \\
&& \, c^\dagger  H c - e \sum_j c_j^\dagger c_j \phi({\bf r}_j).
\end{eqnarray}
The Hamiltonian for the electrons takes a familiar form with a potential applied to the diagonal elements, but
the scalar photons take a strange negative-definite energy.  Such a Hamiltonian of the scalar photons gives rise to the 
``inverted oscillators'' \cite{Glauber86} for the free field modes. 
When $\tilde{c} \to \infty$, due to the existence of the Poisson 
equation, we can think of the total Coulomb energy as being purely due to charges, summed over each pair once, as in Eq.~(\ref{eqqq-Coulomb}).  Alternatively,
we could attribute the energy completely to the field, with a positive energy density of $\frac{1}{2} \epsilon_0 (\nabla \phi)^2$.   The 
Hamiltonian above has a third interpretation: the charge-field interaction is $\sum_j q_j \phi_j$, but this overcounts the
energy by a factor of 2, thus, we need to subtract a half in the field term.   
The canonical quantization means that, in addition to the usual fermionic 
anti-commutation relations, $c_i c_j^\dagger + c_j^\dagger c_i = \delta_{ij}$, 
$c_i c_j + c_j c_i = 0$, and   $c_i^\dagger c_j^\dagger + c_j^\dagger c_i^\dagger = 0$, we impose 
\begin{equation}
\label{eq-phi-pi-commu}
[ \phi({\bf r}),  \Pi_\phi({\bf r}') ] = i\hbar \delta({\bf r} - {\bf r}').
\end{equation}
All the rest of the operators commute, i.e., between $\phi$s, or between $\Pi$s, or between the field and fermionic electron 
operators.  These commutation relations completely define the quantum mechanical problem.   If the commutation relations 
are postulated correctly, we should obtain the correct equations of motion in operator form with the Heisenberg equation.   
In our case, $\ddot{\phi}=[\dot{\phi},\hat H]/(i\hbar)$ gives a wave equation with charges produced by the electrons as source, 
reducing to the Poisson equation in $\tilde{c} \to \infty$, and $i\hbar \dot{c} = [c,\hat H]$ is the Schr\"odinger equation of the 
electron in a potential of $(-e)\phi$. 

We are now in a position to define the scalar photon Green's function,
\begin{equation}
\label{eq-contour-D-def}
D({\bf r}\tau,{\bf r}' \tau') = \frac{1}{i\hbar} \bigl\langle T_\tau \phi({\bf r,\tau}) \phi({\bf r}',\tau') \bigr\rangle,
\end{equation}
here $\bf r$, ${\bf r}'$ are space positions, and $\tau$, $\tau'$ are Keldysh contour times.   $T_\tau$ is the contour 
order super-operator.  The contour time is the pair $\tau = (t,\sigma)$ of real time
and branch index.   The average $\langle\, \cdot\, \rangle = {\rm Tr}(\rho\, \cdot\, )$ is unspecified at the moment;  
it could be a thermal equilibrium or in general of nonequilibrium steady state with a density matrix $\rho$. 
At this point, we need to pause and ask if we should define the Green's function by deviations, 
$\delta \phi = \phi - \langle \phi \rangle$.  For noninteracting free fields, since the Hamiltonian is quadratic in $\phi$,
the average $\langle \phi\rangle$ is zero, the subtraction is not necessary.    For an interacting system with an interaction
term odd in $\phi$, without the subtraction, we do not even have a proper Dyson equation.  In such case $\phi$ should
be understood to be a deviation from the average.  These two definitions differ by a constant.  An additive constant term in 
$D$ independent of the time produces a delta function in the frequency domain, thus in any way does not contribute to transport.  

We briefly summarize some of the key definitions and properties of the Green's functions.  
Readers unfamiliar with the concept should consult the literature on NEGF \cite{Haug08,Jishi13,Wang-review-1,Wang-review-2}. 
For notational brevity, we often omit the ${\bf r}, {\bf r}'$ arguments, and treat them as indexing a matrix.  
Due to the $+$ (forward) and $-$ (backward) branches the contour Green's function $D(\tau, \tau') = D^{\sigma,\sigma'}(t,t')$ gives four Green's functions
in real time, $D^{++} = D^t$ is time ordered, $D^{--}  = D^{\bar{t}}$ is anti-time ordered, $D^{+-} = D^<$ is lesser,
and $D^{-+} = D^>$ is greater.   The four are not linearly independent and are constrained by  $D^t + D^{\bar{t}} = D^> + D^<=D^K$.   The symmetric correlation $D^K$
is known as the Keldysh component. 
The retarded Green's function is $D^r = D^t - D^< = \theta (D^> - D^<)$, where the step function $\theta = \theta(t-t') = 1$ if $t>t'$ and 0 otherwise.
And the advanced is $D^a =D^< - D^{\bar{t}} = - (1-\theta)(D^> - D^<)$, such that $D^> - D^< = D^r - D^a$.  
Letting $1 = ({\bf r}t)$ and $2 = ({\bf r}' t')$, we have the symmetry in time domain as
$D^>(1,2) = D^<(2,1)$, and $D^r(1,2) = D^a(2,1)$.   The Fourier transform into frequency domain is defined by
\begin{equation}
D(\omega) = \int_{-\infty}^{+\infty} dt\, D(t-t') e^{i \omega (t-t')}.
\end{equation}
In frequency domain, we have the Hermitian conjugate  $D^r(\omega)^\dagger = D^a(\omega)$, $D^<(\omega)^\dagger = - D^<(\omega)$.

The reason for introducing the contour ordered Green's function is that it facilitates systematic perturbative expansion at
any temperature, as in a quantum field theory at zero temperature. The Green's function $D$ can be determined with a Dyson equation on contour,  $D  =v + v \Pi D$, if we know the self-energy
$\Pi$, which we can determine by a standard diagrammatic expansion 
\cite{Rammer07,Bruus-Flensberg04}.  To the lowest order of approximation, it is
given by the RPA expression, Eq.~(\ref{eq-RPA-pi}).  The contour version of this equation means, written out
in full, 
\begin{eqnarray}
\label{eq-dyson-Dtau}
D({\bf r}\tau,{\bf r}' \tau') = v({\bf r}\tau,{\bf r}' \tau') + \hspace{13 em} \\
\sum_{j,k} \int d\tau_1 \!\int d\tau_2\, v({\bf r}\tau,{\bf r}_j \tau_1)  \Pi_{jk}(\tau_1, \tau_2)
D({\bf r}_k\tau_2,{\bf r}' \tau'). \nonumber 
\end{eqnarray}
Here we use a mixed representation, while $D$ and $v$ are defined on the whole space, since the electrons are
on a set of discrete sites ${\bf r}_j$,  the self-energy is defined only on the discrete sites.   $v$ is the free
Green's function, i.e., the Green's function when the electrons are absent.   Since the thermal state is essentially
fixed by the electrons, we only need the retarded version of $v$, which can be obtained by an equation 
of motion method.   If we compute the derivative with respect to the first argument of time, $t$, twice, we find
\begin{equation}
\epsilon_0 \left( \frac{1}{\tilde{c}^2} \frac{\partial^2\,}{\partial t^2} - \nabla^2 \right) v({\bf r}t,{\bf r}'t') = \delta({\bf r} - {\bf r}') \delta(t-t').
\end{equation}
We see that $v$ is essentially the Green's function of the wave equation. 
The contour Dyson equation implies a pair of real time equations as the retarded Dyson equation, $D^r = v^r + v^r \Pi^r D^r$,
and the Keldysh equation, $D^< = D^r \Pi^< D^a$,  so  $v^<$ is never needed (this is because in the limit 
$\tilde{c} \to \infty$, the bath at infinity has no effect.  Or in other words, the Coulomb field $\phi$ cannot propagate 
to infinity). 

\subsection{``Poynting scalar''}
How do we describe the energy flux in a pure scalar field theory?    We start from first principles.    The energy density, according
to our separation of the free field and electron-photon interaction, is $u = - \frac{1}{2} \epsilon_0 \left( \dot{\phi}^2/\tilde{c}^2 +  (\nabla\phi)^2\right)$.    From this 
expression, we can obtain a conservation law in differential equation form, 
\begin{eqnarray}
\frac{\partial u}{\partial t} &=& -\epsilon_0 \left(\frac{\dot{\phi}\,\ddot{\phi}}{\tilde{c}^2} + 
 \nabla \dot{\phi} \cdot \nabla \phi \right)  \nonumber \\
 &=&  - \nabla \cdot {\bf j}  - \rho \dot{\phi}.
\end{eqnarray}
Here we define the Poynting scalar as  ${\bf j} = \epsilon_0 \dot{\phi} \nabla \phi$.   The dot denotes partial derivative with 
respect to time.   In obtaining the second line, we note that $\phi$
satisfies a wave equation with the charge density as a source.  Here we treat the expression as a classical equation.   To make
a quantum equivalent, we need to symmetrize the products.     The last term in the expression  $\rho \dot{\phi}$ is for Joule heating. 
If we are calculating the steady-state work, we can perform ``integration by parts,'' so the average Joule heating is also
$-\dot{\rho} \phi$, consistent with the expression given earlier in Sec.~\ref{sec-capacity-physics}.  

We can express the quantum dynamic variable, i.e., quantum field of $\phi$, in terms of the creation and annihilation operators in the
usual way,
\begin{equation}
\label{eq-b-to-phi}
\phi({\bf r}) = \sum_{{\bf k}} \sqrt{\frac{\hbar \tilde{c}^2}{2 \epsilon_0 \tilde\omega_{\bf k}V}}
\left( b_{\bf k} e^{i {\bf k} \cdot{ \bf r}}   + {\rm h.c.}\right),
\end{equation}
except that the commutation relation acquires a minus sign, due to Eq.~(\ref{eq-phi-pi-commu}), namely,
\begin{equation}
[ b_{\bf k}, b_{{\bf k}'}^\dagger] = - \delta_{{\bf k}, {\bf k}'}, \quad  
[b_{\bf k}, b_{{\bf k}'}]  = 0, \quad  [ b_{\bf k}^\dagger, b_{{\bf k}'}^\dagger] = 0.
\end{equation}
Here the wave vector ${\bf k}$ is box quantized in a finite volume of  $V$, and the mode frequency is given by
$\tilde{\omega}_{\bf k} = \tilde{c} | {\bf k} |$. 
With this transformation, the free field term is an inverted oscillator form, 
$\int u\, dV = -\sum_{\bf k} \hbar \tilde\omega_{\bf k}( b_{\bf k} b_{{\bf k}}^\dagger + 1/2)$. 
Because of the minus sign in the commutation relation, and because of the negative-definite nature of
the Hamiltonian, the role of creation and annihilation is reversed.  The ground state is defined by
$b_{\bf k}^\dagger|0\rangle = 0$.   This is because the creation operator still has the meaning of increasing
energy by one unit of $\hbar \tilde\omega_{\bf k}$, and annihilation operator decreasing the energy by one unit,
but due to the negative definiteness, one cannot increase forever.    The implication of this feature is that in order
to remove the zero-point contribution to the Poynting scalar, we must take an anti-normal order.   

Let us explain this point more carefully.   We see that the Poynting scalar is a quadratic form of $\phi$.   If
we expand $\phi$ in terms of $b$ and $b^\dagger$, we get four types of terms,  $bb$, $b^\dagger b^\dagger$,
$b b^\dagger$, and $b^\dagger b$.   The vacuum expectation value $\langle 0 | \cdot | 0 \rangle$ is not zero
for the last type.   The anti-normal order is to swap the last case into the third case, removing the zero point
contribution from the field.   This anti-normal order is also used on general states, so we calculate the 
heat flux using Poynting scalar as \cite{Peng1703arxiv}
\begin{eqnarray}
\langle {\bf j} \rangle &=& \frac{1}{2} \epsilon_0 \bigl\langle\, \vdots\, \dot{\phi} \nabla \phi 
+ \nabla \phi \, \dot{\phi}  \,\vdots\, \bigr\rangle  \nonumber\\
&=& \epsilon_0 \int_0^\infty \frac{d\omega}{\pi} \hbar \omega\, {\rm Re} \nabla_{{\bf r}'} D^{>}(\omega, {\bf r}, {\bf r}')\Big|_{{\bf r}'={\bf r}}, 
\label{eq-scalar-j}
\end{eqnarray}
which we can express in terms of the greater Green's function and integrate over the positive frequencies. 
Here we transform the Green's function into frequency domain, then the derivative with respect to time becomes 
$-i\omega$.    The gradient operator is acting on the second argument of the position.  After the operation, both positions
are set to be equal.   The use of greater instead of lesser Green's function is due to the anti-normal order requirement. 
  
Equation~(\ref{eq-b-to-phi}) is not the solution to the problem but only defines a transformation from $b$ to $\phi$.
However, for the free field which has the time dependence $b_{\bf k}(t) = b_{\bf k} e^{-i \tilde{\omega}_{\bf k} t}$, we
can obtain the free retarded Green's function from the definition,
\begin{eqnarray}
\label{eq-vrt}
v({\bf r}t,{\bf r}'t') &=& \theta(t-t') \frac{1}{i\hbar} \bigl\langle [ \phi({\bf r},t),\phi({\bf r}',t') ] \bigr\rangle  \\
&=& \frac{\tilde{c}^2}{\epsilon_0 }  \sum_{\bf k}
\frac{ \theta(t-t') \sin\bigl(\tilde{\omega}_{\bf k}(t-t')\bigr)}{\tilde{\omega}_{\bf k}V } e^{i {\bf k}\cdot({\bf r} - {\bf r}')} 
\nonumber \\
&=& \frac{1}{4 \pi \epsilon_0 | {\bf r} - {\bf r}' |} \delta(t-t'),\quad \tilde{c} \to \infty. \nonumber 
\end{eqnarray}
If we Fourier transform the expression into $\omega$ space, and then take the limit $\tilde{c} \to \infty$, we
recover the Coulomb interaction in $\bf k$ space, as $v({\bf k}) = 1/(\epsilon_0 k^2)$, and the usual
expression for the Coulomb potential in real space.  
We see also that the free retarded Green's function is independent of the distribution, no matter
what meaning we give to the average.   The free retarded Green's function is determined solely by the
equal-time commutators, unrelated to the distribution controlled by
$\langle b_{\bf k} b_{\bf k}^\dagger \rangle$, which determines $v^<$.     
We do not need the free $v^<$ as mentioned earlier.

\subsection{A parallel plate capacitor as two quantum dots, scalar field}
In this subsection, we treat the parallel plate capacitor problem again, but this time, by the scalar field approach \cite{capacitors}.
Since in a parallel plate capacitor, the field is a function of only one variable, we call it $z$, the transverse direction can
be integrated, giving the area of the capacitor $A$.  As a result, the retarded Dyson equation, in differential form,
$v^{-1} D = 1 + \Pi D$, becomes,
\begin{eqnarray}
-\epsilon_0 A  \left[ \left( \frac{\omega}{\tilde{c}} + i \eta\right)^2 + \frac{\partial^2\ }{\partial z^2} \right]
D^r(z,z',\omega) = \qquad\quad \nonumber \\
 \delta(z-z') + 
 \sum_{\alpha=L,R} \delta(z-z_\alpha) \Pi_{\alpha}(\omega) D^r(z_\alpha,z',\omega).
\end{eqnarray} 
Here we have transformed the equation into frequency domain, and introduced an infinitesimal small damping $\eta$ so 
that the solution is the retarded one.  
This equation can be interpreted as the scalar potential generated by a unit active (external) charge located {at}
$z'$, together with the induced extra charges at the electron sites, $z_L=0$, $z_R=d$, due to the linear response to the 
internal field.  Indeed, the induced charge at site $\alpha$ is given by $\delta q_\alpha = \Pi_\alpha \phi(z_\alpha)$, and $\Pi_\alpha$ is the associated response function. 

To compute the energy current density $j$ using the solution of $D$, we note that the dots are coupled to $D$ only
at $z=0$ and $d$.  As a result, we only need to know $D$ at one of these points.  Also, $D^r(z,z',\omega) = 
D^r(z',z,\omega)$ is symmetric, thus the following solution for $0 < z < d$ is sufficient:
\begin{eqnarray}
D^r( z, 0,\omega) &=&  \left( \frac{\gamma^2-\lambda^2}{\gamma} \Pi_R + 2 \gamma \Omega\right)\frac{1}{\mathcal{D}}, \\
D^r( z,d,\omega) &=& \left((1-\gamma^2)\Pi_L + 2 \Omega\right)\frac{\lambda}{ \gamma \mathcal{D}}{,} 
\end{eqnarray}
where $\tilde{k} = \frac{\omega}{\tilde{c}} + i \eta$,  $\lambda = e^{i\tilde{k}d}$,
$\gamma = e^{i\tilde{k}z}$,  $\Omega = i \epsilon_0 A \tilde{k}$, and  
\begin{equation}
\mathcal{D} = (\lambda^2 - 1) \Pi_L \Pi_R - 2 \Omega\,(\Pi_L + \Pi_R) - 4 \Omega^2.
\end{equation}
To obtain the solution, we set $z'=0$ or $d$, assuming backward moving and decaying wave to the left for $z<0$,
$r e^{-i \tilde{k} z}$, standing wave $A e^{i\tilde{k}z} + B e^{-i \tilde{k}z}$ in the middle segment, $0 < z < d$,
and decaying wave to the right for $z>d$, $t e^{i \tilde{k}z}$.    The wave has to be continuous at $0$ and $d$.   But the first
derivatives are discontinuous with the discontinuities determined by the $\delta$-functions on the right-hand side of the equation.  
This gives four boundary condition matching algebraic equations, uniquely determining the coefficients. 

In the next step, we use the Keldysh equation 
\begin{equation}
D^>(z,z') = \sum_\alpha D^r(z,z_\alpha) \Pi_\alpha^> D^r(z',z_\alpha)^{*},
\end{equation}
where we have omitted $\omega$ argument for simplicity and used $D^a = (D^r)^\dagger$. 
We can now plug $D^>$ into the expression for heat current density, Eq.~(\ref{eq-scalar-j}).   Since our problem is 
quasi-one-dimensional, $\nabla_{{\bf r}'}$ is just $\partial/\partial z'$.   At this point after the space derivative, we can take the limit 
$\tilde{c} \to \infty$ and $\eta \to 0^+$, the result becomes independent of location $z$, and  \cite{capacitors}
\begin{equation}
\langle j \rangle =  \int_0^\infty \frac{d\omega}{\pi A} \mathrm{Re} \Bigl[{i\hbar \omega}{|D_{RL}|^2}\bigl(\Pi_R^> \mathrm{Im}\Pi_L - \Pi_L^>  \mathrm{Im}\Pi_R\bigr)  \Bigr].
\end{equation}
where $D_{RL} = D_{LR} = 1/\bigl(\Pi_L \Pi_R/C - (\Pi_L + \Pi_R)\bigr)$, and $C = A \epsilon_0/d$ is the capacitance.   If
we assume local thermal equilibrium with the fluctuation-dissipation theorem for $\Pi^>_\alpha$, we obtain the same Laudauer
formula of Sec.~\ref{sec-capacity-physics}.   If we adopt a self-consistent calculation, local equilibrium is not a valid assumption, 
then the above equation is suitable for a self-consistent calculation.   We note that the above formula agrees trivially with the
photon version of the Meir-Wingreen formula, Eq.~(\ref{eq-photon-IL-WM}), and it also agrees with the electron version, Eq.~(\ref{MWeq1}),  when the self-consistent iterations are converged. 

\begin{figure}
 \includegraphics[width=0.9\columnwidth]{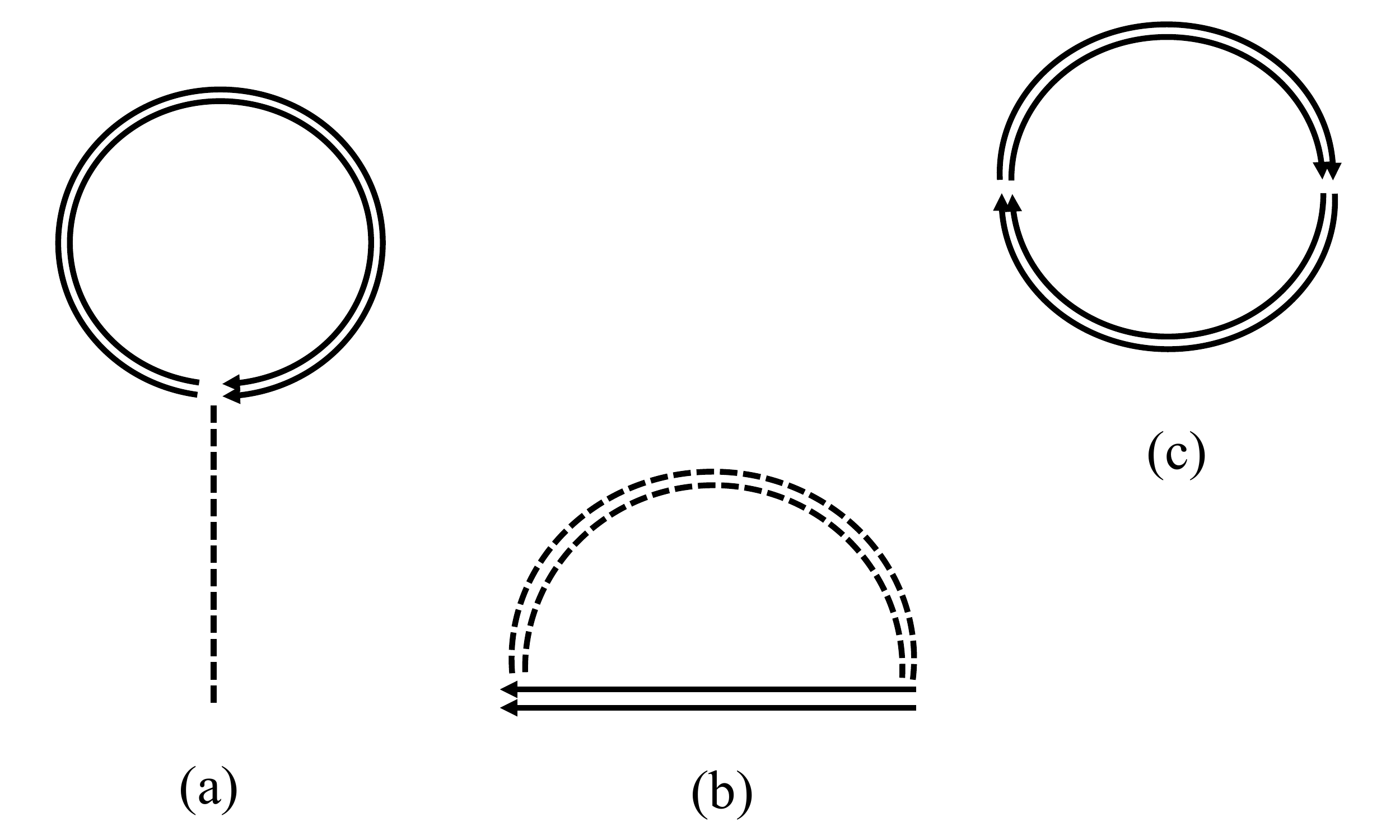}
  \caption{The self-consistent Born approximation diagrams. (a) the Hartree self-energy diagram, (b) the Fock self-energy diagram, (c) the polarization bubble $\Pi$.  The double line with arrows signifies the full electron Green's function $G$, the single dotted line denotes bare Coulomb $v$, while the double dotted line is the screened Coulomb $D$.}
 \label{fig-scba-diag}
\end{figure}

\subsection{Quantum dot model of $\Pi$}
We consider each of the plates as a single quantum dot with maximum charge $Q$, not necessarily the unit charge $e$, with the 
Hamiltonian $ \epsilon_\alpha c^\dagger_\alpha c_\alpha$ for each dot $\alpha = L$ or $R$.   
To set up a $GW$ calculation  (also known as self-consistent Born approximation, see Fig.~\ref{fig-scba-diag}), we start from the retarded Green's function
\begin{equation}
G_\alpha^r(E) = \frac{1}{E - \epsilon_\alpha - \Sigma^r_\alpha(E) - \Sigma_{n,\alpha}^r(E)}.
\end{equation}
For bath self-energies, we take a phenomenological Lorentz-Drude model \cite{Datta00},
\begin{equation}
\Sigma^r_\alpha(E) = \frac{\Gamma_\alpha/2}{ i + E/E_{0,\alpha}}.
\end{equation} 
If $E_{0,\alpha} \to \infty$, it is called a wide-band model, as the coupling with the lead is independent of energy.  Here we 
cut off to the energy scale of $E_{0,\alpha}$.  This gives a better physical picture in real time $t$; the self-energy takes an 
exponential decay form.  

If the dot is not in local thermal equilibrium, we cannot use the fluctuation-dissipation theorem for the electron Green's function, 
and have to use the Keldysh 
equation
\begin{equation}
G_\alpha^{<}(E) = G^r_\alpha(E) \left( \Sigma_\alpha^<(E) + \Sigma_{n,\alpha}^<(E) \right) G^a_\alpha(E).
\end{equation}
The lead, since it serves as a bath, by definition is in equilibrium, so  we can use 
$\Sigma_\alpha^<(E) = - f_\alpha(E) \bigl(\Sigma_\alpha^r(E) - \Sigma_\alpha^a(E)\bigr)$, where
$f_\alpha(E) = 1/(e^{\beta_\alpha(E-\mu_\alpha)}+1)$ is the Fermi function.    The nonlinear interaction self-energy 
$\Sigma_{n,\alpha}$, the Hartree and the Fock term coupling the two dots cannot be in local equilibrium, thus no
fluctuation-dissipation theorem for it.   The nonlinear self-energy can be calculated in real time as
\begin{eqnarray}
\Sigma_{n,\alpha}^<(t) &=& i \hbar Q^2 G_\alpha^<(t) D^<(z_\alpha, z_\alpha,t), \\
\Sigma_{n,\alpha}^>(t) &=& i \hbar Q^2 G_\alpha^>(t) D^>(z_\alpha, z_\alpha,t), \\
\Sigma_{n,\alpha}^r(t) &=& \theta(t) \bigl(\Sigma_{n,\alpha}^>(t) - \Sigma_{n,\alpha}^<(t)\bigr) +  \nonumber \\
&& (- i\hbar) Q^2 \delta(t) \sum_{\alpha'} v(z_\alpha,z_{\alpha'}) G^<_{\alpha'}(t=0).   
\end{eqnarray}
There is no Hartree contribution to lesser and greater components of the interaction self-energy, since the
Hartree term is proportional to $\delta(\tau, \tau')$ on contour.  We also note that the Hartree potential 
should be the unscreened one of $v$.   For the capacitor model, this is 
\begin{equation}
v = \frac{1}{2 \epsilon_0 A \eta} \left( \begin{array}{cc}
1  & e^{-\eta d} \\
e^{-\eta d} & 1 \\
\end{array} \right), \quad \eta \to 0^+,
\end{equation}
which is divergent.  Since the Hartree potential only renormalizes the onsite energy, and charge 
neutrality also requires it to be zero, we drop this term in actual calculation. 
Finally, the scalar photon self-energy or polarizability is calculated according to 
\begin{eqnarray}
\Pi_{\alpha}^>(t) =& -i\hbar Q^2 G_{\alpha}^>(t) G_{\alpha}^<(-t),\\
\Pi_{\alpha}^<(t) =& -i\hbar Q^2 G_{\alpha}^<(t) G_{\alpha}^>(-t),\\
\Pi_{\alpha}^r(t) =& \theta(t) \bigl[\Pi_{\alpha}^>(t) - \Pi_{\alpha}^<(t)\bigr].
\end{eqnarray}
With this new $\Pi$, we need to recalculate $D^r$, and thus new $\Sigma_n$ and $G$, until convergence. 

Since the nonlinear self-energy is easy to calculate in time domain, it is natural we adopt a fast Fourier transform 
method to go between energy and time domain.   However, the errors are sometimes hard to control.   An alternative to the 
Fourier transform is to perform convolution in energy space, never going into time domain.   Energy domain functions
are relatively smooth functions, but time domain function can be highly oscillatory, thus hard to control. 

\begin{figure}
 \includegraphics[width=0.85\columnwidth]{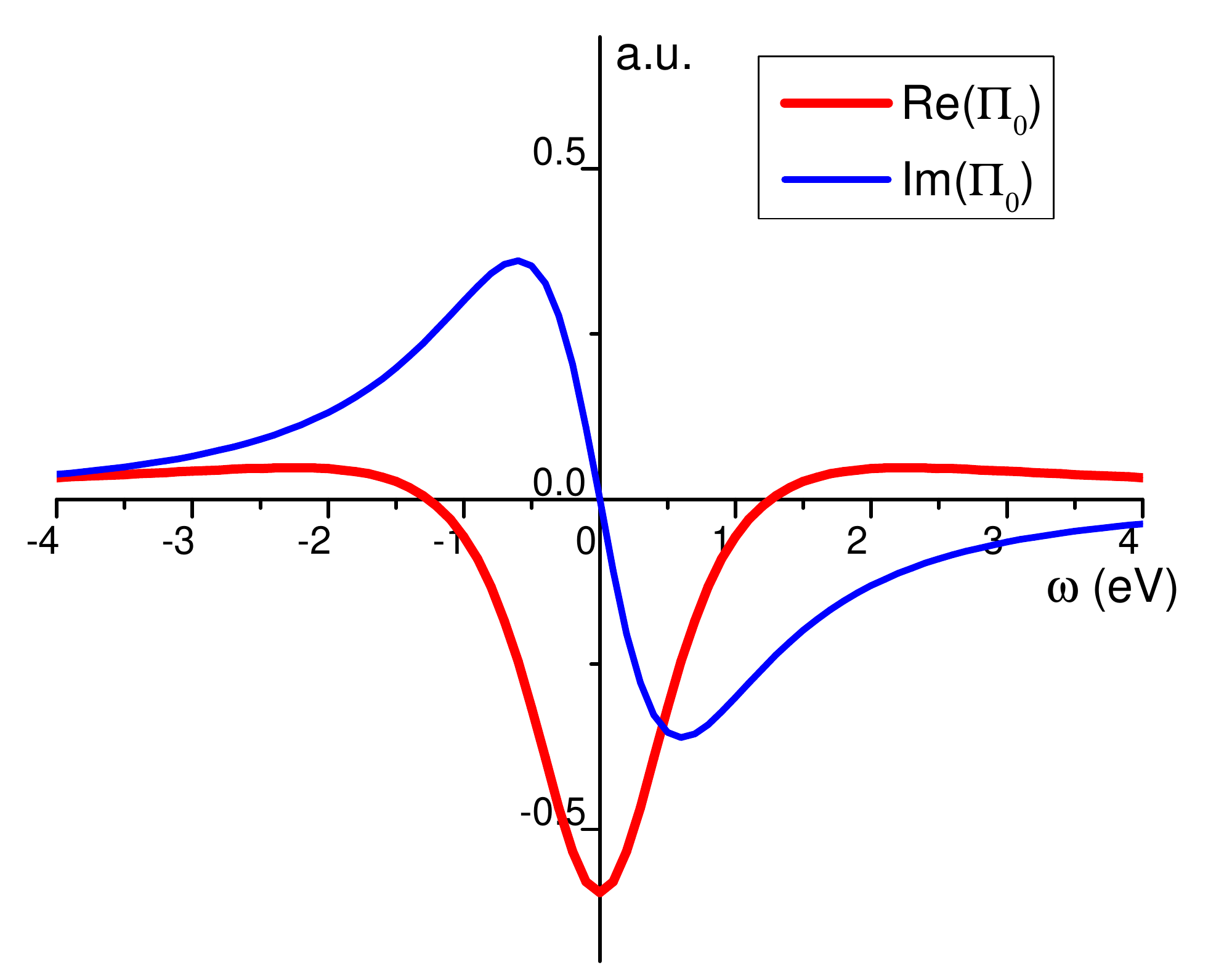}
  \caption{The retarded $\Pi_L$ of the left quantum dot.  For parameter, see next figure~\ref{fig-vsBB}. Note that the real part is symmetry, while the imaginary part is 
  anti-symmetric with respect to the frequency $\omega$.  From \onlinecite{Peng1703arxiv}, Fig.~2.}
 \label{fig-qdot-pi}
\end{figure}

\begin{figure}[htp]  
  \centering
  \includegraphics[totalheight=60mm]{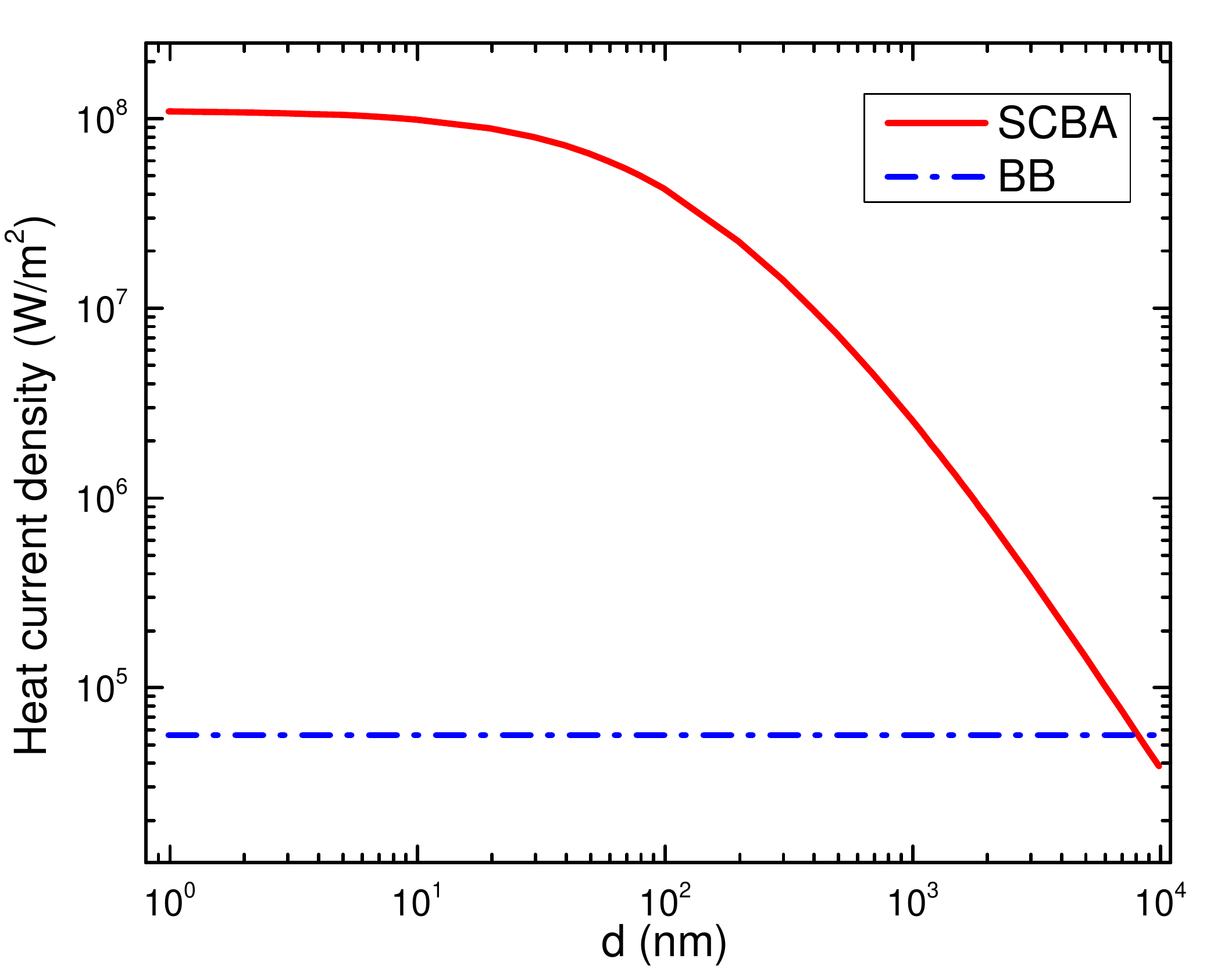}
  \caption{Energy current calculation under SCBA and the blackbody limit in log-log plot. The temperature of dot $L$ is $1000\,$K and dot $R$ is $300\,$K. The chemical potential of dot $L$ is $0\,$eV and dot $R$ is $0.02\,$ eV. Charge $Q=1e$, onsite energy $\epsilon_L = \epsilon_R = 0\,$eV, and plate area is
  $A = 19.2 \times 19.2\,$nm$^2$.  The bath coupling is
  $\Gamma_L = 1\,$eV, $\Gamma_R = 0.5\,$eV, and $E_{0,L} = E_{0,R} = 50\,$eV.   From \onlinecite{Peng1703arxiv}, Fig.~8. }
   \label{fig-vsBB}
\end{figure}

We show in Fig.~\ref{fig-qdot-pi} the real and imaginary parts of the retarded $\Pi$ for the single quantum dot, which is typical of
the photon self-energy. 
Figure \ref{fig-vsBB} presents the distance dependence of heat current density in a double logarithmic plot, calculated with the self-consistent Born approximation.  For large distances, the heat current density decreases like $d^{-2}$. Such a scaling law arises from the capacitor property, as mentioned earlier, which can be manifested in the expression for the transmission coefficient.   Our choice of the parameters should be considered typical. Fig.~\ref{fig-vsBB}
shows a large enhancement of heat transfer mediated by the scalar photons. Comparing the red solid and blue dashed-dotted lines at $10\,$nm, we find that the heat current density is two thousand times larger than the blackbody limit. 
This result demonstrates that the heat transfer channel provided by electron-photon interaction is the dominant one for 
nano-capacitors at small separations.

\section{Heat transport without local equilibrium by current drive}
Earlier in Sec.~\ref{sec-coulomb-interaction} we derived a Landauer-like formula and the Meir-Wingreen formula based on Coulomb
interaction for an electron system.  The Landauer formula assumes the local thermal equilibrium, while the Meir-Wingreen formula is 
a more general formula without the use of local thermal equilibrium.   
As an application of the Meir-Wingreen formula, Eq.~(\ref{eq-photon-IL-WM}), we consider 
two graphene layers separated by a vacuum gap, one of them is driven by a current \cite{Peng1805arxiv}.   
Since the system is under a current drive, 
this is truly a nonequilibrium problem and we cannot use the Landauer formula.   However, something very close to 
it does exist, which is a ``Doppler'' shifted version.  The basic idea is the following:  under the external field drive, the 
layer is nearly in equilibrium, satisfying the Kadanoff-Baym ansatz \cite{Kadanoff62} in the sense,
\begin{eqnarray}
G^< = -f (G^r - G^a),  \\
G^> = (1-f) (G^r - G^a),
\end{eqnarray}
here $f$ is not the equilibrium Fermi function but a shifted one,
\begin{equation}
f(\epsilon) = f^0 - \frac{df^0}{d \epsilon} \Phi \approx f^0(\epsilon - \Phi).
\end{equation} 
Here $f^0$ is the usual equilibrium Fermi function, $f^0 = 1/\big( e^{\beta(\epsilon- \mu)} + 1  \bigr)$, while the nonequilibrium 
distribution $f$ is obtained by deforming the dispersion relation by an amount $\Phi$.   
In principle, we should solve the Boltzmann equation for $f$, but a single-mode relaxation approximation which is equivalent to
the choice of $\Phi_{\bf k}$ below is more practical and simpler to use.  
In order for the ``fluctuation-dissipation'' theorem
above to make sense, the Green's functions are in mode space, i.e.,  it is for $G^{>,<}({\bf k}, E)$ and the spectrum function 
$i (G^r - G^a)$ is essentially a $\delta$-function of $E$ peaked at the electron band $\epsilon_{\bf k}$.    Only the distributions are shifted;
the band stays as it is.    

To the lowest order of approximation, we can expand $\Phi$ by Legendre polynomials, and keep only the most important angular
dependence as 
\begin{equation}
\Phi_{\bf k} = - \hbar {\bf k} \cdot {\bf v}_1 = - \hbar v_1 k \cos \theta.
\end{equation}
Here ${\bf v}_1$ is the drift velocity and $\theta$ is the angle between the wavevector $\bf k$ and the drift velocity.   The
drift velocity is related to the electric current by multiplying it by the carrier charge density.
Using this
version of $G^{>,<}$, we can compute $\Pi^{>,<}$ by the usual formula, which is, in $({\bf q}, \omega)$ space,
\begin{equation}
\Pi^<_{\bf q}(\omega) = \frac{e^2}{iN}\sum_{\bf k} \int \frac{dE}{2\pi} G_{\bf k}^<(E) G_{{\bf k} + {\bf q}}^>(E-\hbar \omega).
\end{equation}
The real space Green's function is related to the Fourier space one by $G_l^{>,<}(t) = \frac{1}{N} \sum_{\bf k} \int \frac{dE}{2\pi \hbar} G_{\bf k}^{>,<}(E) e^{ i ({\bf k} \cdot {\bf R}_l -  Et)/\hbar}$.   Here for simplicity of notation, we assume one atom per unit cell at location of Bravais lattice
point ${\bf R}_l$, and the summation is over the
first Brillouin zone.  $N$ is the number of ${\bf k}$ points.   Using the Kadanoff-Baym ansatz and the identity 
$f^{0}(\epsilon)(1-f^{0}(\epsilon')) = N^{0}\bigl((\epsilon-\epsilon')/\hbar\bigr)\bigl( f^{0}(\epsilon') -f^{0}(\epsilon)\bigr)$, 
one can show that a shifted fluctuation-dissipation theorem also exists, i.e.,
\begin{equation}
\Pi^<_{\bf q}(\omega) = \tilde{N}(\omega) \Bigl(  \Pi^r_{\bf q}(\omega) - \Pi^a_{\bf q}(\omega) \Bigr),
\end{equation} 
and similarly for $\Pi^>$ with $\tilde N + 1$, where we have
\begin{equation}
\tilde{N}(\omega) = \frac{1}{ e^{\hbar (\omega - {\bf q}\cdot {\bf v}_1)/(k_B T)} -1}. 
\end{equation}
This is a Doppler shifted Bose distribution \cite{Duppen16,Svintsov19,Morgado17,Shapiro17}.    The linearity of $\Phi_{\bf k}$ with respect to $\bf k$ is important here as 
$\epsilon - \Phi  - \epsilon' + \Phi' = \hbar \omega -\Phi_{\bf k} + \Phi_{{\bf k} + {\bf q}} = \hbar (\omega - {\bf q} \cdot {\bf v}_1)$ depends only on ${\bf q}$, 
otherwise $\tilde N$ cannot be factored out of the summation over $\bf k$.
Note that due to the replacement $f^0 \to f$, the retarded $\Pi^r$ for graphene is no longer Doppler shifting 
the equilibrium one, but needs to be computed afresh.    Plugging in these formulas satisfying a modified version
of the fluctuation-dissipation theorem, we still have a Landauer formula, just like Eq.~(\ref{eq-second-Laudauer}), except that the Bose distribution 
function $N^{0}(\omega)$ is replaced by the Doppler shifted one of $\tilde{N}(\omega) = N^{0}(\omega - {\bf q} \cdot {\bf v}_1)$. 
However, since the original Meir-Wingreen formula is an integration of the frequency $\omega$ from $-\infty$ to $+\infty$, and the Doppler shift
breaks the symmetry of $\omega \to -\omega$,  the Landauer version also needs to integrate from $-\infty$ to $+\infty$ just 
like the Meir-Wingreen version.  This 
ends our theory for near-field heat transfer under current drive. 

To make a concrete calculation, we need to simplify our Caroli expression for transmission,  
${\rm Tr}(D^r \Gamma_R D^a \Gamma_L)$, using the fact that we have a lattice 
symmetry.    Here the trace is over the collection of all electron sites.   Since trace is invariant with respect to a similarity transform,
it is more efficient if we perform the trace in wave-vector space for the transverse directions using lattice symmetry.   Then 
the trace on the sites becomes a sum over the wavevectors.  To simplify the treatment, we ignore possible local inhomogeneous effect and treat each site ${\bf R}_l$ as on 
a Bravais lattice.   What does this amount to is for the self-energy $\Pi$ we consider all the charges in a unit cell as one single unit, by summing over all the charges in the unit cell as if it is located at the lattice site.    The error introduced is negligible when the distance $d$ involved is much larger than the lattice constant.   The lattice symmetry implies a translation invariance which means that correlation
functions such as $\Pi$ and $D^r$ are function of the relative distance between two points.   We can make a discrete Fourier transform
into the two-dimensional $\bf q$ wavevector space, still keeping $z$ the transport direction in real space, as 
\begin{equation}
D({\bf q}, z, z') = \sum_{l} D({\bf R}_l, z; {\bf 0}, z') e^{ -i {\bf q} \cdot {\bf R}_l}.
\end{equation}
Here ${\bf R}_l = l_1 {\bf a}_1 + l_2 {\bf a}_2$ is a two-dimensional lattice vector running through index $l=(l_1, l_2)$ on the crystal
lattice sites, $\bf q$ runs over the reciprocal space in the first Brillouin zone, laying in the plane of graphene.   This convention of discrete Fourier transform
ensures that the dimensions in real space and in $\bf q$ are the same, i.e.,  the inverse capacitance, $[D]=[\Pi^{-1}] = [U/e^2]$.   
As a result of this transformation for $D^r$ as well as for $\Pi$, we can work in $\bf q$ space,
in which each value of $\bf q$ is block-diagonal and we can focus on one particular $\bf q$.   The three-dimensional problem reduces
to a quasi-one-dimensional problem.    

We note that in our original definition of the scalar photon Green's function, Eq.~(\ref{eq-contour-D-def}), it is defined on the
continuum for any real space $\bf r$.   Fortunately, in the Caroli formula, we only need to know the values of $D$ on the lattice
sites where electrons exist.  As a result, we don't need to solve the Dyson equation covering the whole space $\bf r$ but just over these 
discrete points $({\bf R}_l, z)$ with $z = 0$ for the left and $d$ for the right sheet of graphene.  The Dyson equation still takes 
the form $D^r = v + v \Pi D^r$, or $(D^r)^{-1} = v^{-1} - \Pi$, here $\Pi = \Pi^r$ is the retarded version.  
The free Green's function for the transverse directions in $\bf q$ space and $z$ direction in real space is obtained by solving the
Poisson equation in mixed representation with a $\delta$ source, as 
\begin{equation}
\left(  q^2 - \frac{\partial^2\,}{\partial z^2}\right) v({\bf q}, z,z') = \frac{1}{\epsilon_0\Omega} \delta(z-z').
\end{equation}
Here $q = |{\bf q}|$ is the magnitude of the wave vector, $\Omega$ is the area of one unit cell.  This extra $\Omega$ 
factor in the denominator on the right-hand side has to do with the fact that our $\bf q$ is not continuous, but discretized 
on a grid and our Fourier transform is the discrete version.  Putting it in another way, $1/\Omega$ is the value of the
discretized two-dimensional Dirac $\delta$-function in the transverse direction, Fourier transformed in $\bf q$ space.  In the space of
$\bf q$, and $z$ taking 0 and $d$,  the matrices  are $2 \times 2$, as 
\begin{equation}
(D^r)^{-1} = \left[ \frac{1}{2 \epsilon_0 q \Omega} \left( \begin{array}{cc}
1 &  e^{-qd} \\
e^{-qd} & 1  \end{array}
\right) \right]^{-1}\!\!\! - 
\left( \begin{array}{cc}
\Pi_L & 0 \\
0 & \Pi_R \end{array}
\right).
\end{equation} 
For the first term as $v^{-1} = C$, the inverse can be worked out to give
$C_{11}  = C_{22} = 2 \epsilon_0 q \Omega/(1- e^{-2qd})$, and $C_{12} = C_{21} = - e^{-qd} C_{11}$.   We note that if we take the limit $q \to 0$, we obtain the same matrix as before
with a capacitance in this case as $\epsilon_0 \Omega/d$, i.e., the effective area of the capacitor is the area of a unit cell. 
The explicit expression for $D^r$ matrix elements can be easily worked out, e.g., 
\begin{equation}
D^{r}_{LR} = D^{r}_{RL} = { -C_{21} \over (C_{11} - \Pi_L) (C_{22} - \Pi_R) - C_{12} C_{21} }. 
\end{equation}
The advanced version is obtained by Hermitian conjugate.  We define the reflection coefficient as 
\begin{equation}
r_\alpha = \frac{v_\alpha \Pi_\alpha}{ 1- v_\alpha \Pi_\alpha} = v_\alpha \chi_\alpha, \quad \alpha = L, R.
\end{equation}
Here, $v_\alpha = 1/( 2 \epsilon_0 q \Omega)$ is the Coulomb potential in two-dimension in $\bf q$ space, 
$\chi_\alpha$ is charge-charge correlation or susceptibility, while $\Pi_\alpha$ is the self-energy or polarizability. 
$\epsilon_\alpha = 1- v_\alpha \Pi_\alpha$ is the dielectric function.   Using the reflection coefficients with some algebra, 
we can simplify the transmission as \cite{ilic}
\begin{equation}
T_{\bf q}(\omega) = { 4\, {\rm Im\,} r_L\, {\rm Im\,} r_R\, e^{-2 q d} \over
\left| 1 - r_L r_R\, e^{-2qd}  \right|^2}.   
\end{equation}
Of course, $r_\alpha$ is a function of the wave-vector $\bf q$ as well as the angular frequency $\omega$ which we have suppressed.
One might be curious about why $r_\alpha$ is called a reflection coefficient.   Indeed, it does have to do with the wave reflection.   In the 
traditional approach to near-field heat transfer of Polder and van Hove \cite{Polder71,Pendry99}, one solves a wave scattering problem with transmission and
multiple reflections between the plates.  Ignoring the $s$-wave polarization, which is small at near distance, and focusing on the
$p$-wave (electric field is in the plane of incidence) and at the non-retardation limit of $c \to \infty$, one obtains exactly the same result as above using the fluctuational electrodynamics.   Here our approach is based on NEGF, which in some way is simpler.
 
Finally, to obtain the total heat current, one sums over all the modes $\bf q$ in the first Brillouin zone with $N$ sampling points,   
and integrates over the frequency, i.e.,  $I_L = \int_{-\infty}^{+\infty} d\omega/(4\pi) \, \hbar \omega \sum_{\bf q} T_{\bf q}(\omega) (\tilde{N}_L
-\tilde{N}_R)$.
The number of $\bf q$ 
points is related to the actual area of the plates by $A = N \Omega$.   To obtain the heat transfer per unit area, we divide by the
area, i.e.,  $(1/A) \sum_{\bf q} \cdots = \int d^2{\bf q}/(2\pi)^2 \cdots$.    It is important that the points of $\bf q$ are 
in the first Brillouin zone, i.e., the Wigner-Seitz cell with the $\Gamma$-point at the cell center.   This is because although the materials
have lattice periodicity with $\Pi_{\bf q}$ a periodic function in the reciprocal space, the photon free Green's function $v$ does not
has such periodicity, our forcing $D^r$ running on lattice sites is an approximation. 

\begin{figure}[htp] 
  \centering
  \includegraphics[width=\columnwidth]{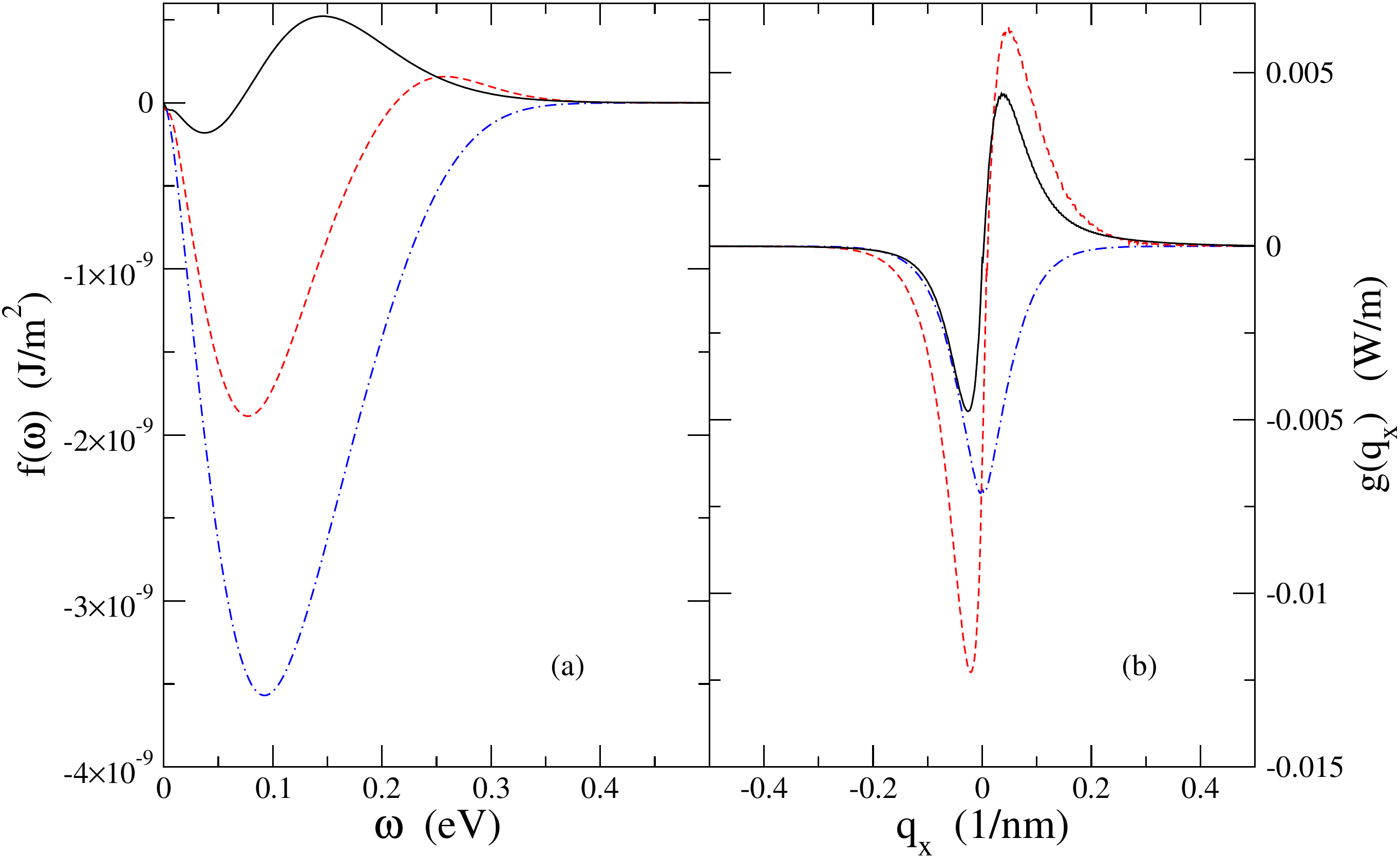}
  \caption{ (a) Integrated spectral transfer function $f(\omega)$ as a function of frequency and (b) 
  $g(q_x)$ as a function of wave vector in the driven direction, with different drift velocities: no drift (blue dash-dot line), total heat
  current density $I_L/A=-0.84\,$MW/m$^2$; $v_1=5.0 \times 10^5 \ \mathrm{m/s}$ (red dash line), $-0.30\,$MW/m$^2$; and $v_1=9.0 \times 10^5 \ \mathrm{m/s}$
  (black solid line), $+0.09\,$MW/m$^2$.  The temperatures are $T_L=300\,$K and $T_R=320\,$K. The chemical potential of graphene $\mu$ is set as 0.1\,eV. Gap distance $d$ is set as 10\,nm.  The damping parameter is $\eta = 9\,$meV.  From \onlinecite{Peng1805arxiv}, Fig.~2.}
 \label{DL-figure2}
\end{figure} 	

\begin{figure}[htp] 
  \centering
  \includegraphics[width=0.85\columnwidth]{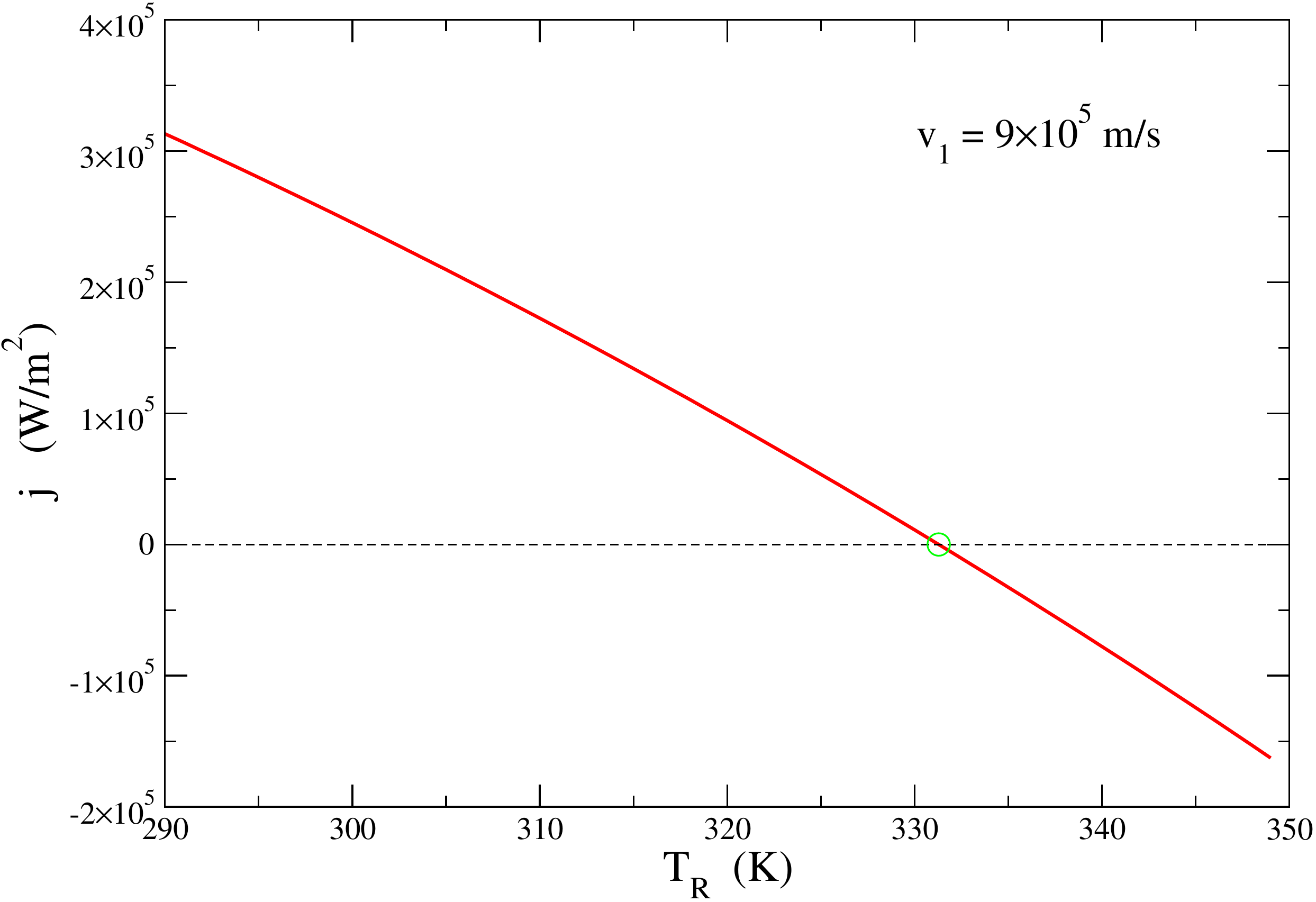} 
  \caption{Heat current density from left to right layer as a function of $T_R$ (temperature of the right layer) with drift velocity $v_1=9.0 \times 10^5 \ \mathrm{m/s}$ (red solid line).  The dotted line is a reference line for zero current density. The green circle indicates the point for ``off temperature''. The chemical potential of graphene $\mu$ is set as 0.1\,eV. Temperature of left layer of graphene $T_L$ is set as 300\,K. Vacuum gap distance $d$ is set as 10\,nm.   From \onlinecite{Peng1805arxiv}, Fig.~3.}
 \label{DL-figure3}
\end{figure} 

After this long preparation, in Fig.~\ref{DL-figure2}, we illustrate the effect of drift velocity to heat transfer in the two-layer
graphene setup.   We emphasize that the current drive not only changes the Bose distribution by a Doppler shift, but also, the 
self-energy $\Pi$ or in turn the reflection coefficient $r$ needs to be recalculated anew.   The changes in $\Pi_\alpha$ or $r_\alpha$
are necessary so that the integrand for the heat transfer is not divergent or at least still integrable at the point when 
$\omega - {\bf q} \cdot {\bf v}_1 = 0$.   Details are in \onlinecite{Peng1805arxiv}. 
 Here we drive with velocity $v_1$ in $x$ direction for the left layer (called 1 in the original paper), 
and compute the heat transfer out of the left layer.   Although the drive breaks symmetry in frequency for each given $\bf q$, after
integration over all $\bf q$, the even symmetry in $\omega$ is restored.  Figure~\ref{DL-figure2}(a) plots the $\bf q$ integrated result $f(\omega)$ for fixed
$\omega$.  Further integration over frequency gives the total heat transfer.   We notice originally without drift, heat is flowing
from right to left (negative value on the plot) since the right is hotter.  But as the drift velocity increases, the heat transfer reverses sign, going from cold to hot.   
This is understandable as the left layer of graphene is no longer in local thermal equilibrium, and it is not a broken down of the second 
law of thermodynamics.   Figure~\ref{DL-figure2}(b) demonstrates the effect of drift to the distribution of the total integrand over $q_x$
integrated over $q_y$ and $\omega$.   Without drift, the distribution in $q_x$ is symmetric with respect to $q_x=0$, while drift breaks this symmetry, and it turns
into having both positive and negative contributions when $v_1$ is large, causing a cancellation effect for the total heat transfer. 
In Fig.~\ref{DL-figure3}, the reversal of the heat transfer direction is clearly shown when the temperature of the right-side sheet is
varied.    There is a particular balance temperature where heat transfer is zero even though the temperatures of the two sheets are 
different.  Above this temperature, we have a net cooling effect for the right side due to
the current driven on the left side.   We comment that driving a conductor with current will produce Joule heat as well as electro-luminescence, the 
effects of electron-phonon and electron-transverse-photon interactions.   
This extra heat has not been taken into account in our theory. 
 
\section{Full counting statistics for energy transfer}
At the nanoscale due to thermal agitation, the measured results themselves are fluctuational quantities \cite{Herz20,Wise-arxiv22}. 
The full counting statistics \cite{Wang11prb,Tang18} here means that we compute not only energy but also high order moments in a transport setup.   
To compute the higher moments, it is convenient we compute the total heat $Q$ of a fixed duration in a two-time measurement of the left bath
$\hat H_L$.   The energy of the left bath is measured at an earlier time $t_0$ and then measured again at time $t$,  the decrease in 
energy is the transfer of the heat to the right out of the left bath.    According to the standard measurement interpretation of quantum mechanics,
the result of  a measurement is an eigenvalue of $\hat H_L$.    We will be interested in the long time $(t-t_0)\to \infty$ which gives
a simpler result.    Other protocols of measurements are also possible, but the two-time measurement results in a simpler
mathematics, although it is not clear how this measurement  of energy of the bath can be done experimentally when the bath is 
supposed to be infinitely large.   For the long-time result, the details of measurement protocols should not matter.

We consider the system as two blocks of metal with a Hamiltonian $\hat H_L + \hat H_R + \hat H_\gamma + \hat V = \hat H_0
+\hat V$.    $\hat H_\gamma$ is the negative-definite free scalar photon Hamiltonian.  The
last term $\hat V$ is the Coulomb interaction in the scalar field form,  $\hat V = \sum_{jkl} M_{jk}^l c^\dagger_j c_k \phi_l$. 
Only parts of the sites near the interface between the left and right blocks have the Coulomb interaction.   Deep into the 
baths, we set the Coulomb interaction to zero (due to screening).  
We prepare the system to be in the decoupled product initial state given by the density matrix
\begin{equation}
\hat \rho(t_0) \propto  e^{-\beta_L (\hat H_L - \mu_L \hat N_L)}\, e^{-\beta_R (\hat H_R - \mu_R \hat N_R)}\, e^{-\beta_\gamma \hat H_\gamma}, 
\end{equation}
here the left and right baths are in the grand-canonical ensembles and the last factor is for the scalar photons in canonical ensemble.   We
can set $\beta_\gamma = 0$ without loss of generality.    

We can prove a very general formula by defining a ``partition function'' or moment generating function as
\begin{equation}
\label{eq-Zxi}
Z(\xi) = \left\langle T_\tau  e^{-\frac{i}{\hbar} \int d\tau \hat V^x(\tau)} \right\rangle_{\hat H_0}.
\end{equation} 
Here, the exponential is contour ordered, and the integral is over the contour from $t_0$ on the upper branch to $t$ and then back 
to $t_0$ from the lower branch.   $\hat V^x$ is the interaction term, but Heisenberg  evolved by $\hbar x$ in the interaction
picture by the left bath, i.e.,
\begin{equation}
\hat V^x = e^{ix \hat H_L} \hat V e^{-ix \hat H_L}.
\end{equation}
The amount of $x$ is contour dependent, it is $-\xi/2$ on the upper $(+)$ branch and $\xi/2$ on the lower $(-)$ branch.   With
this definition of $Z$, the $n$-th moment of $Q$ is obtained by $n$-th derivative of $Z$,
$\langle Q^n \rangle = \partial^n Z(\xi) /\partial (i\xi)^n$ evaluated at $\xi=0$ and the $n$-th order cumulant is
\begin{equation}
\langle\langle Q^n \rangle\rangle = { \partial^n \ln Z(\xi) \over \partial (i\xi)^n }\Big|_{\xi=0}.
\end{equation}   
The advantage of working with the cumulants instead of moments is that they are all linear in $t-t_0$ at long time.
The first order moment and cumulant are the same, 
$\langle Q \rangle = \langle \langle Q \rangle\rangle = \partial \ln Z/\partial (i\xi) = (t-t_0)I_L$, $I_L$ is the current. 
The second cumulant is just the variance or the fluctuations of the current,  $\langle \langle Q^2 \rangle\rangle = \langle Q^2 \rangle - \langle Q \rangle^2 = (t-t_0)\sigma^2$. 

To show the validity of Eq.~(\ref{eq-Zxi}), let $\hat H_L |\varphi_a\rangle = a |\varphi_a\rangle$, here $a$ is the eigenvalue
of the isolated left side with the eigenstate $|\varphi_a\rangle$.    If a measurement is performed at time $t_0$ obtaining the eigenvalue
$a$, and then measured again at time $t$ obtaining the eigenvalue $b$, the generating function is
\begin{eqnarray}
Z(\xi) &=& \sum_{a,b} e^{i(a-b)\xi} P(b,a)  \\
&=& \sum_{a,b} e^{i(a-b)\xi} {\rm Tr}\Bigl[ \hat \rho(t_0)  \Pi_a U(t_0,t) \Pi_b U(t,t_0)  \Bigr]. \nonumber
\end{eqnarray}
Here $P(b,a)$ is the probability of being in state $b$ at time $t$ given that it is in state $a$ at an initial time $t_0$, which can be
expressed by the evolution operator $U$ of the full Hamiltonian and the projectors of the respective states,
\begin{equation}
\Pi_a = | \varphi_a \rangle \langle \varphi_a|. 
\end{equation} 
We see that $Z(0)=1$ due to the probability normalization.  Taking the derivative with respect to $(i\xi)$ $n$ times, and then
set $\xi$ to 0, we obtain the expectation value of $Q^n = (a-b)^n$ over the probability $P(b,a)$. 
we have $[\Pi_a, \hat H_L]  = 0$, 
and because of the choice of the product initial state, we also have $\bigl[\Pi_a, \hat \rho(t_0)\bigr] = 0$.  Since
\begin{equation}
\sum_a e^{ia \xi} \Pi_a = e^{i\xi \hat H_L},\qquad 
\sum_b e^{-ib \xi} \Pi_b = e^{-i\xi \hat H_L},
\end{equation}
we can express $Z(\xi)$ as 
\begin{eqnarray}
Z(\xi) & = & {\rm Tr}\Bigl[ \hat \rho(t_0)  e^{i\xi \hat H_L} U(t_0,t) e^{-i\xi \hat H_L} U(t,t_0)  \Bigr] \nonumber \\
&=& {\rm Tr}\Bigl[ \hat \rho(t_0)  U^{\frac{\xi}{2}}(t_0,t) U^{-\frac{\xi}{2}}(t,t_0)  \Bigr].
\end{eqnarray}
Here we have split the exponential factors into two halves and used a cyclic permutation of the trace.   The superscript on $U$ denotes
an extra Heisenberg evolution with $\hat H_L$, i.e.,  $U^x = e^{i x \hat H_L} U  e^{-i x \hat H_L}$.  This extra $x$ dependence
can be transferred from $U$ into the Hamiltonian,  to give $\hat H^x = \hat H_0 + \hat V^x$.   Here the noninteracting Hamiltonian
$\hat H_0$ is unaffected as $\hat H_L$ commutes with the three free terms, $\hat H_L$, $\hat H_R$, and $\hat H_\gamma$. 
At this point, we transform the expression into the interaction picture, given Eq.~(\ref{eq-Zxi}) with the average evaluated with respect
to $\hat H_0$.   In the interaction picture, the effect of the Heisenberg evolution is to shift the time argument of the left side
as $c_L(\tau) = c_L(t-\hbar \xi/2)$ on the upper branch, and $c_L(t+\hbar \xi/2)$ on the lower branch.   This in turn means a corresponding shifts of time argument for the 
self-energy $\Pi_L$.

\begin{figure}[htp] 
  \centering
  \includegraphics[width=1.0\columnwidth]{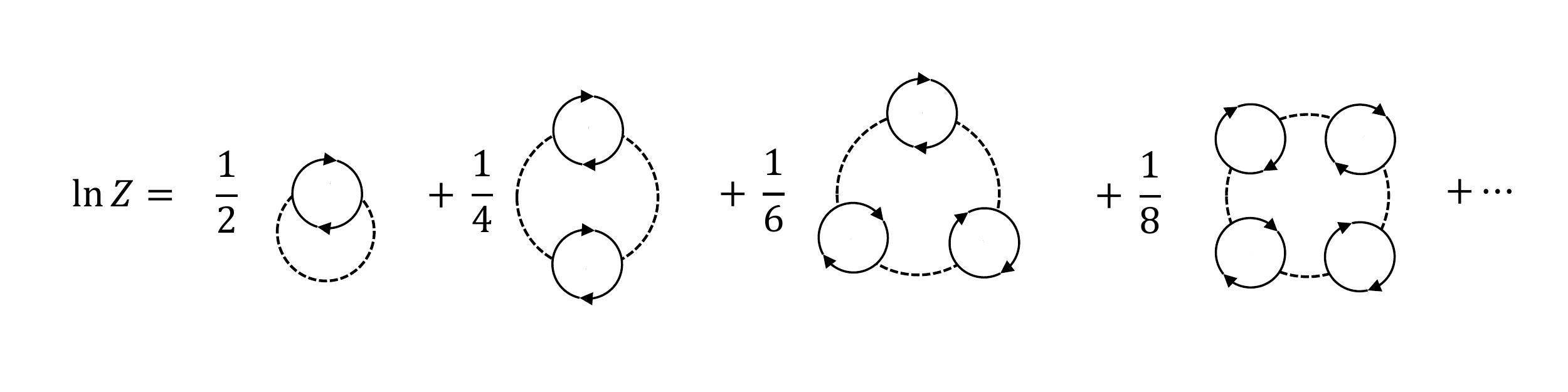} 
  \caption{The first few terms of diagrammatic expansion of ${\rm ln} Z$.}
 \label{fig-log-Z}
\end{figure} 

The rest of the steps follow a standard diagrammatic expansion.   We work at the level of RPA for the Coulomb interaction.
Since the photon Hamiltonian is quadratic in the scalar field $\phi$, an odd number of $\hat V$ evaluates to zero.  Thus the
lowest order of nonzero term, when the exponential is expanded, is $1/(2(i\hbar)^2) \int d\tau_1 \int d\tau_2 \langle
T_\tau \hat V(\tau_1) \hat V(\tau_2)\rangle_0$.  Applying Wick's theorem, we can write the result as 
$\frac{1}{2} {\rm Tr}_{\tau j}(v \Pi^{(0)})$, here $v$ is the bare Coulomb in $j$ and $\tau$ space which contains a 
$\delta$ function in the $\tau$ variable as Coulomb interaction is instantaneous, and $\Pi^{(0)}$ is the RPA bubble 
diagram result.   We will drop the superscript $(0)$ for notational simplicity.   If we continue the expansion to higher orders,
and collect only these bubble diagrams that form a ring  (see Fig.~\ref{fig-log-Z}), we can sum the diagrams as a logarithm due to the $1/n$ factor at the 
$n$-th order. These diagrams are the same as the RPA result for the grand potential in a Matsubara Green's function
formulation \cite{Mahan00}.  Finally, we have the RPA expression for the generating function as
\begin{equation}
\ln Z(\xi) = - \frac{1}{2} {\rm Tr} \ln \bigl( I - v (\Pi_L^x + \Pi_R)\bigr).
\end{equation}
Here the trace is performed in the combined contour time $\tau$ and site $j$ space, that is,
${\rm Tr}( \cdot ) = \sum_j \int d\tau\, (\cdot)$. 
$I$ is the identity is this space. To aid the Fourier transform 
in the full time domain, it is convenient to take the limit $t_0 \to -\infty$.    We eliminate the bare Coulomb $v$ in favor
of the screened Coulomb $D$ by introducing $\Pi^A_L = \Pi^x_L - \Pi_L$,  then 
$I - v(\Pi_L^x + \Pi_R) = v( v^{-1} - \Pi_L - \Pi_R) - v \Pi_L^A = v(D^{-1} - \Pi^A_L) = 
v D^{-1} (I - D \Pi_L^A)$.    Since the factor $vD^{-1}$ does not depend on $\xi$, it is an additive constant to $\ln Z$ and
will not contribute to the derivatives, so we drop it and redefine the generating function as
\begin{equation}
\ln Z(\xi) = - \frac{1}{2}  {\rm Tr} \ln\left( I - D \Pi_L^A\right). 
\end{equation}
If we Taylor expand in $\xi$, the linear term gives the current $I_L$.  After simplifying from the contour time to real time, and 
then Fourier transform to frequency domain for the Green's functions and self-energies, we recover the Meir-Wingreen formula,
Eq.~(\ref{eq-photon-IL-WM}).     If we expand up to second order in $\xi$, we obtain the variance.  Explicit formula for the variance of heat transfer has been 
derived by \onlinecite{Herz20} within a scattering operator formalism. 
 If we use local equilibrium, we can express the generating function in terms of a matrix
$D^r \Gamma_R D^a \Gamma_L$ which is identical to the usual Levitov-Lesovik formula \cite{Agarwallapre12,Tang18,Levitov93}. 

\section{Density functional theory calculation based on Coulomb interaction}

So far, our electron models have been in the tight-binding form where the electrons are allowed to sit only at a discrete set of sites.   This has
the computational efficiency advantage of dealing with finite-dimensional Hermitian matrices, enough to capture the solid-state band structure 
for a lattice, for example.
In fact, our systems can be any structure unrestricted by lattice periodicity, e.g., tips of two triangles made of graphene \cite{Tang19}.    
Assuming that electrons sit only at certain discrete places is an approximation.    The density functional theory (DFT) \cite{Parr89} offers 
a parameter-free approach to real complex materials.  In DFT approach 
the electrons are treated as distributed in the whole space continuously.    Another feature of the plane-wave based DFT
is that we must work on a periodic unit cell or super-cell.  As a result, we must consider the fluctuations of the electron density within
a cell.   The theory developed earlier based on the Green's function $D$ and the self-energy $\Pi$ needs some revisions, but it is
only of a technical nature and no new conceptual difficulty remains.

The definition $D$ remains the same as given by Eq.~(\ref{eq-contour-D-def}), whether the electrons are treated as discrete or 
continuous degrees of freedom.   For a continuum of electron density, $\rho$, it is convenient to define the contour version $\Pi$  as
\begin{equation}
\label{eq-contour-dPi}
\Pi({\bf r}\tau; {\bf r}'\tau') = \frac{1}{i\hbar} \bigl\langle T_\tau \rho({\bf r},\tau)\rho({\bf r}',\tau') \bigr\rangle_{\rm ir}.
\end{equation}
Here the subscript $\rm ir$ means we take only the irreducible diagrams in a Feynman-diagrammatic expansion with the
Coulomb interaction.  The irreducible ones are those that cannot be cut into two disconnected pieces
by a single bare scalar photon line $v$.   Without $\rm ir$, the charge-charge correlation is $\chi = \Pi \epsilon^{-1}$; 
$\epsilon = 1 - v \Pi$ is the longitudinal dielectric function.  Under the 
random phase approximation, we just take the lowest order bubble diagram. 
We can lump the charges in a cell to a point.  Assuming all the cells having the same volume $\Omega$, the relation between
a continuous description and the earlier discrete description for $\Pi$ is  $\Pi_{ij} = \Omega^2 \Pi({\bf r}_i, {\bf r}_j)$.   Here
${\bf r}_i$ is a point in the cell $i$.  Exactly where it is in the cell does not matter provided that the function varies with the position 
smoothly enough.   From this relation, we can see that the Dyson equation in a continuous charge description is obtained by replacing the summations  by integrals over space, i.e.,  in Eq.~(\ref{eq-dyson-Dtau}), we replace  $\sum_{j,k} \int d\tau_1 \int d\tau_2 v({\bf r}\tau,{\bf r}_j \tau_1)  
\Pi_{jk}(\tau_1, \tau_2) D({\bf r}_k\tau_2,{\bf r}' \tau')$ by the integral
\begin{displaymath}
\int \!\! d^3{\bf r}_1\! \int \!\! d\tau_1\! \int \!\! d^3{\bf r}_2\! \int \!\! d\tau_2 v({\bf r}\tau,{\bf r}_1 \tau_1)  \Pi({\bf r}_1\tau_1,{\bf r}_2 \tau_2)
D({\bf r}_2\tau_2,{\bf r}' \tau').
\end{displaymath}

The quantities like the above expressed in real space $\bf r$ are not convenient for actual computation.  The translation invariance 
in a crystal means we should evoke the convolution theorem in Fourier transform so that it becomes multiplication in $\bf q$ space.  
However, if we also take into account the local inhomogeneity inside a cell, the Fourier transform is slightly more complicated for
which we elaborate below. 

For a periodic system, the single-electron wave function satisfies the Bloch theorem \cite{Ashcroft-Mermin76}, 
$\phi({\bf r}) = e^{i {\bf k} \cdot {\bf r}} u({\bf r})$; that is, a plane wave modulated by a periodic function, 
$u({\bf r}) = u({\bf r} + {\bf R})$.   Here ${\bf R} = l_1 {\bf a}_1 + l_2 {\bf a}_2 + l_3 {\bf a}_3$ specifies the 
Bravais lattice sites by the unit cell vectors ${\bf a}_i$ and integers $l_i$.  This fact implies that the average electron density is a cell-periodic function, 
\begin{equation}
\langle \rho({\bf r})\rangle  = \langle \rho({\bf r} + {\bf R}) \rangle,
\end{equation}
since under the Kohn-Sham DFT framework, the electron density is simply the sum of  $|\phi({\bf r})|^2$ of the occupied bands. 
The operator $\rho$ itself before the thermodynamic average does not have this cell periodicity.  A periodic function can be Fourier expanded as
\begin{equation}
\langle \rho({\bf r})\rangle  = \sum_{\bf G} \rho_{\bf G}\, e^{ i {\bf G} \cdot {\bf r} },
\end{equation}
here the capital letter $\bf G$ is the reciprocal lattice vector, running over all the values ${\bf G} = n_1 {\bf b}_1 + n_2 {\bf b}_2 + 
n_3 {\bf b}_3$, $n_i$ are integers and ${\bf a}_i \cdot {\bf b}_j = 2\pi \delta_{ij}$, $i,j=1,2,3$.  The Fourier coefficients are obtained by
$\rho_{\bf G} = (1/\Omega) \int_\Omega \langle \rho({\bf r}) \rangle e^{-i {\bf G} \cdot {\bf r}} d^3{\bf r}$, 
integrating over one unit cell with cell volume $\Omega$.  

The charge-charge correlation involves two positions, possibly in different unit cells.   For simplicity, let us consider the static 
correlation $f({\bf r}, {\bf r}') = \langle \rho({\bf r}) \rho({\bf r}') \rangle$.   The lattice cell translation means that, if we shift
both positions by a common Bravais lattice vector, the correlation between the new pair should be the same, i.e., we have
\begin{equation}
f({\bf r} + {\bf R}, {\bf r}' +{\bf R}) = f({\bf r}, {\bf r}').
\end{equation} 
Note that this is different from a continuous translation symmetry of $f$ being a function of the difference ${\bf r} - {\bf r}'$ only. 
We will show that the cell translation symmetry can be Fourier expanded with double series $\bf G$ and ${\bf G}'$ and a Fourier
integral in the first Brillouin zone, as 
\begin{equation}
\label{eq-doubleG}
f({\bf r},{\bf r}') = \sum_{{\bf G}, {\bf G}'} \int_{\rm 1BZ}\! \frac{d^3 {\bf q}}{(2\pi)^3} 
\tilde{F}_{{\bf G}{\bf G}'}({\bf q}) e^{i ({\bf G} + {\bf q}) \cdot{\bf r} - i ({\bf G}' + {\bf q}) \cdot {\bf r}'}. 
\end{equation}
In the above, if we keep only the ${\bf G}={\bf G}' = {\bf 0}$ term, we obtain a continuous translation symmetry,
$f({\bf r}, {\bf r}') = f({\bf r}-{\bf  r}')$; this is a long-wave approximation.  The extra nonzero $\bf G$-vector terms
reflect the local inhomogeneity within a cell. 
To proof this result, first we can choose a Bravais vector such that the first argument is in the first cell,  and define 
$F({\bf r}, {\bf \Delta}) = f({\bf r}, {\bf r} + {\bf \Delta})$.  Here $\bf \Delta$ is still arbitrary running over the whole space. 
If we consider $F$ as a function of $\bf r$ fixing the difference ${\bf \Delta} = {\bf r}' - {\bf r}$, $F$ is a periodic function of 
$\bf r$.  So, we can write 
\begin{equation}
F({\bf r}, {\bf \Delta}) = \sum_{\bf G} c_{\bf G}({\bf \Delta}) e^{i {\bf G} \cdot {\bf r}}. 
\end{equation}
The Fourier coefficients $c_{\bf G}$ are still a function of $\bf \Delta$ which varies continuously in the full space. 
For the variable $\bf \Delta$ we can make a Fourier integral transform into $\bf q$.  This is 
\begin{equation}
c_{\bf G}({\bf \Delta}) = \int \frac{d^3 {\bf q}}{(2\pi)^3} \tilde{f}_{\bf G}({\bf q})\, e^{- i {\bf q} . {\bf \Delta}}.
\end{equation}
The integral over the full Fourier space $\bf q$ can be split into pieces of reciprocal space cell ${\bf G}'$ with 
a change of variable to each cell by ${\bf q} \to {\bf G}' + {\bf q}$.   Then the new $\bf q$ variable varies in the
first Brillouin zone only.  After some regrouping and simplification, noting that  ${\bf \Delta} = {\bf r}' - {\bf r}$, and
defining $\tilde{F}_{{\bf G}{\bf G}'}({\bf q}) = \tilde{f}_{{\bf G}-{\bf G}'}({\bf G}' + {\bf q})$, we obtain the desired
result, Eq.~(\ref{eq-doubleG}).   Tracing back the steps, we can compute the Fourier expansion coefficient as
\begin{eqnarray}
\label{eq-doubleG-back}
\tilde{F}_{{\bf G}{\bf G}'}({\bf q}) &=& \sum_{{\bf R}} \frac{1}{\Omega} \int_\Omega d^3{\bf r}\int_\Omega d^3{\bf r}'
f({\bf r} + {\bf R},{\bf r}') e^{-i \varphi}, \nonumber\\
\varphi &=& ({\bf G} + {\bf q}) \cdot{\bf r} - ({\bf G}' + {\bf q}) \cdot {\bf r}' +  {\bf q} \cdot {\bf R}. 
\end{eqnarray} 
Here both of the integral variables $\bf r$ and ${\bf r}'$ are in the first unit cell.   With the ${\bf G}$, ${\bf G}'$, and $\bf q$
variables for the correlation functions, we also have a convolution theorem.   That is, the expression of type $v \Pi D$ in
the Dyson equation can be written as a matrix multiplication indexed by ${\bf G},{\bf G}'$ for each given $\bf q$.  Similarly,
a trace by integration over position $\bf r$ can now be expressed as a trace in $\bf G$ as a matrix trace and integration of $\bf q$ 
in the first Brillouin zone. 

\subsection{Adler-Wiser formula}
We now give a formula for the retarded $\Pi^r$ expressed by the Kohn-Sham or independent single-particle orbitals known as 
the Adler-Wiser formula \cite{Adler62,Wiser63}.  
The retarded formula can be computed according to $\Pi^r(t) = \theta(t) \bigl(\Pi^>(t) - \Pi^<(t)\bigr)$, and then Fourier transformed
into frequency domain.   The lesser and greater components can be read-off from the contour expression, Eq.~(\ref{eq-contour-dPi}), as
$\Pi^>(t) = \langle \rho(t) \rho(0) \rangle/(i\hbar)$ and 
$\Pi^<(t) = \langle \rho(0) \rho(t) \rangle/(i\hbar)$.   We remind the reader here that $\rho$ is a quantum operator, which
can be expressed in the quantum field as $\rho = (-e) \Psi^\dagger \Psi$, where $\Psi$ is space and time dependent which we 
expand in the mode space,
\begin{equation}
\Psi({\bf r}t) = \sum_{nk\sigma} c_{n{\bf k}\sigma}(t)\, \phi_{n{\bf k}\sigma}({\bf r}).
\end{equation}
Here the eigenmodes are labeled by the band index $n$, the wave vector $\bf k$, and spin $\sigma = \uparrow, \downarrow$.
However, for a spin-independent problem, the net effect of spin degeneracy is simply multiplying the final expression of the 
polarizability by a factor of 2.   In the following, we will treat our electrons as spinless and then keep a factor of 2 for
$\Pi^r$.   The Kohn-Sham wave function $\phi_{n{\bf k}}$ must be normalized to 1 in the whole system of volume 
$V = N \Omega$ in order to give the correct electron density.  
In mode space, the Hamiltonian is diagonal, thus the time-dependence for the annihilation operator is
simply the free evolution, $c_{n{\bf k}}(t) = c_{n {\bf k}}\, e^{- i \epsilon_{n{\bf k}}t/\hbar}$, here $\epsilon_{n{\bf k}}$
is the electron band energy.   

In evaluating the density-density correlation, we encounter terms of the form $\langle c^\dagger c\, c^\dagger c\rangle$  which
we apply Wick's theorem to factor into product of two $c$'s.   Noting that $\langle c c\rangle$ or $\langle c^\dagger c^\dagger\rangle$ is 
0, the remaining terms are related to the Fermi function, i.e.,  $\langle c_{j}^\dagger c_l\rangle = \delta_{jl} f_j$, and 
$ \langle c_{j} c_l^\dagger \rangle = \delta_{jl} (1-f_j)$.   Here we have used a short-hand notation $j\equiv (n{\bf k})$,
 $l\equiv (n'{\bf k}')$. 
With some algebra, we can express the retarded scalar photon self-energy in the frequency domain as
\begin{equation}
\Pi^r({\bf r},{\bf r}', \omega) = 
2 e^2 \sum_{jl} { (f_j -f_l)  \phi_j({\bf r}) \phi_j^{*}({\bf r}') \phi_l({\bf r}')\phi_l^{*}({\bf r})
\over  \epsilon_j - \epsilon_l - \hbar \omega - i \eta}.
\end{equation}    
Here the extra small damping $i \eta$ in the denominator is necessary, so that the poles in a complex frequency plane is
below the real axis, and the inverse transform has the property $\Pi^r(t) = 0$ when $t<0$.  This $\eta$ parameter also has
a physical meaning.  We can interpret it as the inverse of electron lifetime of the quasi particle.   Finally, using our general 
transformation formula for the $({\bf r},{\bf r}')$ correlation to the ${\bf G}, {\bf G}', {\bf q}$ space, the self-energy 
becomes
\begin{eqnarray}
\Pi^r_{{\bf G}{\bf G}'}({\bf q}, \omega)\,= \hspace{16em}  \\
  \frac{2e^2}{N\Omega} 
\sum_{j,l} 
{ \langle l|e^{-i({\bf G} + {\bf q})\cdot \hat {\bf r}} |j \rangle  \langle j|e^{i({\bf G}' + {\bf q})\cdot \hat {\bf r}} |l \rangle  
\bigl(f_j -f_l\bigr)
\over  \epsilon_j - \epsilon_l - \hbar \omega - i \eta}.\nonumber
\end{eqnarray}
Here we define the matrix element $\langle l |\hat{A} |j \rangle \equiv \int_{N\Omega} d^3 {\bf r}\, \phi_l^* \hat A \phi_j$; the
integration is over the full volume with $N$ unit cells.   This is the Adler-Wiser formula.    Note that the single particle operator
$\hat A$ is a shift of the momentum, so the matrix elements are zero unless the momentum $\bf k$ in $j$ is related to that in $l$ by
${\bf k} = {\bf k}' + {\bf q}$ modulo $\bf G$.    Efficient evaluation of this expression with massively parallel algorithms has been 
implemented in the well-known BerkeleyGW package \cite{bgw1,bgw2}, but only for zero temperature. 

\subsection{Solving the Dyson equation}
Here we consider the parallel plate geometry for near-field heat transfer.   Since the first-principles codes require periodic supercell
while the heat transfer problem is intrinsically a non-periodic problem in the transport direction, there is a fundamental conflict.  
We can imagine putting two slabs of materials into the simulation cell with sufficient vacuum gaps.  Although the transverse directions
are intrinsically periodic, the transfer direction, if we still use periodic boundary condition, may requires large vacuum gap to void
artificial interactions.   As a result, it is best we treat each slab separately by DFT and combine the results \cite{Quek19}.   We also assume the periodicity
in the transverse directions for the two slabs is the same, otherwise, how to combine them is a problem.   Finally, if the slab is
thick, we work in a mixed representation of ${\bf G}_{\perp}$ and $z$, and consider the 
self-energy in the form 
$\Pi^r_{{\bf G}_\perp{\bf G}'_\perp}({\bf q}_\perp, z,z',\omega)$.  Here the $z$ dependence is obtained by Fourier transforming $G_z,G'_z,q_z$ into
real space,  and ${\bf G}_\perp = (G_x, G_y)$ lays in the plane.    

Very likely the above approach is still too computationally intensive; so far no one was able to do a calculation for near-field heat transfer.   In the following,
we make further approximation, this is to treat the slab as infinitely thin, in such a way that the density of the electrons is  
confined strictly in 2D at location $z=0$ and $d$.   In this way, the $z$ variables become discrete.    We can define surface charge
density by integrating over $z$ of the volume density,  $\sigma = \int \rho\, dz$, and define the surface version of $\Pi$.   A careful analysis shows that this surface version can be obtained from the volume version in $\bf q$ space just by multiplying the supercell length 
$c$ in $z$ direction, i.e., \cite{Rasmussen16},
\begin{equation}
\Pi^{\rm 2D}_{{\bf G}_\perp{\bf G}'_\perp}({\bf q}_\perp,\omega) = 
c\, \Pi^{\rm 3D}_{({\bf G}_\perp,0),({\bf G}'_\perp,0)}(({\bf q}_\perp, 0),\omega). 
\end{equation}
Here in 3D the wave vector ${\bf q}= ({\bf q}_\perp, q_z)$ and ${\bf G}=({\bf G}_\perp, G_z)$, and for both of them 
the $z$-components are set to 0 on the right-hand side of the equation.   This is a convenient formula to use as existing 
DFT codes are for 3D problems. 

The Dyson equation needs to be transferred into a mixed representation, that is, ${\bf r}_\perp = (x,y)$ variables are transformed to 
${\bf G}_\perp$ and ${\bf q}_\perp$ and $z$ variable stays.   Using our general transformation, Eq.~(\ref{eq-doubleG}) and
(\ref{eq-doubleG-back}), specialized to 2D, we obtain 
\begin{eqnarray}
\label{eq-dyson-Dqperp}
D({\bf q}_\perp, \omega) &= &  v({\bf q}_\perp) + \\
&&  v({\bf q}_\perp) 
\left( \begin{array}{cc} 
 \Pi_0({\bf q}_\perp, \omega) & 0 \\
0 & \Pi_1({\bf q}_\perp, \omega)
\end{array} \right)
D({\bf q}_\perp, \omega) . \nonumber 
\end{eqnarray}
Here $v$ and $D$ are matrices in the combined $z,{\bf G}_\perp$ space, $z = 0$ and $d$ only.  ${\bf G}_\perp$ runs over
an energy cut-off choice.   If we take only ${\bf G}_\perp = {\bf G}'_\perp = {\bf 0}$, our matrices $v$ and $D$ will be $2\times 2$ which
gives a result where local inhomogeneity is ignored.   $\Pi_0$ and $\Pi_1$ are the 2D polarizability matrix located at $z=0$ and
$d$, respectively.   This equation for the retarded $D$ is easily solved, in the form $(I-v\Pi)D = v$, by calling numerical packages 
such as LAPACK \cite{lapack99}.    
 
Before closing this long theory session, we also need to transform the bare Coulomb Green's function into mixed representation.  
In real space $\bf r$ and time domain $t$, it is given by Eq.~(\ref{eq-vrt}).   We note that the bare Coulomb potential is a
function of the difference ${\bf r} - {\bf r}'$ only.  This means in the $\bf G$ representation, it is a diagonal matrix.      
Fourier transforming into full 3D space, we get $v({\bf q}) = 1/(\epsilon_0 q^2)$.   A two-dimensional expression is obtained
if we perform an inverse Fourier transform for $q_z$ back to real space $z$ using the Cauchy residue theorem, 
\begin{equation}
\label{eq-vGG}
v_{{\bf G}_\perp{\bf G}'_\perp}(z,z',{\bf q}_\perp) = 
\delta_{{\bf G}_\perp,{\bf G}'_\perp} { e^{- | {\bf q}_\perp + {\bf G}_\perp| |z-z'|} \over 2 \epsilon_0 | {\bf q}_\perp + {\bf G}_\perp|}.
\end{equation}
Finally, the Caroli formula remains the same with sum over $\bf q$ in the first Brillouin zone and trace over
${\bf G}_\perp$ as matrix index. 

\subsection{Example calculation of multiple-layer graphene}

Using the first-principles method introduced above, the near-field heat flux between the monolayer graphene has been performed in Ref.~\onlinecite{zhu1}. As the first-principles method can be easily applied to different systems without further model treatment \cite{zhu2,zhu3}, we
present here calculations of the heat flux between two parallel graphene sheets with finite layers. We start from the ground state calculations by using DFT as implemented in QUANTUM ESPRESSO \cite{qe1,qe2}. The norm-conserving pseudopotential generated by the Martins-Troullier method \cite{martin} with the Perdew-Burke-Ernzerhof exchange-correlation functional \cite{pbe} in the generalized gradient approximation has been adopted. The Kohn-Sham wave functions are expanded using the plane-wave basis set with a 60 Ry energy cut-off. The Fermi-Dirac smearing with a 0.002 Ry smearing width is employed to treat the partial occupancies. The in-plane lattice constants are 
$a = b = 2.46\,$\AA\ and the inter-layer distance for multiple-layer graphene is $3.40\,$\AA. To avoid interactions from the neighboring lattice in the $z$ direction, a large lattice constant of $c = 24.6\,$\AA\ is set in the $z$ direction of the unit cell.

The scalar photon self-energy $\Pi$ of each side is calculated on top of the ground state band structure by using the BerkeleyGW package \cite{bgw1,bgw2}. A $90\times90\times1$ Monkhorst-Pack \cite{mp} grid is used to sample the first Brillouin zone for the nonlocal polarizability, while the long-wave ($q\to0$) polarizability is obtained from a much finer $300\times300\times1$ grid. To avoid divergence of the Coulomb potential, we use a small value of $q=10^{-5}$ a.u.\ in the calculation of contributions from the long-wave polarizability. The energy broadening factor $\eta$ is set to $0.05\,$eV, which corresponds to an electron relaxation time of $10^{-14}$ s \cite{ilic}. We neglect the local
field effects that are important only for systems with inhomogeneous geometry \cite{zhu4}. Then we solve the Dyson equation, Eq.~(\ref{eq-dyson-Dqperp}), and calculate the transmission coefficient from the Caroli formula. Lastly, we integrate over frequencies to get the heat flux.

\begin{figure}
	\includegraphics[width=8.6 cm]{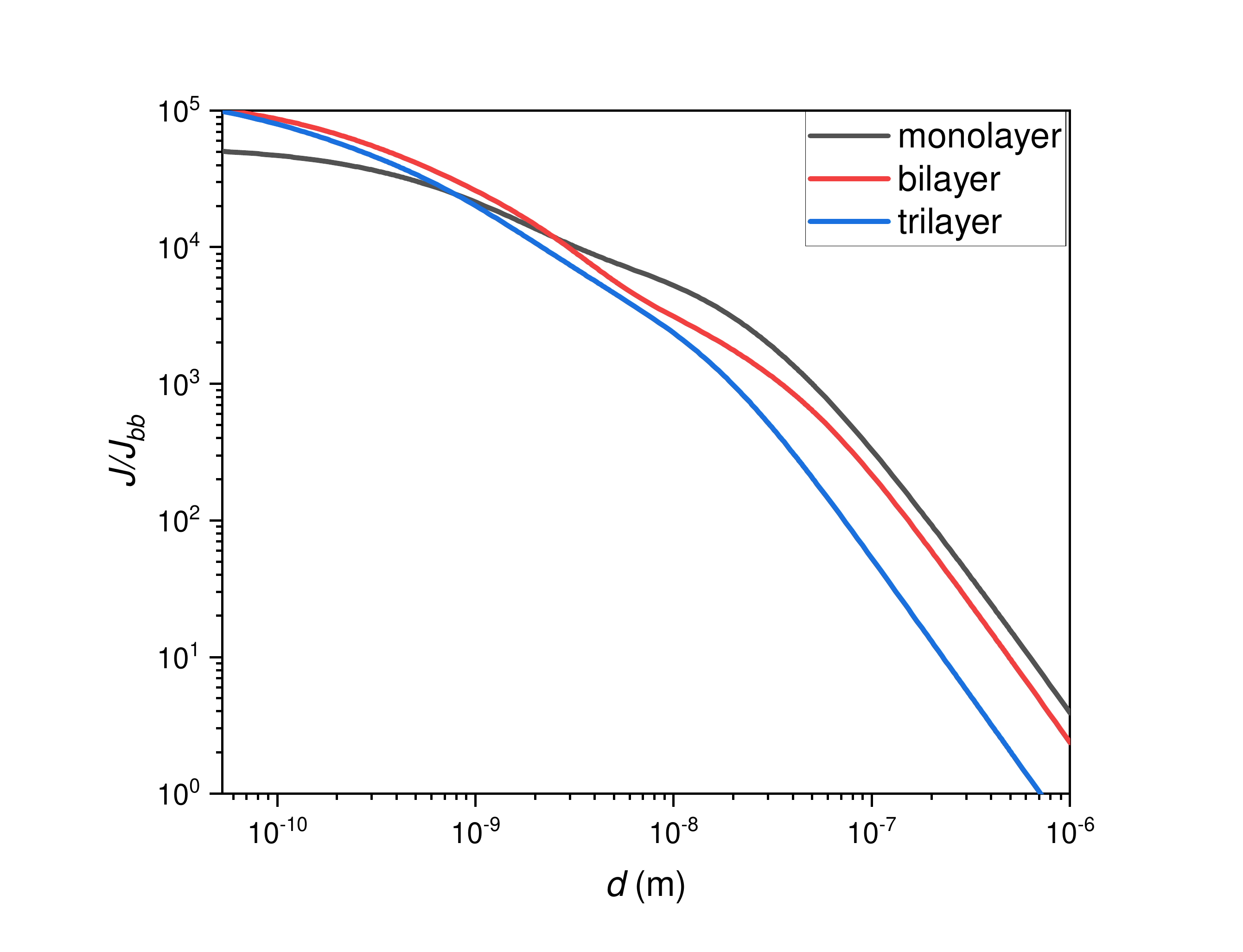}
	\caption{Distance dependence of the near-field heat flux ratio between two parallel single and multiple-layer graphene sheets. The temperatures of two sides are fixed at $T_L=$ 1000 K and $T_R=$ 300 K.}
	\label{fig_graphene_dft}
\end{figure}

In Fig.~\ref{fig_graphene_dft}, we show the calculated heat flux of two parallel single and multiple-layer graphene sheets as a function of gap sizes. The vertical coordinate is the ratio of the calculated near-field heat flux to the black-body radiative heat flux given by the Stefan-Boltzmann law $J_{bb}=\sigma(T_0^4-T_1^4)$, with $\sigma \approx 5.67\times10^{-8} \,$W/(m$^2$\,K$^4$). As shown, the near-field heat flux is remarkably larger than that of the black-body radiation for all three systems. For monolayer graphene, a converged ratio around $5\times10^4$ is shown which agrees well with a previous report that used a tight-binding method to calculate the density response function of graphene \cite{jiang17}. The saturation of heat flux in the extreme near field originates from the nonlocal effect of wave vectors, which is a typical behavior of thermal radiation mediated by $p$-polarized evanescent waves \cite{nonlocal}. Without spatial dispersion, the heat flux calculated from a local response function shows a $1/d$ dependence at short separation, which agrees with the previous report \cite{pablo}.  At extreme small distances, the heat flux between bilayer graphene approximately doubles the value of the monolayer graphene. However, heat flux between trilayer graphene sheets even becomes slightly smaller than that of the bilayer graphene. This may be due to the fact that we treat each side as infinitely thin in Eqs.~(105) and (106) for simplicity. 
With further increases of the sample layers, this treatment is not good in the extreme near field as we assume the gap size should be larger 
than the inter-layer distance.

With an increase of the gap size, the heat flux decreases monotonically for all three systems. Nevertheless, the magnitude of the heat flux is smaller for the multiple-layer graphene.  We suspect that increasing the layer number decreasing the energy transfer is due to a screening between the layers. At the distance between $7\,$\AA\ to $3\,$nm, the influence from the finite layer is not significant and all three systems show similar results. Due to the exponential factor that appears in the 2D Coulomb potential in Eq.~(\ref{eq-vGG}), the long-wave ($q\to0$) contribution becomes dominant at large distances. When $d > 100\,$nm, the heat fluxes for all three systems show an asymptotic dependence of $1/d^2$, which is consistent with the result of near-field heat flux between parallel plate capacitors \cite{capacitors}. 

\newpage

\part{Vector photon and Coulomb gauge}

So far in part I, we have focused on the scalar potential and ignored the vector potential in the 
electrodynamics.   The picture of the Coulomb interaction is a valid approximation when
the length scale is shorter than some typical length scale of order micrometers or less at room temperature, but is certainly
not correct for long distances.   We know that the electromagnetic waves can propagate to infinity, 
but only from the transverse part of the field.   Due to charge neutrality, the Coulomb interaction
decays much faster with distance and cannot have any effect at infinity. 
 In this part, we treat the energy
transport taking into account the full electromagnetic field contributions.   We study the thermal 
radiation from a cluster of objects modeled as a collection of tight-binding electrons.    This is more
than just the ideal blackbody radiation, which is independent of the details of the materials.  Here again,
we characterize the systems with a version of $\Pi$, but it is now a tensor associated with the vector field $\bf A$,
or the current-current correlation at the RPA level of approximation. 

\section{General formulation with transverse vector field}

\subsection{Lagrangian and Hamiltonian, gauge invariance}

To add the contribution from the transverse vector field $\bf A$, we start from the Lagrangian of the 
scalar field version, Eq.~(\ref{eq-scalar-L}), by the Peierls substitution \cite{Peierls33,Graf95}
of the tight-binding Hamiltonian $H$, and an 
extra transverse field piece, obtaining
\begin{eqnarray}
L &=& \int dV \frac{\epsilon_0}{2}\left( - \frac{\dot{\phi}^2}{\tilde{c}^2} + (\nabla \phi)^2 \right) +
\nonumber \\
&&  \frac{1}{2} \int dV \left[ \epsilon_0 \left(\frac{\partial {\bf A}}{\partial t}\right)^2 
-\frac{1}{\mu_0} \left(  \nabla \times {\bf A} \right)^2 \right] + 
\nonumber \\
&& i \hbar\, c^\dagger  \frac{dc}{dt} - \sum_{j,l}c^\dagger_j H_{jl} c_l 
\exp\left( -i\frac{e}{\hbar}\int_l^j {\bf A}\cdot d{\bf r} \right) + \nonumber \\
&& e \sum_j c_j^\dagger c_j \phi({\bf r}_j).
\label{eq-vector-L}
\end{eqnarray}
Here the second line is from the ``kinetic'' and ``potential'' energy of the free transverse field.
The word ``transverse'' means that the vector field satisfies $\nabla \cdot {\bf A} = 0$.  The meaning
is clearer if this equation is Fourier transformed into the wave-vector space, which is 
$i {\bf q} \cdot {\bf A}_{\bf q}=0$, which says that the direction of $\bf A$ is perpendicular to the 
direction of the wavevector, thus transverse.   From the
Lagrangian above, we recover the Hamiltonian as (taking the limit of $\tilde{c} \to \infty$)
\begin{eqnarray}
\hat H &=&
\frac{\epsilon_0}{2} \int dV \left[ - (\nabla \phi)^2 + \left(\frac{\partial {\bf A}}{\partial t}\right)^2 
+c^2  \left(  \nabla \times {\bf A} \right)^2 \right] + 
\nonumber \\
&& \!\!\!\!\!\!\!\!\!\!\!\! \sum_{j,l}c^\dagger_j H_{jl} c_l 
\exp\left(\! -i\frac{e}{\hbar}\int_l^j \!\! {\bf A}\cdot d{\bf r} \right)\!  - e \sum_j c_j^\dagger c_j \phi({\bf r}_j).
\quad\quad\>
\label{eq-H-gauge-inv}
\end{eqnarray}
Here the integral on the exponential is a line integral from site $l$ to site $j$ in a straight path.   We check
that the Hamiltonian is gauge invariant in the sense, that if we make the replacement, 
\begin{eqnarray}
\phi & \to & \phi + {\partial \chi \over \partial t}, \\
{\bf A} & \to & {\bf A} - \nabla \chi, \\
c_j & \to &  c_j \exp\left( i\frac{e}{\hbar} \chi_j \right),
\end{eqnarray}
the result will remain the same independent of $\chi$, 
where $\chi_j = \chi({\bf r}_j,t)$ is an arbitrary smooth function of space and time.  In a sense, the requirement of the gauge invariance 
uniquely fixes the form of the electron-photon interaction.   It is noted that the Peierls substitution form of a tight-binding
model has a fundamental limitation \cite{JjLi20} as it cannot describe transitions among electronic
states at the same location.  We use it for its simplicity, and it is a good starting point to describe metals. 

\subsection{Quantization, current operator, and Green's functions}

We now discuss the quantization of the electromagnetic field.   Because of the transverseness of the 
vector field, we see that the scalar and vector fields are not coupled, thus we can quantize $\phi$ as in 
part I.  For the vector field, from the Lagrangian, the conjugate momentum is 
\begin{equation}
{\bf \Pi} = { \delta L \over \delta \dot{\bf A}} = \epsilon_0 \dot{\bf A}.
\end{equation} 
However, due to the transverse nature of the field $\bf A$, the three components, $A_x$, $A_y$, and
$A_z$ cannot be treated as independent quantities, thus we cannot postulate commutation relation
in the usual way.   The true degrees of freedom are demonstrated more clearly in the Fourier space 
after the transformation.   This is the standard approach in Coulomb gauge \cite{Tannoudji89}.   To make a long story 
short, we just give the commutation relation as
\begin{equation}
\label{eq-A-commutator}
\bigl[ A_\mu({\bf r}), \Pi_\nu({\bf r}') \bigr] = i \hbar\, \delta^\bot_{\mu\nu}({\bf r} - {\bf r}'),
\end{equation}
where the right-hand side is the transverse $\delta$-function defined by an inverse Fourier transform
\begin{equation}
\delta^\bot_{\mu\nu}({\bf r}) \equiv \int \frac{d^3 {\bf k}}{(2\pi)^3}
\left( \delta_{\mu\nu}- { k_\mu k_\nu \over |{\bf k}|^2} \right) e^{i {\bf k} \cdot {\bf r}}.
\end{equation}
Here $\mu$, $\nu$ take the Cartesian directions $x$, $y$, or $z$, and $\delta_{\mu\nu}=1$ 
if $\mu=\nu$ and 0 otherwise, which is the usual Kronecker delta.   It is worth noting that the
transverse $\delta$-function is nonlocal and decays in space as $1/r^3$ at long distance. 

Together with the earlier commutation relations for $\phi$ and $c_j, c_j^\dagger$, the problem is 
completely specified.   We can apply the Heisenberg equations of motion for $c_j$, $\phi$ and $\bf A$,
obtaining a Schr\"odinger equation for $c_j$ with a Peierls substituted $H$ and extra external potential due to 
$\phi$, and a Poisson equation for $\phi$ (after taking the 
limit $\tilde{c} \to \infty$) with the usual charge density $\rho = (-e) \sum_{j} c_j^\dagger c_j 
\delta({\bf r} - {\bf r}_j)$ as the source.  Finally, the equation for $\bf A$ is 
\begin{equation}
\label{eq-wave-A-jperp}
\frac{1}{c^2} \frac{\partial^2 {\bf A}}{\partial t^2} - \nabla^2 {\bf A} = \mu_0\, {\bf j}_\bot({\bf r}).
\end{equation}
Here the transverse current is
\begin{equation}
{\bf j}_\bot({\bf r}) = \frac{1}{i\hbar} \left[ {\bf \Pi}({\bf r}), 
\sum_{j,l}c^\dagger_j H_{jl} c_l 
\exp\left( -i\frac{e}{\hbar}\int_l^j \!\! {\bf A}\cdot d{\bf r} \right)  \right].
\end{equation}
These are consistent with Maxwell's equations. 

Due to the presence of the vector potential on the exponential, the commutator is hard to compute explicitly.   But the integral is proportional to the 
lattice spacing $a$, which is small.   In the continuum limit, we take $a$ to zero, as a result, we need
to keep only to the second order in the expansion.   The third and higher orders vanish in 
the continuum limit.   But again, the transverse $\delta$-function causes
some complication.  Formally, we can write
\begin{equation}
{\bf j}_\bot({\bf r}) = P_\bot {\bf j} = 
\int d{\bf r}' {\bf \delta}^\bot({\bf r} - {\bf r}') \cdot {\bf j}({\bf r}').
\end{equation}
Here ${\bf \delta}^\bot$ is the $3\times 3$ tensor.  There are two terms to the current $\bf j$, 
a paramagnetic term independent of the vector potential, and a diamagnetic term proportional to 
$\bf A$, just like the electron-photon interaction in a first quantization formulation.    The explicit form
depends on how one approximates the line integral.   Here we adopt  the trapezoidal rule for the 
integral, 
\begin{equation}
\int_l^j {\bf A} \cdot d{\bf r} \approx \frac{1}{2} ({\bf A}_j + {\bf A}_l) \cdot \bigl({\bf R}_j - {\bf R}_l\bigr),
\end{equation}
where ${\bf R}_l$ and ${\bf R}_j$ are the respective locations of the two sites.  The field is evaluated at these sites.
Using this approximation, we can give an explicit formula for the paramagnetic term as
\begin{eqnarray}
\label{eq-current-jr}
{\bf j}({\bf r}) &=& \frac{1}{2} \sum_{j,l}  {\bf I}_{jl} \bigl(  \delta({\bf r} - {\bf R}_j) + 
\delta({\bf r} - {\bf R}_l) \bigr), \\
{\bf I}_{jl} &=& \frac{(-e)}{i \hbar} c_j^\dagger H_{jl} c_l \bigl({\bf R}_j - {\bf R}_l\bigr)= (-e) \hat {\bf V}_{jl}.
\end{eqnarray} 
When an electron hops from the site $l$ to $j$, it is not really clear where the current is located.  It 
could be attributed to the middle of the sites, or one of the sites.   Here we take an average of current
being associated with site $j$ or $l$.   The local total current obtained by volume integrating the
current density around the two sites, ${\bf I}_{jl}$, has a more useful interpretation; it is the velocity of the electron when hopping
from site $l$ to $j$, times the charge of electron, $(-e)$.  

By Taylor expanding the Peierls substituted Hamiltonian, and using the same trapezoidal approximation
for the line integral, we can write the interaction part of the Hamiltonian as 
\begin{equation}
\hat H_{\rm int} = \sum_{jkl\mu} c^\dagger_j M_{jk}^{l\mu} c_k A_{l\mu} = - 
\sum_{l} {\bf I}_l \cdot {\bf A}_l, 
\end{equation} 
where we can express the tensor $M$ in terms of the velocity matrix as 
$M_{jk}^{l\mu} = (e/2) \bigl( \delta_{jl} + \delta_{kl}\bigr) V_{jk}^\mu$.  The index $\mu$ labels
the Cartesian directions, $x$, $y$, or $z$.   We also introduce
the volume integrated current around the site $l$, as defined by the above equation, which
we can write compactly as $I_{l\mu} = - c^\dagger M^{l\mu}c$, where $c$ is a column vector
of the annihilation operators, $c^\dagger$ is a row vector of creation operators, and $M^{l\mu}$
is a Hermitian matrix in the electron site space.   The current associated with the site is useful
to define the current-current correlation on a discrete lattice.  Finally, we can also obtain
the current density, Eq.~(\ref{eq-current-jr}), by the functional derivative of the interaction
Hamiltonian with respect to the vector field by  ${\bf j}({\bf r}) = -
\delta \hat H_{\rm int}/\delta {{\bf A}({\bf r})}$.

The diamagnetic term is usually not important as it is higher order, also, it contributes only a
purely real, diagonal $\Pi$, so it is not dissipative.  In the continuum limit, we have a simple
expression for the current ${\bf j}_A = -  e^2(n/m) {\bf A}$, which is the London 
equation \cite{deGennes99}. 
Here $n$ is the electron density and $m$ is its mass.  On lattice, if we 
use the trapezoidal approximation, we get a complicated form of the type 
$H_{\rm int}^{A^2} = \frac{1}{2} \sum_{l,l',\mu,\nu,j,k} c^\dagger_j N_{jk}^{l\mu,l'\nu} c_k 
A_{l\mu} A_{l'\nu}$.   

Finally, based on the vector field $\bf A$ and the current ${\bf I}_l$ we can define two similar contour
ordered Green's functions, call them $D$ and $\Pi$ as they play very similar role as in the scalar photon
theory, except that $D$ and $\Pi$ will have additional directional indices as $3 \times 3$ tensors or dyadic. 
\begin{eqnarray}
D_{\mu\nu}({\bf r}\tau; {\bf r}'\tau') &=& \frac{1}{i\hbar} \bigl\langle
T_\tau A_\mu({\bf r},\tau) A_\nu({\bf r}',\tau')\bigr\rangle,  \\
\Pi_{l\mu,l'\nu}(\tau; \tau') &=& \frac{1}{i\hbar} \bigl\langle
T_\tau I_{l\mu}(\tau) I_{l'\nu}(\tau')\bigr\rangle_{\rm ir}.  \label{eq-Pi-local}
\end{eqnarray}
Here $D$ is defined in the full space of $\bf r$, while $\Pi$ is only on the discrete sites.  In computing
$\Pi$ under the random phase approximation, we take only the lowest order in such an expansion,
ignoring the $A^2$ terms in the Hamiltonian.  With the electron-photon interaction of the form
$-\sum {\bf I} \cdot {\bf A}$, we still have the standard Dyson equation of the form, $D = d + d \Pi D$,
except that the sizes of the matrices are 3 times larger.

\subsection{Free Green's function}
We need to determine the free photon Green's function $d$ associated with the transverse 
vector field $\bf A$ (earlier for the scalar field $\phi$ it was called $v$).
This can be done with the standard procedure of quantization of the field.   We can express
the vector field with the annihilation and creation operator as \cite{Loudon3rd,Mahan00} 
\begin{equation}
{\bf A}({\bf r}) = \sum_{{\bf q}\lambda} \sqrt{\frac{\hbar}{2 \epsilon_0 \omega_{\bf q} V}}\;
{\bf e}_{{\bf q}\lambda} a_{{\bf q}\lambda} e^{ i {\bf q}\cdot{\bf r}} + {\rm h.c.}.
\end{equation}
where $\omega_{\bf q} = c |{\bf q}|$ is the free photon dispersion relation, ${\bf e}_{{\bf q}\lambda}$
are the two mutually perpendicular unit polarization vectors indicated by the index $\lambda$, which are also orthogonal to $\bf q$.  This condition gives a transverse field for $\bf A$, i.e., ${\boldmath \nabla} \cdot {\bf A} =0$.   $a_{{\bf q}\lambda}$ is the associated annihilation operator for the mode ${\bf q},\lambda$. 
These are the standard bosonic operators satisfying $[a_{{\bf q}\lambda},a_{{\bf q}'\lambda'}]=0$,
 $[a_{{\bf q}\lambda},a_{{\bf q}'\lambda'}^\dagger]=\delta_{{\bf q}{\bf q}'}\delta_{\lambda\lambda'}$. 
 The field exists in a
finite volume $V$ with periodic boundary conditions, thus the 
wave vectors $\bf q$ are discrete.   This equation together with a corresponding equation for
the conjugate field ${\bf \Pi}({\bf r})$ is viewed as a variable transformation
between $a_{{\bf q}\lambda}$ and ${\bf A}({\bf r})$.    This is the correct transformation if the commutation relation between $\bf A$ and ${\bf \Pi}$, Eq.~(\ref{eq-A-commutator}), is fulfilled and the Hamiltonian is diagonal, 
$\hat H_\gamma = \sum_{{\bf q}\lambda} \hbar \omega_{{\bf q}\lambda} ( a_{{\bf q}\lambda}^\dagger 
a_{{\bf q}\lambda} + 1/2)$.   Indeed, these claims can be verified.  

For the free field, the time dependence for the annihilation operator is trivially
$a_{{\bf q}\lambda}(t) = a_{{\bf q}\lambda} e^{ - i \omega_{\bf q}t}$.   Using this result, 
plugging into the definition of the retarded Green's function, we obtain,  
\begin{eqnarray}
d_{\mu\nu}^r({\bf r},t) &=& \frac{1}{\epsilon_0 V} \sum_{\bf q} 
\left( \delta_{\mu\nu} - {q_\mu q_\nu \over q^2}\right) e^{i {\bf q} \cdot {\bf r}} d_{{\bf q}}(t),\\
d_{{\bf q}}(t) &=& - \theta(t) {\sin(\omega_{\bf q}t) \over \omega_{\bf q}} e^{-\eta t}.
\end{eqnarray}
The factor in the brackets takes care of the transverseness of the Green's function, which is the 
transverse projector in $\bf q$ space.    The transverse projector appears because two of the
polarization vectors and the unit vector ${\bf q}/q$ form an orthonormal basis.   We can use
the completeness relation to eliminate the reference to the polarization vectors, ${\bf e}_{{\bf q}\lambda}$.   The second line defines the retarded Green's function for a single mode
in time domain.   In frequency domain, it is  $1/\bigl( (\omega + i \eta)^2 - \omega_{\bf q}^2\bigr)$. 
If we take the volume to infinity, the summation can be turned into an integral in $\bf q$.  The final expression in real space
and frequency domain can be obtained with the help of the residue theorem with the contour integral \cite{Arfken7th} on 
the complex plane of $q$, given as \cite{Keller11}
\begin{eqnarray}
\label{eq-d-leftright-arrow}
\stackrel{\leftrightarrow}{d^r}\!({\bf r},\omega) &=& 
-\frac{1}{4\pi \epsilon_0 c^2 r} \Bigg\{ e^{i \frac{\omega}{c} r}\,
\bigl( \stackrel{\leftrightarrow}{\bf U} - \hat{\bf R}\hat{\bf R} \bigr) +  \\
&& \biggl[ - \frac{e^{i\frac{\omega}{c}r}}{ i\frac{\omega}{c}r} + 
\frac{ e^{i\frac{\omega}{c}r}-1}{\left( i \frac{\omega}{c}r \right)^2}\biggr]
\bigl( \stackrel{\leftrightarrow}{\bf U} - 3 \hat{\bf R}\hat{\bf R} \bigr) \Bigg\}.  \nonumber
\end{eqnarray} 
Here, we express the $3\times 3$ tensor in the Cartesian directions in the dyadic notation,
$r = |{\bf r}|$ is the magnitude of the vector, and  $\hat {\bf R} = {\bf r}/r$ is a unit vector in the radial
direction.   $\stackrel{\leftrightarrow}{\bf U}$ is the unit or identity matrix. 

Another equivalent way to obtain the retarded free Green's function is from the equation it must
satisfy.   We observe that the transverse vector field $\bf A$ satisfies a wave equation with the 
transverse current as a source, Eq.~(\ref{eq-wave-A-jperp}).   The Green's function
maps the current to the vector field, ${\bf A} = - \stackrel{\leftrightarrow}{d^r}\! \cdot {\bf j}$.  
The Green's function satisfies the same equation as the vector potential, but with a transverse
$\delta$ function as a source.  When the wave equation and the transverse $\delta$ function is
expressed in $\omega$ and $\bf q$ space, we have a simple result for the Green's function as,
\begin{equation}
\stackrel{\leftrightarrow}{d^r}\!\!({\bf q}, \omega) = 
{ \stackrel{\leftrightarrow}{\bf U}\! - \hat{\bf q} \hat{\bf q} \over \epsilon_0 \bigl(  (\omega + i\eta)^2 - c^2 q^2 \bigr) },  
\label{eq-d-modespace}
\end{equation}
where $\hat{\bf q} = {\bf q}/|{\bf q}|$ is a unit vector. 
Fourier integral transforming into real space $\bf r$, we obtain the explicit formula of Eq.~(\ref{eq-d-leftright-arrow}). 

In computing transport quantities in a planar geometry, such as between graphene sheets or surface of a lattice,
we need the free Green's function in mixed representation, i.e., the transverse direction is in wave vector 
space, ${\bf q}_\perp$, but the transport direction, say $z$, is in real space.   The retarded Green's function in
the mixed representation can be obtained by inverse Fourier transform $q_z$ back to real space. 
\begin{eqnarray}
d^{r,\mu\nu}({\bf q}_\perp, \omega, z, z')  = \hspace{4.7cm} \\
 \int_{-\infty}^{+\infty}\!\! {dq_z \over 2\pi} \,
{ \left( \delta_{\mu\nu} - \frac{q_\mu q_\nu}{q^2}\right) 
e^{i q_z( z - z')} \over a^2 \epsilon_0 \Bigl( (\omega + i0^+) ^2 - c^2 q_\perp^2
- c^2 q_z^2\Bigr) },   \nonumber 
\end{eqnarray}
where $q^2 = | {\bf q} |^2 = q_\perp^2 + q_z^2$.   Here $a^2$ is the area of a unit cell, since we will
consider a discrete set of ${\bf q}_\perp$.  This integral can be performed using the residue theorem.  
After somewhat lengthy and tedious calculation, we obtain \cite{Peng1607arxiv},
\begin{eqnarray}
d^{r,\alpha\beta} &=& \delta_{\alpha\beta}\, d - q_\alpha q_\beta F, \quad \alpha, \beta = x,y\\
d^{r,\alpha z} &=& d^{r,z\alpha} = {\rm sgn}(z-z')q_\alpha ( B - A)/C, \\
d^{r,zz} &=& q_\perp^2 F.
\end{eqnarray}
We have introduced the shorthand notations $A=e^{i \tilde{q}_z |z-z'|}$,
$B=e^{-q_\perp|z-z'|}$,  $d = A/(a^2 \epsilon_0 2 i c^2 \tilde{q}_z)$,
$F= (A/\tilde{q}_z + i B/q_\perp)/C$, $C=a^2\epsilon_0 2 i \omega^2$, and $\tilde{q}_z = \pm \sqrt{ 
[(\omega + i0^+)/c]^2 - q_\perp^2}$, where the sign is chosen such that
${\rm Im}\, \tilde{q}_z > 0$.  We note that the free Green's function is transverse in the sense
$ i q_x d^{r,x\beta} + i q_y d^{r,y\beta} + \frac{\partial\ }{\partial z} d^{r,z\beta}=0$ for
$\beta = x,y,z$.  
  
\section{Poynting vector and energy transport}
At a place in vacuum without material presence, the energy can be transferred through the 
electromagnetic waves.    This is described by the Poynting vector, ${\bf S} = {\bf E} \times {\bf B}/\mu_0$,
in electrodynamics.    The meaning of the Poynting vector is made clear through Poynting's theorem
which reflects the energy conservation \cite{Griffiths17},
\begin{equation}
{\partial u \over \partial t} + \nabla \cdot {\bf S} = - {\bf E} \cdot {\bf j},
\end{equation} 
here $u = 1/2\, ( \epsilon_0 E^2 + B^2/\mu_0)$ is the field energy density.  The right-hand side is the
Joule heating contribution where $\bf j$ is the electric current density.   If we consider a volume $V$ and integrate
over the volume, using Gauss's theorem, the divergence over the volume becomes a surface integral,
so 
\begin{equation}
I = \int d{\bf \Sigma} \cdot {\bf S}
\end{equation}
is the energy flux going out of the volume where $d{\bf \Sigma}$ is the surface element with an outward norm.
 
In order to use this classical expression for the energy current, we need to consider several additional features
in a quantum theory.   We need to perform an ensemble average by expressing the Poynting vector
in terms of the Green's function $D$.   We need to worry about the operator order as $\bf E$ and $\bf B$
in general are not commuting quantities.   We also need to remove the zero-point motion contribution to
the Poynting vector, as intuitively we do not expect that the zero-point motion of the electromagnetic waves transfers energy when objects are stationary  (see, however \onlinecite{XZhang19}).  Lastly, we need to worry about our choice of gauge.    Since we have used the Coulomb gauge
here, we have ${\bf E} = - \nabla \phi - \partial {\bf A}/\partial t$, here the vector field is transverse.
This split of $\bf E$ into a scalar potential term and a transverse vector potential term means we can
also split the Poynting vector into two corresponding terms, ${\bf S} = {\bf S}_\parallel + {\bf S}_\perp$,
where ${\bf S}_\parallel  = - \nabla \phi \times {\bf B}/\mu_0$ and ${\bf S}_\perp = - \partial {\bf A}/\partial t 
\times (\nabla \times {\bf A})/\mu_0$.   The longitudinal contribution ${\bf S}_\parallel$ can be
transformed back to a volume integral by the divergence theorem and the curl of $\bf B$ is then 
related to the electric current.   In steady state, the average of a time derivative is zero.  Using this
property and Maxwell's equations, we can show that ${\bf S}_\parallel$ outside matter is the same as the `Poynting
scalar' discussed earlier, plus a cross correlation term between the longitudinal and
transverse fields, $-\epsilon_0 \phi \,\partial^2 {\bf A}/\partial t^2$.   In this section, we shall focus on the transverse contribution
${\bf S}_\perp$.  

As for the remaining issues, in general for a product  $AB$ of two Hermitian operators, $A$ and $B$,  the result is not Hermitian, and its expectation is not guaranteed to be real.   Thus, we will replace the product by
the symmetrized version,  $(AB + BA)/2$.    To remove the zero-point motion contribution, we need to
impose a normal order \cite{Janowicz03,Loudon3rd}, so the final form is $:(AB + BA):/2$.   We elaborate this point in the next
subsection in some detail.
  
\subsection{Operator order}
Given a Schr\"odinger picture Hermitian operator $A$, the Heisenberg operator $A(t)$ is assumed to be defined for 
all $t$, by $A(t) = e^{i\hat{H}t/\hbar} A\, e^{-i\hat{H}t/\hbar}$.  We Fourier decompose the operator as
\begin{eqnarray}
A(t) &=& A^{+}(t)  + A^{-}(t) \\
&=&  \int_0^{+\infty} \frac{d\omega}{2\pi} \tilde{A}(\omega) e^{-i\omega t} + 
\int_{-\infty}^0  \frac{d\omega}{2\pi} \tilde{A}(\omega) e^{-i\omega t}.  \nonumber 
\end{eqnarray}
$A^{+}(t)$ is the positive frequency part of $A(t)$ by integrating over the positive frequencies in the
Fourier space, and $A^{-}(t)$ is called the negative frequency part of $A(t)$.    The positive part of the 
frequency is associated with annihilation operators and negative one with creation operators \cite{Bogoliubov82}.  The normal order or anti-normal order is defined in terms of $A^{\pm}(t)$.  The
fundamental assumptions for the operators are
\begin{equation}
A^{+}(t) |0\rangle = 0, \quad  \langle 0|A^{-}(t) = 0,
\end{equation}
where $|0\rangle$ is the vacuum state.   What we have in mind is the free photon field, but we assume
it is generally valid.   Also, since $A(t)$ is Hermitian, $A^{+}(t) = \bigl[ A^{-}(t) \bigr]^\dagger$.

Consider a steady-state Green's function formed by two operators, $A$ and $B$, and defined as 
\begin{equation}
D_{AB}^>(t) = \frac{1}{i\hbar} \langle A(t) B(0)\rangle,\quad B(0) = B.
\end{equation}
The decomposition of the positive and negative frequency parts of $A(t)$ naturally leads to positive and negative frequency 
parts of $D_{AB}^>(t)$, thus we must have
\begin{eqnarray}
\frac{1}{i\hbar} \langle A^{+}(t) B\rangle &=& \int_0^{+\infty} \frac{d\omega}{2\pi} \tilde{D}_{AB}^{>}(\omega) e^{-i\omega t}, \\
\frac{1}{i\hbar} \langle A^{-}(t) B\rangle &=& \int_{-\infty}^0 \frac{d\omega}{2\pi} \tilde{D}_{AB}^{>}(\omega) e^{-i\omega t}.
\end{eqnarray}
That is, the positive frequency part  $A^+(t)$ contributes to positive $\omega$ in $D^>$ only.   Similarly,
$D^<_{AB}(t) = \langle B A(t) \rangle/(i\hbar)$ can be decomposed analogously.   The normal order
is defined by
\begin{eqnarray}
: A^{-} B^{+} : &=& A^- B^+, \\
: B^+ A^- : &=& A^- B^+, \\
: A^+ B^+ : &=& A^+ B^+, \\
: A^- B^- : &=& A^- B^-, 
\end{eqnarray}
i.e., the right-most operator should be the annihilation operator if it is not already so.   The normal order has no effect
if it is already normal ordered.   We assume that the normal order has no effect if both are $+$, or $-$.   The normal ordering is a linear operation.  So we 
can express the equal-time correlation as 
\begin{eqnarray}
\langle : A B : \rangle &=& \langle [A^+, B^+] \rangle + \langle A^- B + B A^+\rangle, \\
\langle : B A : \rangle &=& \langle [B^-, A^-] \rangle + \langle A^- B + B A^+\rangle.
\end{eqnarray}
Our basic assumption is that the positive (negative) frequency part contains only annihilation (creation) 
operators, and they commute.   So the first term is zero for whatever meaning of the average.  
As a result, we have, consisting with \onlinecite{Agarwal75}, 
\begin{eqnarray}
\langle : AB: \rangle &=& \langle :BA:\rangle = \langle A^- B + BA^+ \rangle \nonumber \\
&=& {\rm Re\ } i\hbar  \int_0^{\infty} \frac{d\omega}{\pi} \tilde{D}_{AB}^{<}(\omega). 
\label{eq-positive-wDless}
\end{eqnarray}
Here we use the Fourier decomposition at time $t=0$ and use a symmetry of the Green's function,
$\tilde D_{AB}^>(\omega) = \tilde D_{BA}^<(-\omega)$.   We see that the two terms,  $A^- B$ and
$B A^+$, are Hermitian conjugate of each other, so the result is explicitly real.  
The last formula, Eq.~(\ref{eq-positive-wDless}),
has the effect of removing the zero-point motion contribution by taking integral only for the positive
frequencies for the lesser Green's function, remembering that due to the fluctuation-dissipation theorem, 
$D^<$ is proportional to the Bose function $N(\omega)$ in thermal equilibrium which decays to 0 at high
frequency exponentially.   If this were $D^>$, we have $N(\omega) + 1$, which may run into problem
of divergence when integrated over the positive frequencies. 
We note that a symmetrization of the operator product is not necessary as the normal order automatically
makes the result symmetric with respective to the product order. 

\subsection{Average transverse Poynting vector}
With the preparation in the above subsection of a formula to express the equal-time correlation of a 
normal ordered product as a positive frequency integral of the lesser Green's function in the frequency domain,
we can work out an expression of the average of the Poynting vector,
\begin{eqnarray}
\langle S_\perp^i\rangle &\!\!=\!\!& \frac{1}{\mu_0} \langle : ({\bf E}_\perp \times {\bf B})_i: \rangle\!\!\!\!\!\!\!\! \\
&\!\!=\!\!& \frac{1}{\mu_0} \langle  \bigl[: (-\dot{\bf A}) \times (\nabla \times {\bf A}) : \bigr]_i \rangle \nonumber \\ 
&\!\!=\!\!& - \frac{1}{\mu_0}\!\!\!\sum_{ijklm}\!\! \epsilon_{ijk} \epsilon_{klm}\!\! \left[ 
\frac{\partial\ }{\partial t} \frac{\partial\ }{\partial x'_l} \langle : \! A_j({\bf r}, t) A_m({\bf r}', t')\!
:\rangle    
\right]_{\stackrel{t'=t}{{\bf r}' ={\bf r}}} \nonumber  \\
&\!\!=\!\!& {\rm Re}\!\! \sum_{ijklm}\!\! \epsilon_{ijk} \epsilon_{klm} \int_0^\infty\!\! \frac{d\omega}{\mu_0 \pi} \hbar \omega \left[ 
\frac{\partial\ }{\partial x'_l} D^<_{mj}({\bf r}', {\bf r}, \omega) 
\right]_{{\bf r}' ={\bf r}}. \nonumber
\end{eqnarray} 
Here for the Poynting vector, we look for the $i$-th Cartesian component, expressed in terms of the
transverse vector field $\bf A$.   The average Poynting vector depends on explicitly the location $\bf r$ which
we have suppressed in notation.  Since we are interested in steady-state average, the Poynting vector does
not depend on time.    The vector cross products are written out explicitly with the Levi-Civita symbol,
$\epsilon_{ijk}$, which is $\epsilon_{xyz}=1$ and antisymmetric for each permutation of any two indices,
e.g., $\epsilon_{ijk} = - \epsilon_{jik}$.
In order to express the final result in terms of an $AA$ correlation, i.e., the $D^<$ photon Green's function, we
need to pull the time and space derivatives outside the average by changing the variable $\bf r$ to 
${\bf r}'$ and $t$ to $t'$, and changing them back to $\bf r$ and $t$ after the derivatives are performed.   We use Eq.~(\ref{eq-positive-wDless}) in the last step, identifying $A_m({\bf r}',t')$ as the operator $A$ and $A_j({\bf r},t)$ as $B$.   Notice that time derivative in time domain becomes $(+i\omega)$ in the 
frequency domain as $t$ is the second argument in $D^<$. 

To compute the total energy current, we need to make a dot product with the  surface norm ${\bf n}$ and to  integrate
over the surface.  We can simplify the formula a bit by summing over the index $k$ in the product of Levi-Civita symbols,
\begin{equation}
\sum_{k} \epsilon_{ijk} \epsilon_{klm} = \delta_{il}\delta_{jm} -  \delta_{im}\delta_{jl}.
\end{equation}
Using this identity, we can write the surface integral for the energy transfer as
\begin{equation}
\label{eq-S-Dlss}
I = \int\!\! d\Sigma\! \int_0^\infty \!\!\!\frac{d\omega}{\mu_0 \pi} \hbar \omega {\rm Re}
\Bigl[  ({\bf n} \cdot {\bf \nabla}') {\rm Tr}(D^<) - {\rm Tr}( {\bf\nabla}' {\bf n} \cdot D^< )\Bigr].
\end{equation}
Here ${\bf\nabla}'$ is the gradient operator acting on the first argument of $D^<$, the combination
${\bf\nabla}' {\bf n}$ is a dyadic, i.e., as a $3\times3$ matrix with component $\partial/\partial x_i'\, n_j$
as a differential operator acting on $D^<$,
${\bf n}$ is the unit norm of the surface element, $d\Sigma$ is the magnitude of the surface 
element, the dot $\cdot$ is a scalar product in the first term
and matrix multiplication in the second term.  The trace is over the matrix indices. 

\subsection{Radiation at far field}
We consider a cluster of some materials of finite sizes and compute the radiation at far field.   At sufficiently far 
distances from the matter, the longitudinal part from ${\bf S}_\parallel$ decays to zero and only the transverse
field in ${\bf S}_\perp$ propagates to infinity.   So for the energy radiation to infinity, we only need to
compute the contribution from the above formula.   In the far field, we have another simplification which 
we can use, that is, the electromagnetic waves at far field are spherical waves of the form $e^{ i \frac{\omega}{c} R}$,
where $R =|{\bf R}|$ is the distance to the coordinate origin.   Since the photon Green's function satisfies
the same equation as the field, the Green's function also takes this form.   As a result, the gradient operator
${\bf \nabla}'$ can be replaced by the vector $i \frac{\omega}{c} \hat{\bf R}$, here $\hat{\bf R} = 
{\bf R}/R$ is the unit vector from the origin to the observation point $\bf R$.    We can choose a large
sphere of radius $R$ to perform the surface integral as a solid angle integration,  $d\Sigma = d\Omega R^2$. 
Using this observation and replacing the general surface norm $\bf n$ by $\hat{\bf R}$, we obtain
\begin{equation}
\label{eq-URRDls}
I = {\rm Re} \int_0^\infty\!\!\! \frac{d\omega}{c\mu_0 \pi} (i\hbar \omega^2) \int d\Omega R^2 
{\rm Tr}\Bigl[ ( \stackrel{\leftrightarrow}{\bf U} - \hat {\bf R} \hat {\bf R}) D^< \Bigr].
\end{equation}
Here $ \stackrel{\leftrightarrow}{\bf U}$ is the unit matrix,  $P= \stackrel{\leftrightarrow}{\bf U} - \hat {\bf R} \hat {\bf R}$ is the transverse projector, having
the property $P^2 = P$.  Two matrices $P$ and $D^<$ are multiplied and then trace is taken. 

We can express $D^<$ by the Keldysh equation as  $D^< = D^r \Pi^< D^a$, where $D^r$ is the retarded
Green's function, and $D^a = (D^r)^\dagger$ is the advanced version, and $\Pi^<$ is the material
property we call self-energy (with respect to the photons).   Here the three matrices are multiplied,
which implies a sum over the sites as well as directions, i.e,
\begin{equation}
D^<_{ij}({\bf r}, {\bf r'}, \omega) = \sum_{ll'\alpha\beta} D^r_{i\alpha}({\bf r}, {\bf r}_l,\omega) \Pi_{l\alpha,l'\beta}^{<}(\omega)
D^a_{\beta j}({\bf r}_{l'}, {\bf r}',\omega) .
\end{equation} 
We still have to solve the Dyson equation for the retarded Green's function,  $D^r = d^r + d^r \Pi^r D^r$.  
However, for the far field problem, the corrections to the free Green's function are rather small since 
$\Pi^r D^r \sim (v/c)^2$ is of the order of the ratio of electron velocity to the speed of light squared.  As a result, it is sufficient and is a good approximation just using the free Green's 
function $d^r$.  Also, since ${\bf r} = {\bf R} \to \infty$, it does not matter where the atoms are located with 
respect to $\bf R$, so we also set all the positions ${\bf r}_l$ in the argument of the Green's function at the
origin ${\bf 0}$.   We call this a monopole approximation. Then our approximate Green's function is
\begin{equation}
 D^r({\bf R}, {\bf r}_l, \omega) \approx
- { 1 \over 4\pi\epsilon_0 c^2 R } e^{i\frac{\omega}{c} R} ( \stackrel{\leftrightarrow}{\bf U} - \hat {\bf R} \hat {\bf R}).
\end{equation}
Putting this result into Eq.~(\ref{eq-URRDls}), using the cyclic property of trace, and the fact that $P$ is a
projector, $P^3=P$, we find 
\begin{equation}
I = {\rm Re} \int_0^\infty\!\! d\omega\, \frac{i\hbar \omega^2}{16\pi^3 \epsilon_0 c^3} \int d\Omega\,
{\rm Tr}\Bigl[ ( \stackrel{\leftrightarrow}{\bf U} - \hat {\bf R} \hat {\bf R}) \Pi^<(\omega) \Bigr].
\end{equation}
Here the total $\Pi^<(\omega) = \sum_{l,l'} \Pi^<_{ll'}$ is the sum over all the sites.  Since the only angular 
dependence is in $\hat{\bf R}=(\sin\theta \cos\phi, \sin\theta \sin\phi, \cos\theta)$, the integration over the solid angle divided by $4\pi$ is equivalent to
an average over the projector.   We can verify by a direct integration that
\begin{equation}
\overline{\hat {\bf R} \hat {\bf R}} = 
\frac{1}{4\pi} \int d\Omega\,  \hat {\bf R} \hat {\bf R} = \frac{1}{3} \stackrel{\leftrightarrow}{\bf U}.
\end{equation}
Using this simple result, we obtain the final expression for energy radiation as \cite{Zhang20prb}
\begin{equation}
\label{eq-radiation-farpiless}
I = \int_0^\infty\!\! d\omega\, \frac{-\hbar \omega^2}{6\pi^2 \epsilon_0 c^3} {\rm Im\,} {\rm Tr} \bigl(\Pi^<(\omega)\bigr).
\end{equation}

As a simple application of this formula, we reproduce the textbook result of dipole radiation.   Consider a charge 
moving according to $z(t) = a \cos(\omega_0 t) = \frac{1}{2} a ( e^{-i\omega_0 t} + e^{+i\omega_0 t})$
in $z$ direction.  $a$ is the amplitude of the oscillation with frequency $\omega_0$.    The velocity is 
$v(t) = dz(t)/dt = \frac{1}{2} (-i\omega_0 a)( e^{-i\omega_0 t}- e^{+i\omega_0 t})$.   We need the
total current (current density integrated over the volume), which is $J(t) = q v(t)$, here $q$ is the charge
of the particle.   The self-energy $\Pi$ is the current-current correlation.  As we are dealing with a classical charge, and athermal, 
we do not distinguish $\Pi^<$ with $\Pi^>$ and just call it $\Pi$.   $\Pi$ is a 3 by 3 matrix, but since the 
particle is moving in $z$ direction, we only have the $\Pi_{zz}$ component nonzero, which is, in real 
time,
\begin{equation}
\Pi(t,t')_{zz} = \frac{q^2}{i\hbar} \bigl\langle v(t) v(t') \bigr\rangle.
\end{equation}
We see if we plug in the formula for the velocity, we do not have a time-translationally invariant result.  This 
is because we assume that the oscillator has no damping.   If we introduce a damping to the oscillator, then the
$t-t'$ dependence remains but $t+t'$ will be damped out at long time.  So performing the average has the
effect of getting rid of the $t+t'$ dependence.   Fourier transform the result into frequency domain, we find
\begin{equation}
\Pi_{zz}(\omega) = { 2\pi (\omega_0 p)^2 \over i 4\hbar} \bigl(\delta(\omega + \omega_0) + \delta(\omega - \omega_0) \bigr),
\end{equation}  
here $p = qa$ is the dipole moment.  The correlation is $\delta$ peaked at $\pm \omega_0$.   
Putting this expression in the radiation power formula, we find
\begin{equation}
I = {\omega_0^4 p^2 \over 12 \pi \epsilon_0 c^3}.
\end{equation}
This is the result in electrodynamics for dipole radiation \cite{Jackson3rd}.

\begin{figure}
  \centering
  \includegraphics[width=\columnwidth]{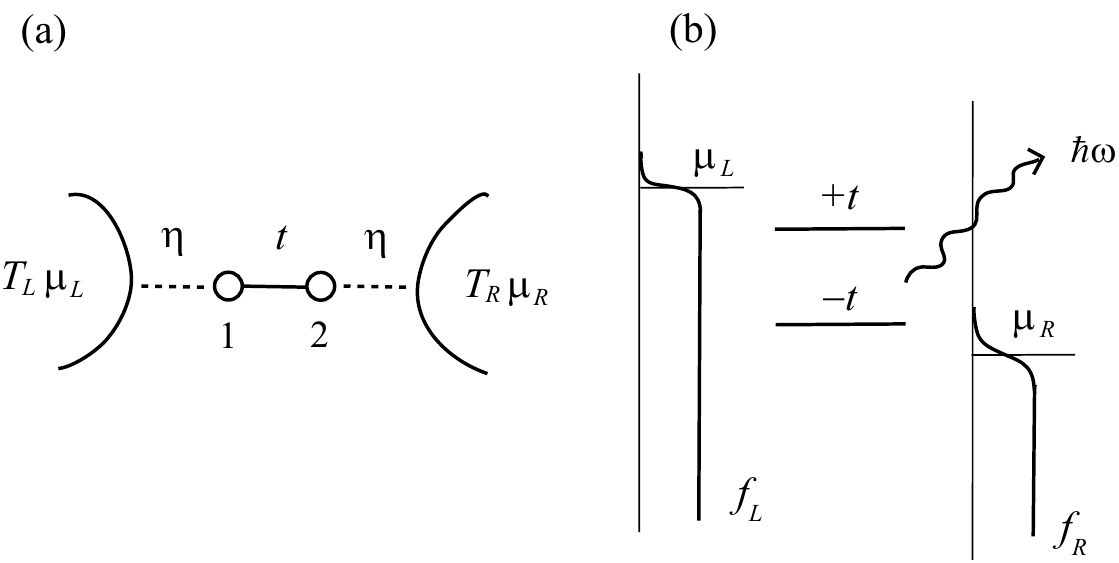}
  \caption{(a) Two-dot model connected to the left and right baths with strength $\eta$. electron hops between two sites with hopping parameter $t$.  (b) Energy levels of the two-dot model and relation to the bath distribution functions.  When electron jumps from the excited state to the ground state, it releases energy $2t = \hbar \omega$.}
 \label{fig-twodots}
\end{figure}

\section{a two-dot model}  
The simple example given above does not involve thermal equilibrium (or nonequilibrium).   
In this section, we give a different example where the material system is modeled explicitly.   Obviously,
an isolated electric dipole cannot oscillate forever.   In order to have a steady state 
established, we must supply continuously energy.   This energy is pumped in through the effect of
the baths.   Here we consider a two-dot model consisting of two electron sites, see Fig.~\ref{fig-twodots}.   An electron can hop
from the left bath into the left dot, and then it may hop again to the right dot, and eventually hop to
the right bath.   This is a toy model for electroluminescence --- the emission of light due to 
electric current \cite{Kuhnke17,ZhangZQ-lum}.  The hopping between the dots generates
electric current, which couples the electron with photons in space, generating  radiation.   A special situation
is that when the temperatures and chemical potentials of the two baths are equal, then it is a thermal radiation problem.
The role of the baths is to supply and to dump the electrons, and we assume that they do not couple to the
field. 

The double-dot Hamiltonian before coupling to the baths is given by 
\begin{equation}
\hat{H}_C = -t (c^\dagger_1 c_2 + c^\dagger_2 c_1) = (c_1^\dagger\; c_2^\dagger)H_C 
\left( \begin{array}{c} 
c_1 \\ c_2 
\end{array}
\right).
\end{equation}
Here $c_1^\dagger$ creates an electron at site 1, and $c^\dagger_2$ at site 2. 
The isolated center has four many-body eigenstates with no particles,  energy $E_1=0$, and
the vacuum state $\Psi_1 = |0\rangle$, one particle at the ground state of the one-particle state with
energy $E_2 = -t$ and $\Psi_2 = \frac{1}{\sqrt{2}}\bigl( c^\dagger_1 +  c^\dagger_2\bigr)|0\rangle$ 
in a symmetric combination, and one particle at excited state with $E_3 = +t$ with an antisymmetric combination,
$\Psi_3 = \frac{1}{\sqrt{2}}\bigl( c^\dagger_1 -  c^\dagger_2\bigr)|0\rangle$, and finally,
$E_4 = 0$ and $\Psi_4 = c^\dagger_1 c^\dagger_2 |0\rangle$ with both sites occupied. 
The energy levels of the isolated center give us a clear picture of when light will be emitted 
based on energy conservation.   If an electron flows from left to right without hopping between the
one-particle states, it cannot emit photon.   The process that emits photon is the one that an electron comes 
at the excited state and stays there for sufficiently long time, then it spontaneously jumps to the
ground state and loses energy to the space at infinity.

We can consider the two sites as part of a 1D chain exposed to the coupling to photon field.  In such a strong system-bath 
coupling regime, the emission turns out rather weak as most of the time, the electrons simply travel
through the wire without emission.  So a weak coupling of the double-dot to the baths is preferred.   We
use the simplest possible bath coupling as the wide-band model, where the coupling is constant 
independent of the energy.  With the effect of the baths, the retarded Green's function of the electron
is given by
\begin{equation}
(E - H_C - \Sigma^r) G^r = I,
\end{equation}
where $\Sigma^r = \Sigma_L^r + \Sigma_R^r$ is the self-energy due to the baths.   In the wide-band 
approximation, we take 
\begin{equation}
\Sigma_L^r = \left( \begin{array}{cc}
- i\eta  & 0\\ 
0  & 0 
\end{array}
\right), \quad
\Sigma_R^r =\left( \begin{array}{cc}
0 & 0 \\
0 & - i \eta 
\end{array}
\right),
\end{equation}
such that the left dot (1) is coupled to the left bath, and right dot (2) to the right bath.  We note that
the damping parameter is related to the usual notation $\Gamma_L \equiv i( \Sigma_L^r - \Sigma_L^a)$
by $\eta = (\Gamma_L)_{11}/2$, and similar for $\Gamma_R$.   With this treatment of the baths, it is completely
equivalent to the usual replacement of $E \to E + i\eta$ to the free electron Green's function.   The 2 by 2 
matrix elements for the Green's function are obtained from a matrix inverse $(E+i\eta - H_C)^{-1}$ as 
\begin{eqnarray}
G_{11}^r = G_{22}^r &=& { E + i\eta \over (E+i\eta)^2 - t^2 },\\
G_{12}^r = G_{21}^r &=& { -t \over (E+i\eta)^2 - t^2 }. 
\end{eqnarray}

Having obtained the retarded Green's function and clarified the bath self-energies, our next step is to use
them to obtain the photon self-energy $\Pi^<$ as it enters explicitly for the radiation power.  Initially,
the self-energy as defined by Eq.~(\ref{eq-Pi-local}) is location dependent.   But in calculating the total
radiation, a long-wave approximation was used as the typical thermal or even optical wave lengths are
much larger comparing to the atomic distances.  This leads to a total $\Pi^<$ which is only directional
dependent as a 3 by 3 matrix.   While the local current is expressed by the coupling matrix $M^{l\mu}$,
summing over $l$ gives the velocity matrix times the electron charge, i.e.,
\begin{equation}
I_\mu = \sum_l I_{l\mu} = (-e) c^\dagger V^\mu c,\quad \mu=x,y,z.
\end{equation}  
Under the random phase approximation (RPA), we treat the electrons as free electrons not coupled to
the field (but can be coupled to the baths through bath self-energy).   Since they are free electrons, 
we can apply Wick's theorem, taking care that the electron operators are anti-commuting under the 
contour order sign.   A general term in $\Pi(\tau, \tau')$ takes the form  $\langle T_\tau c^\dagger_1
c_2 c^\dagger_3 c_4\rangle$.   Equal-time decomposition gives a constant in time due to time 
translational invariance.   This constant can be absorbed into a redefinition of $D$, thus it does not
appear as a self-energy term.  Alternatively, the constant results in a $\delta$ function in frequency domain which does not contribute to transport.    For a normal metal, the particle non-conserving terms,
$\langle T_\tau c^\dagger_1 c^\dagger_3\rangle$ or $\langle T_\tau c_2 c_4\rangle$, is zero.  This left with only one possible combination, 
$c_2 c^\dagger_3$ and $c_1^\dagger c_4$.  After swapping into the correct order for the Green's
function, which produces a minus sign, we find that the contour ordered $\Pi$ can be expressed in terms of 
the electron Green's function as
\begin{equation}
\Pi_{\mu\nu}(\tau, \tau') = -i\hbar e^2 {\rm Tr} \bigl[ V^{\mu} G(\tau,\tau') V^\nu G(\tau',\tau) \bigr].
\end{equation}
Here the trace is over the electron site space, and the contour ordered electron Green's function is 
$G_{jk}(\tau, \tau') = \bigl\langle T_\tau c_j(\tau) c_k^\dagger(\tau') \bigr\rangle/(i\hbar)$.   The lesser
component is then given by 
$\Pi_{\mu\nu}^<(t) = -i\hbar e^2 {\rm Tr} \bigl[ V^{\mu} G^<(t) V^\nu G^>(-t) \bigr]$, 
here time-translational invariance is assumed.  Since the radiation power formula requires the frequency
domain one, we Fourier transform into $\omega$, obtaining
\begin{equation}
\label{eq-Pils}
\Pi_{\mu\nu}^<(\omega) = -i e^2\int_{-\infty}^{+\infty}\!\! \frac{dE}{2\pi} {\rm Tr} \bigl[ V^{\mu} G^<(E) V^\nu G^>(E-\hbar \omega) \bigr],
\end{equation}

Given the Hamiltonian matrix element as $H_{jk}$, the velocity matrix elements can be constructed from
it as 
\begin{equation}
V_{jk}^\mu = \frac{1}{i\hbar} H_{jk} (R_j^\mu - R_k^\mu).
\end{equation} 
On a continuum, the velocity is the rate of change of position, ${\bf v} = d{\bf r}/dt = 
[{\bf r}, H]/(i\hbar)$.  Taking the matrix element with the states where the position operator is diagonal, we 
obtain the above in a tight-binding representation. 
For the two-dot model, the electron can move only in one direction with a spacing $a$, call it $x$, then 
\begin{equation}
V^x = \left( \begin{array}{cc}
0  & - i \frac{at}{\hbar} \\
i \frac{at}{\hbar} & 0
\end{array}
\right).
\end{equation}
Plugging this result into the general formula, Eq.~(\ref{eq-Pils}), we have
\begin{eqnarray}
\Pi^<_{xx}(\omega) &=& i e^2 \left(\frac{at}{\hbar}\right)^2 
\int_{-\infty}^{+\infty}\!\! \frac{dE}{2\pi} \Big[  G_{12}^<(E) G_{12}^>(E-\hbar \omega) + \nonumber \\
&&  G_{21}^<(E) G_{21}^>(E-\hbar \omega)  - G_{11}^<(E) G_{22}^>(E-\hbar \omega) \nonumber \\
&& - G_{22}^<(E) G_{11}^>(E-\hbar \omega)  \Big].
\end{eqnarray}
The integral can be performed approximately if we use the fact that $\eta$ is small.  In this limit, the
spectrum of the system is two sharp peaks at $\pm t$.  The lesser Green's function is obtained from the
Keldysh equation by 
\begin{equation}
G^{<}(E) = G^r(E)  \left( \begin{array}{cc}
2 i f_L \eta & 0 \\
0 & 2 i f_R \eta 
\end{array}\right) 
G^a(E).
\end{equation}
Here $f_L$ and $f_R$ are the Fermi functions associated with the left and right bath.  $G^>$ is 
obtained by replacing $f$ by $f-1$.   The calculation
becomes somewhat tedious, but in the limit $\eta \to 0$, it simplifies greatly.  Skipping some details, the final
result for power after integrating over $E$ and $\omega$ is 
\begin{equation}
I = \frac{1}{3\pi} \frac{e^2 v_0^2 t^2}{\hbar^2 \epsilon_0 c^3} (f_L + f_R) (2 - f_L' - f_R'), 
\end{equation} 
here $v_0 = at/\hbar$, $f_L \equiv f_L(t) = 1/(e^{\beta_L (t-\mu_L)}+1)$ is the electron occupation 
at the excited state from the left bath, and $f_L' \equiv f_L(-t)$ is the electron occupation at the ground state 
from the left bath.  $f_R$ and $f_R'$ are similarly defined.   At high bias with $\mu_L \to +\infty$ and
$\mu_R \to -\infty$, we have $f_L = f_L' = 1$ and $f_R = f_R' = 0$.   In this limit, we see that the power
obtained is identical to the dipole oscillator result if we identify the frequency by
$\hbar \omega = 2t$ and a dipole moment with $p=ea/2$.  This value of dipole moment is consistent
with the matrix element $\langle \Psi_{+}|(-e) \hat{x} | \Psi_{-}\rangle$ between the excited and ground state
of the position operator.  The radiation power at high bias can also be written as $\hbar\omega/\tau$ where
$1/\tau$ is the spontaneous decay rate as in the Weisskopf-Wigner theory \cite{Weisskopf30}.
  
At the beginning of this section, we mentioned that the energy emitted is supplied by the electron baths.
This point can be made quantitatively.    We denote the energy out of the left bath as $I_L$, and that
of right bath as $I_R$.   We can also think of the infinity as a bath, but it only absorbs energy, $I_\infty = -I\leq 0$.
Conservation of energy means $I_L + I_R + I_\infty = 0$.   This conservation can be checked explicitly.
We can calculate the energies of the left and right baths from the Meir-Wingreen formula, 
Eq.~(\ref{MWeq1}).   Here we must consider the coupling of electrons with the field, thus the
electron Green's function should be the one with the nonlinear self-energy due to photons, i.e.
$G_n =  G + G \Sigma_n G_n$, where $G$ is the free electron ones.   The electron-photon couplings
are weak, so we use again the lowest order expansion approximation, i.e.,  
$G_n^< \approx G^< + G^r \Sigma_n^< G^a$.   We also note ${\rm Tr}(G^> \Sigma^< - G^< \Sigma^>)
= 0$, where $\Sigma^{>,<}$ is the sum of the left and right bath self-energy.   We are not interested in
the energies that come out of the left bath and go into the right bath.   If we compute the total,
$I_L + I_R$, the first term from the free electron $G^<$ vanishes, and we shall focus on the second
term only.   The nonlinear self-energy $\Sigma_n^<$ is similar to the scalar photon theory case and is
due to the Fock diagram.   The Fock diagram needs $D^<$ which is obtained from the Keldysh equation.
Omitting some calculation details, the energy that is lost to space from left  bath is, 
\begin{eqnarray}
I_L &=& { e^2 v_0^2 t^2 \over 6 \pi \hbar^2 \epsilon_0 c^3} \Big[ 
f_L (2 - f_L' - f_R') \nonumber \\
&& +\, (1- f_L')(f_L + f_R)
\Big]. 
\end{eqnarray}
$I_R$ is obtained by swapping $L \leftrightarrow R$, and the sum is equal to $I$ obtained earlier.

\newpage

\part{Full theory in temporal gauge}

In part II, we discussed the electromagnetic interactions with electrons in the transverse gauge by the vector potential, and in part I, 
we take care of the Coulomb interaction with the scalar potential.   In principle, taking both together we have  a full theory of the electrodynamics.  The transverse gauge, also known as the Coulomb gauge, is a standard approach in condensed matter physics for
ease of quantization.  However, it is not the most economical, as we have to calculate  the 
charge-charge correlation, current-current correlation, or possibly a cross correlation between charge and current as a 4 by 4 matrix for $\Pi$.  Since charge
and current are related by the continuity equation, these correlation functions are related.   In the standard
fluctuational electrodynamics (FE) of Polder and von Hove type \cite{Rytov53,Polder71}, it is usually formulated by the $\bf E$ and $\bf B$ fields which are gauge
independent quantities.  It is then possible to reformulate our vector-field based theory in a different gauge, known as the  
$\phi=0$ gauge or temporal gauge \cite{Heisenberg30,Creutz79}, which is more directly related to the gauge independent FE theory.   Since we have
banished the scalar field, we only need to consider a general vector field $\bf A$ without the transverse requirement, and we only
need to compute the current-current correlation.    The drawback of the $\phi=0$ gauge is that we need to impose additional condition 
on the quantum states so that Gauss's law $\nabla \cdot {\bf E} = \rho/\epsilon_0$ is satisfied.   This extra complication seems not 
to hinder our formulation as we usually never consider the states explicitly in an NEGF formulation.

\section{Hamiltonian, etc.}
In the $\phi=0$ gauge, the Lagrangian is the same as in Eq.~(\ref{eq-vector-L}), except that the scalar field $\phi$ is set to 0, and the transverse condition on $\bf A$ is not imposed.  The Hamiltonian, by a similar step, is the same with $\phi=0$, i.e.,
 \begin{eqnarray}
\hat H &=&
\frac{\epsilon_0}{2} \int dV \left[ \left(\frac{\partial {\bf A}}{\partial t}\right)^2 
+c^2  \left(  \nabla \times {\bf A} \right)^2 \right] + 
\nonumber \\
&&\sum_{j,l}c^\dagger_j H_{jl} c_l 
\exp\left(\! -i\frac{e}{\hbar}\int_l^j \!\! {\bf A}\cdot d{\bf r} \right).
\end{eqnarray}
Note that the Hamiltonian, Eq.~(\ref{eq-H-gauge-inv}), is gauge invariant.  Here we commit a gauge choice by setting $\phi = 0$. 
The gauge is not completely fixed as we can still make transformation ${\bf A} \to {\bf A} - \nabla \chi$,  $c_j \to \exp(ie\chi_j/\hbar) c_j$ with
a time-independent $\chi$. 
Since the vector field here is the full one, the canonical commutation relation uses the normal Dirac $\delta$ function instead of the transverse
delta function, 
\begin{equation}
\label{eqA-Pi-comm}
\bigl[ A_\mu({\bf r}), \Pi_\nu({\bf r}') \bigr] = i \hbar\, \delta_{\mu\nu}\delta({\bf r} - {\bf r}'),
\end{equation}
where the conjugate momentum $\Pi_\nu = \epsilon_0 \dot{A}_\nu = - \epsilon_0 E_\nu$.    The Heisenberg equation of motion for the vector field is
\begin{equation}
v^{-1} {\bf A} =  -\epsilon_0\left( \frac{\partial^2 {\bf A}}{\partial t^2} + c^2 \nabla \times (\nabla \times {\bf A}) \right)  =-{\bf j}.
\end{equation}
Here we define $v^{-1}$ as the $3\times 3$ matrix in the directional index and as a differential operator defined by the middle term acting
on $\bf A$.  It turns out that its inverse is the free Green's function, i.e.,  ${\bf A} = -v {\bf j}$.   The awkward minus sign is
needed so that $v$ is consistent with the NEGF definition of the retarded Green's function by the commutator,
$\theta(t) \bigl\langle [A_\mu({\bf r},t), A_\nu({\bf 0},0)]\bigr\rangle/(i\hbar)$.    Formally, the current density on the right-hand side is given by a
functional derivative with respect to $\bf A$ or a commutator of the conjugate variable of $\bf A$ to the Peierls substitution term,
\begin{eqnarray}
{\bf j} &=& - \frac{\delta\,}{\delta {\bf A}} \sum_{jl} c_j^\dagger H_{jl} c_l \exp\left( - \frac{ie}{\hbar} \int_l^j {\bf A} \cdot d{\bf r} \right) \\
&=& \frac{\epsilon_0}{i\hbar} \left[ \dot{{\bf A}}({\bf r}), \sum_{jl} c_j^\dagger H_{jl} c_l \exp\left( - \frac{ie}{\hbar} \int_l^j {\bf A} \cdot d{\bf r} \right) \right]. \nonumber
\end{eqnarray} 

This all looks easy in the $\phi=0$ gauge.   However, the catch is that Gauss's law is missing from the Lagrangian, so we have to
impose it as an extra condition.  The question is when and where.  It can be shown that Gauss's law,  if it is satisfied at one 
particular time, is also satisfied at all times, i.e.,  $[ \nabla \cdot {\bf E} - \rho/\epsilon_0, \hat {H}] =0$.   This means that we can
impose the requirement as an initial condition, i.e., initial states.   The physical states  $\Psi$ must be selected such that
$( \nabla \cdot {\bf E} - \rho/\epsilon_0)\Psi =0$ \cite{Fradkin21}.   Alternatively, we can impose the condition on the operator $\bf A$.  
The problem with $\bf A$ is that the Fock state space created by it is larger than physically allowed.  In 
the following, in determining the free Green's function below, we will apply Gauss's law
when solving for $\bf A$. 

\subsection{Free field Green's function}
The solution to the Green's function in the absence of matter depends on the choice 
of boundary conditions.   In full space,  
the easiest way to obtain the retarded Green's function in the temporal gauge is to solve the equation of motion,
$v^{-1}{\bf A} = - {\bf j}$, in Fourier space.   To incorporate Gauss's law, we note $\nabla \times (\nabla \times 
{\bf A}) = \nabla (\nabla \cdot {\bf A}) - \nabla^2 {\bf A}$.   Using $\nabla \cdot {\bf E} = \rho/\epsilon_0$,
${\bf E} = - \partial {\bf A}/\partial t$,  together with charge conservation, $\partial \rho/\partial t + 
\nabla \cdot {\bf j}=0$, in frequency domain, the divergence of $\bf A$ can be expressed as proportional to the
divergence of the current.  We move this term to the right-hand side, thus it becomes a source term of a wave equation.   We obtain the frequency and wave-vector
space equation with the replacement $\partial /\partial t \to -i\omega$, $\nabla \to  i{\bf q}$, as 
\begin{equation}
{\bf A}(\omega, {\bf q}) = - {\stackrel{\leftrightarrow}{\bf U} - {\bf q}{\bf q}/(\omega/c)^2 \over \epsilon_0 (\omega^2 - c^2 q^2)} \cdot {\bf j}(\omega, {\bf q}). 
\end{equation} 
The coefficient in front of ${\bf j}$ is the Green's function in $(\omega, {\bf q})$ space.  
This is nearly the same as Eq.~(\ref{eq-d-modespace}) for the transverse gauge except that 
the numerator is not a projector.  To be qualified as retarded,
we also need a displacement for the frequency, $\omega \to \omega + i\eta$, with an infinitesimal positive $\eta$.
The Green's function in real space is obtained by inverse Fourier transform, as
\cite{Novotny06,Keller11}
\begin{eqnarray}
\label{eq-v-real-space}
v^r({\bf r},\omega) &=& 
-\frac{e^{i \frac{\omega}{c} r}}{4\pi \epsilon_0 c^2 r} \Bigg\{ 
\bigl( \stackrel{\leftrightarrow}{\bf U} - \hat{\bf R}\hat{\bf R} \bigr) +  \nonumber \\
&& \biggl[ - \frac{1}{ i\frac{\omega}{c}r} + 
\frac{1}{\left( i \frac{\omega}{c}r \right)^2}\biggr]
\bigl( \stackrel{\leftrightarrow}{\bf U} - 3 \hat{\bf R}\hat{\bf R} \bigr) \Bigg\}. 
\end{eqnarray} 
Here $\stackrel{\leftrightarrow}{\bf U}$ is the identity dyadic,  $\hat{\bf R} = {\bf r}/r$ is the radial direction unit vector.  This is almost the same as  Eq.~(\ref{eq-d-leftright-arrow}) for the Coulomb gauge, except that they differ at the last term at $1/r^3$.   
Our notation $v^r$
is related to the usual dyadic Green's function by $v^r = - \mu_0 \stackrel{\leftrightarrow}{\bf G}$.  The mixed representation is useful for planar geometry,  for which we Fourier transform the variable 
in $z$ direction back to real space, with the result
\begin{eqnarray}
v^r(\omega, {\bf q}_\perp, z,z') &=& 
\frac{e^{i q_z|z-z'|}}{2iq_z \epsilon_0 c^2 a^2} \times \\
&& \left(
\begin{array}{ccc}
1-\frac{q_x^2}{(\omega/c)^2} &  -\frac{q_x q_y}{(\omega/c)^2} &  - s \frac{q_x q_z}{(\omega/c)^2} \\
  -\frac{q_y q_x}{(\omega/c)^2} &  1-\frac{q_y^2}{(\omega/c)^2} &  - s \frac{q_y q_z}{(\omega/c)^2} \\
- s\frac{q_z q_x}{(\omega/c)^2} &  - s\frac{q_z q_y}{(\omega/c)^2} &  1- \frac{q_z^2}{(\omega/c)^2} 
\end{array}
\right). \nonumber
\end{eqnarray}
Here $s$ is the sign of $(z-z')$, and $q_z = \sqrt{(\omega/c)^2 - q^2_\perp}$ for the propagating mode
when $\omega/c > q_\perp$ and $q_z = i  \sqrt{q^2_\perp - (\omega/c)^2}$ for the evanescent mode when 
$\omega/c < q_\perp$,  $q_\perp = |{\bf q}_\perp|$.   We assume ${\bf q}_\perp = (q_x,q_y)$ takes a discrete 
set of values in the first Brillouin zone of a square lattice with lattice constant $a$.   

\section{A unified theory for energy, momentum and angular momentum transfer}

\subsection{Conservations of energy, momentum, and angular momentum}
The three conservation laws, energy, momentum, and angular momentum, are due to the symmetries of time translation, space displacement 
and rotation.    We have already discussed the equation describing the energy conservation through the Poynting theorem, namely,
$\partial u/\partial t + \nabla \cdot {\bf S} = - {\bf E} \cdot {\bf j}$.  We now consider the other two, momentum and angular momentum conservations. 
The momentum conservation is related to the momentum density of the field.   The Poynting vector $\bf S$ is the energy flux, i.e., the 
energy current per unit cross-section area.  So, the magnitude of the Poynting vector is $S=uc$ where $c$ is the speed of light, since photons
move with the speed of light.   The relation between energy of a photon and momentum is $cp = \epsilon$ since photon is a massless
relativistic particle.   This means that the momentum density is given by $u/c$, or in terms of the Poynting vector, is
$S/c^2$, or in vector form, ${\bf S}/c^2=\epsilon_0 {\bf E} \times {\bf B}$.    We compute the rate of change of the momentum density.   
With the help of Maxwell's equations, we find \cite{Jackson3rd} 
\begin{equation}
\label{eq-P-consv}
-\frac{1}{c^2} {\partial {\bf S}\over \partial t} + \nabla \cdot \stackrel{\leftrightarrow} {\bf T} \,= \rho {\bf E} + {\bf j} \times {\bf B},
\end{equation}
here $\stackrel{\leftrightarrow}{\bf T} = \epsilon_0 {\bf E}{\bf E} + \frac{1}{\mu_0} {\bf B}{\bf B} - u {\stackrel{\leftrightarrow}{\bf U}}$
is Maxwell's stress tensor, and the right-hand side is the Lorentz force applied to object per unit volume, $\bf f$.   The conservation equation associated
with the angular momentum is similarly obtained by noting that  ${\bf l} = {\bf r} \times {\bf S}/c^2 = \epsilon_0 {\bf r} \times 
({\bf E} \times {\bf B})$ is the angular momentum density.    Taking again the rate of change of ${\bf l}$ and using Maxwell's equations,
we have \cite{Barnett02,Keller11}
\begin{equation}
\label{eq-L-consv}
-{\partial {\bf l}\over \partial t} - \nabla \cdot (\stackrel{\leftrightarrow} {\bf T} \times {\bf r})  = {\bf r} \times  {\bf f},
\end{equation}
where ${\bf f} = \rho {\bf E} + {\bf j} \times {\bf B}$.   We can obtain the angular momentum conservation equation (\ref{eq-L-consv})
from the momentum conservation, Eq.~(\ref{eq-P-consv}), by multiplying it by ${\bf r} \times\,$, and using the fact that 
$\stackrel{\leftrightarrow}{\bf T}$ is symmetric.  As a result, we can `pull' the divergence operator $\nabla \cdot\,$ out,  i.e.,
${\bf r} \times (\nabla \cdot  \stackrel{\leftrightarrow}{\bf T}) = - \nabla \cdot (\stackrel{\leftrightarrow}{\bf T} \times {\bf r})$.  
Note that when a tensor $\stackrel{\leftrightarrow}{\bf T}$ is cross-multiplied by $\bf r$, the result is still a tensor, with 
component $(\stackrel{\leftrightarrow}{\bf T} \times {\bf r})_{ij} =  \sum_{kl} T_{i k} x_l \epsilon_{klj}$.  When a tensor is dotted with 
another vector, the result is a vector, e.g.,  $(\nabla \cdot  \stackrel{\leftrightarrow}{\bf T})_i = \sum_{j} 
\partial /\partial x_j T_{ji}$.  For  nonsymmetric tensor, such as $\stackrel{\leftrightarrow} {\bf T}\! \times {\bf r}$, dot from the left is 
different from dot from the right.  We have a symmetry $-{\bf \Sigma} \cdot (\stackrel{\leftrightarrow}{\bf T} \times {\bf r}) = 
({\bf r} \times \stackrel{\leftrightarrow}{\bf T}) \cdot {\bf \Sigma}$.

From the conservation equations, we have the physical interpretation that the Poynting vector $\bf S$ gives the energy flux,
and the Maxwell stress tensor $\stackrel{\leftrightarrow}{\bf T}$  integrated over an enclosing surface with outward norm is the force applied to the body, and 
$-\!\stackrel{\leftrightarrow} {\bf T}\! \times {\bf r}$ the torque applied to the body.    
Since we are mainly interested in steady-state average, the average of a rate of change of a finite quantity is zero.   Since averaging 
and taking partial derivative commute, we find  $\langle \partial a /\partial t \rangle = \partial \langle a \rangle /\partial t = 0$.   Using
this result, together with the conservation equations, we can compute the energy, momentum, and angular momentum transfer
in two ways  --- by integrating over the surface enclosing some body or by a volume integral of the expressions on the right-hand side
of the equations.  

These classical expressions are changed to quantum operators in a quantum theory, with symmetrized product if necessary, and also earlier
in part II we argue for a normal order in order to remove the zero-point motion contribution.   However, here we have a second
thought for the normal ordering.   The reason is that it is precisely the zero-point motion that gives rise to the Casimir force.   
If we continue using normal order, we would not be able to predict a Casimir force.    In fact, we need to use a symmetric order,
$\frac{1}{2} \langle (A B + BA)\rangle$.   This means that the Green's function will be the Keldysh one of $D^K = D^> + D^<$ that 
enters the expressions for physical observables of the transport quantities.   In order to show that this is not {\it ad hoc} and arbitrary,
we will say that the symmetric order is fundamental, and show explicitly, for energy transport and also for non-confined geometries such
as radiation to infinity, the zero-point motion contribution cancels by itself.    This will be demonstrated in Sec.~\ref{sec-XIA}.   With these considerations,
the power emitted, the force applied, and torque applied to a body $\alpha$ enclosed with a surface are
\begin{eqnarray}
I_\alpha &=& \oint_{\bf \Sigma} {\bf S} \cdot d{\bf \Sigma} = \int_{V} (-{\bf E} \cdot {\bf j}) dV,\\
{\bf F}_\alpha &=& \oint_{\bf \Sigma}  \stackrel{\leftrightarrow} {\bf T} \cdot\, d{\bf \Sigma} =\int_V {\bf f}\, dV,\\
{\bf N}_\alpha &=& \oint_{\bf \Sigma} {\bf r} \times \stackrel{\leftrightarrow} {\bf T} \cdot\, d{\bf \Sigma} = \int_V  {\bf r} \times {\bf f}\, dV.
\end{eqnarray}
Here $V$ is the volume enclosed by a simply connected surface $\bf \Sigma$.   We have used Gauss's theorem to change the divergence
over the volume into a surface integral with outward norm. 

\subsection{Expressing transport quantities by Green's functions}

We comment on the surface integral expressions first.  We have already given a general expression for heat transfer in terms of the Green's function $D$ by 
Eq.~(\ref{eq-S-Dlss}), except that now $D$ is defined by the full ${\bf A}$ without imposing the transverse requirement.   Also, since
we have decided to use the symmetric operator order, $D^<$ there should be replaced by the Keldysh version $D^K/2 = (D^< + D^>)/2$.  
The rest remain.    For the force and torque with Maxwell's stress tensor, we can write down similar expressions, but it is messy and not particularly
illuminating.   However, the expressions simplify if we take the long distance far field limit by integrating over a sphere of radius $R$.  
Instead of trying to derive these formulas explicitly here, we will give the results as a special case of the more general Meir-Wingreen
formula, after we have elaborated on the concept of bath at infinity. 

For the volume integrals, we remove the explicit charge $\rho$ dependence in favor of the current $\bf j$.   Due to charge conservation, 
we can compute the charge from the current.   The Lorentz force per unit volume is ${\bf f} = \rho {\bf E} + {\bf j} \times {\bf B}
= - \rho\, \partial {\bf A}/\partial t + {\bf j} \times ({\bf\nabla} \times {\bf A})$.  In steady state, 
$\langle \partial (\rho {\bf A})/\partial t \rangle = \langle (\partial \rho/\partial t) {\bf A} \rangle + \langle \rho \partial {\bf A}/\partial t\rangle = 0$.   We can move the time derivative from the vector potential to charge with a minus sign.   Using the continuity equation,
$\partial \rho/\partial t = - {\bf\nabla} \cdot {\bf j}$, we can write the force as (valid after taking average)
\begin{eqnarray}
{\bf f} &=&  - ( {\bf\nabla} \cdot {\bf j} ) {\bf A} + {\bf j} \times ( {\bf\nabla} \times {\bf A}) \nonumber \\
&=& - \sum_{\mu} \partial_{\mu} (j_\mu {\bf A}) + \sum_{\nu} j_\nu {\bf\nabla} A_\nu.
\label{eq-f-A}
\end{eqnarray} 
Here we have used the triple cross-product formula and combined the charge term with one of the cross-product terms.   
The index $\mu$ or $\nu$ takes the Cartesian directions, $x$, $y$, or $z$.  Note that the first term is a divergence.   If we integrate over 
a volume large enough to enclose the object where at the surface there is no current, then the first term is zero.   We can then use for the 
total force calculation inside the integral as
\begin{equation}
{\bf f} = \sum_{\nu}  j_\nu {\bf\nabla} A_\nu.
\end{equation}
However, the first term in Eq.(\ref{eq-f-A}) is no longer a divergence in the torque calculation in ${\bf r} \times {\bf f}$.   By
an integration by parts in the space index $\mu$, we can move the derivative to $\bf r$.   The same argument that at the surface
there is no electric current can be used to eliminate the surface contribution.  This gives the integrand for the total torque as
\begin{equation}
{\bf r} \times {\bf f} = {\bf j} \times {\bf A} + \sum_{\nu} j_\nu ({\bf r} \times {\bf\nabla}) A_\nu.
\end{equation}
Here the first term is interpreted to be the spin part of the contribution to angular momentum transfer, as it is independent of a choice
of the coordinate origin, while the second term is interpreted as due to orbital angular momentum \cite{Barnett16}. 

Now we have transformed each of the integrands for the three transport quantities in terms of the current density $\bf j$ and vector
potential $\bf A$.  The rest of the steps can be dealt with in a unified way.   First, we define a new Green's function as an intermediate
quality by
\begin{equation}
F_{\mu\nu}({\bf r}\tau;{\bf r}'\tau') = \frac{1}{i\hbar} \bigl\langle  T_\tau A_\mu({\bf r},\tau) j_\nu({\bf r}',\tau') \bigr\rangle. 
\end{equation}
Using this Green's function, which reflects the interaction between the field and matter, we can write 
\begin{eqnarray}
I_\alpha &=&{\rm Re} \int_0^\infty \frac{d\omega}{2\pi} \hbar \omega \,{\rm Tr} \left[ F^K(\omega) \right], \\
{\bf F}_\alpha &=&{\rm Re}  \int_0^\infty \frac{d\omega}{2\pi} i \hbar \,{\rm Tr} \left[ \nabla_{{\bf r}} F^K(\omega) \right], \\
{\bf N}_{\alpha} &=&{\rm Re}  \int_0^\infty \frac{d\omega}{2\pi}   {\rm Tr} \Big[ i \hbar\, {\bf r} \times \nabla_{{\bf r}} F^K(\omega)  
-{\bf S} F^{K}(\omega) \Big]. \qquad
\end{eqnarray}
Note that with the symmetric order of any two Hermitian operators, we have $\frac{1}{2} \langle AB + BA\rangle =
{\rm Re} \int_0^\infty i \hbar D_{AB}^K(\omega)\, d\omega/(2\pi)$, as the correlation defined in time domain by
$i\hbar D_{AB}^K(t) = \langle A(t) B + BA(t)\rangle$ is real.  As a result, in frequency domain the real part is symmetric, and imaginary
part antisymmetric.   By integrating from $-\infty$ to $+\infty$ only the real part of $D_{AB}^K(\omega)$
survives.   We have used the symmetry to write the integrals with the positive frequency only.  In the above expressions, we should view the Keldysh component $F^K$ as a matrix indexed by both position $\bf r$ as well as direction $\nu$.   The 
trace is in the combined space, i.e.,  ${\rm Tr}[ \cdots] = \int_{V_\alpha} dV\sum_{\nu} \cdots$.   The volume integral covers
the object $\alpha$ only.   The gradient operator 
${\bf\nabla}_{\bf r}$ acts on the first argument of $F^K = F^K({\bf r}, {\bf r}', \omega)$ which is associated with the argument of
the vector field $\bf A$.   The factor in the energy current, $\hbar \omega$, is due to the time derivative with respect to $\bf A$, Fourier 
transformed to the frequency domain.   The first term of the torque expression,  $\hat {\bf L} =  {\bf r}\times \bigl( \frac{\hbar}{i}  
{\bf\nabla}_{\bf r}\bigr)$, is the single particle orbital angular momentum operator in the position space, while $S^\mu_{\nu\gamma}
=(-i\hbar)\epsilon_{\mu\nu\gamma}$ is the spin operator in the Cartesian direction acting on $F^K$. 

We need to connect our $F$ back to our earlier Green's function of the field $D$ and the materials properties $\Pi$.   In fact, such a 
relation does exist, it is
\begin{equation}
F = -D \Pi_\alpha.
\end{equation}
This equation should be best viewed as defined on the Keldysh contour, and is a convolution in space as well as Keldysh contour.  
In the next subsection, we will prove this result and point out an additivity assumption needed for its validity.   Here, if this expression is
assumed, then the Keldysh component is obtained by the Langreth rule,  $-F^K = D^r \Pi_\alpha^K + D^K \Pi_\alpha^a$.   With 
this, we obtain the Meir-Wingreen formulas for the transport quantities as \cite{ZhangYM22prb} 
\begin{eqnarray}
\label{meir-wingreen-I}
I_\alpha &=& \int_0^\infty \frac{d\omega}{2\pi} (-\hbar \omega) \,{\rm Re\, Tr}\bigl(D^r \Pi_\alpha^K + D^K \Pi_\alpha^a\bigr),\quad \\
{\bf F}_\alpha &=&  \int_0^\infty \frac{d\omega}{2\pi} \,{\rm Re\, Tr} \Bigl[\hat{\bf p}\bigl(D^r \Pi_\alpha^K + D^K \Pi_\alpha^a\bigr)\Bigr] , \\
\label{meir-wingreen-N}
{\bf N}_{\alpha} &=&  \int_0^\infty \frac{d\omega}{2\pi}   {\rm Re\, Tr} \Bigl[ \hat{\bf J}\bigl(D^r \Pi_\alpha^K + D^K \Pi_\alpha^a\bigr) \Bigr] . 
\end{eqnarray} 
Here $\hat{\bf p} = \frac{\hbar}{i}{\bf\nabla}$ is the momentum operator, and $\hat{\bf J} = {\bf r}\times \hat{\bf p} + {\bf S}$ is
the total angular momentum operator.

\begin{figure}
\centering
\includegraphics[width=0.8\columnwidth]{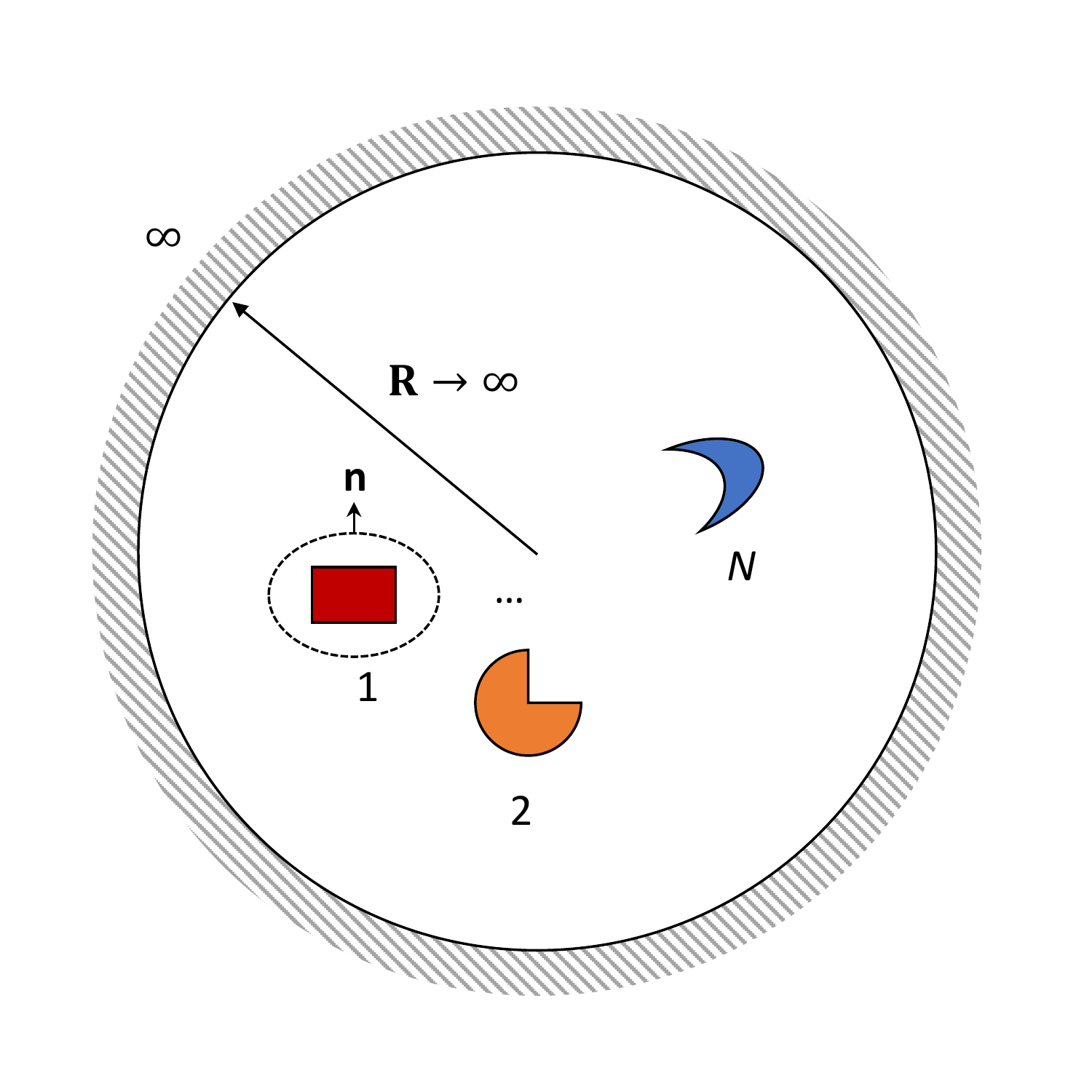}
\caption{$N$ objects of arbitrary shapes, each experiencing energy radiation, force and torque exerted to. There is one more special
object called $\infty$ which is the space outside the sphere of radius $R$.   The norm vector $\bf n$ is a unit vector pointing outwards 
from the enclosing surface.
}
\label{fig:objects}
\end{figure}

\subsection{Properties of the Meir-Wingreen formulas}
In this subsection, we investigate some of the properties implied by the Meir-Wingreen formulas for energy, momentum, and angular momentum transfers.
We first discuss the conservation laws.    We assume $N$ well-separated objects with $\alpha=1$, 2 \dots, $N$ localized in a bounded region,
see Fig.~\ref{fig:objects}.    In order to account for the loss of the transported quantities to infinity, it is necessary to introduce one more
object $(N+1)$ as the `object' at infinity.  The conservations of the physical observables are then
\begin{equation}
\sum_{\alpha=1}^{N+1} I_\alpha = 0,\quad
\sum_{\alpha=1}^{N+1} {\bf F}_\alpha  = {\bf 0}, \quad
\sum_{\alpha=1}^{N+1}{\bf N}_\alpha  = {\bf 0}.
\end{equation}  
These equations are obviously true, as the volume integrals can also be obtained by surface integrals enclosing the objects.  Each
surface separating the objects is used twice with opposite sense of the ``outward'' norm.   Since the $\alpha$ dependence is only
through the self-energy $\Pi_\alpha$, it is necessary to introduce also a self-energy for the infinity.  With the total self-energy 
$\Pi = \sum_\alpha^{N+1} \Pi_\alpha$, the conservations are satisfied for the three quantities if 
\begin{equation}
D^r \Pi^K + D^K \Pi^a = D^r \Pi^K (I + D^a \Pi^a) = 0.
\end{equation}
Here to get the second expression we have used the Keldysh equation $D^K = D^r \Pi^K D^a$.  Since the factor $D^r \Pi^K$ cannot 
be zero in general, we require that  $I + D^a \Pi^a = 0$.   This is indeed valid if we recall the Dyson equation is 
\begin{equation}
\label{eq-dyson-inf}
D^r = v^r + v^r \sum_{\alpha}^N \Pi_\alpha^r D^r, 
\end{equation}
which we can also write as $(v^r)^{-1} D^r = I + \sum_{\alpha}^N \Pi_\alpha^r D^r$.   If we identify the self-energy at infinity (or of the environment) 
as the differential operator \cite{Krueger12prb}, 
\begin{equation}
\label{eq-Pi-inf-v}
\Pi_\infty^r = -(v^r)^{-1},
\end{equation} 
and move the $-\Pi_\infty^r D^r$ term to the right side, we find $I + \Pi^r D^r=0$.  Here $\Pi^r$ is a sum from object 1 to 
$N+1$ with $\Pi^r_{N+1} \equiv \Pi_\infty^r$. 
Taking the Hermitian conjugate of this equation, we obtain the needed identity.   We note that in the Dyson equation, the object at infinity is only
an absorbing boundary condition as quantities transported to infinity cannot come back.    The self-energies in the Dyson equation on
the right-hand side in Eq.~(\ref{eq-dyson-inf}) does not include an ``object'' at infinity if the equation is solved in an unbounded domain.   
It is useful to think of each of the objects, including the
object at infinity as a bath, supplying energy, momentum, and angular momentum for the photon fields. This derivation presents to us an 
explicit expression for the bath at infinity as a differential operator.   In a later section, we give an algebraic expression defined on the
surface of a sphere of radius $R\to \infty$. 

We can give an alternative argument for the self-energy of the environment (bath at infinity) as $-(v^r)^{-1}$.   We recall that the 
contour ordered Green's function $D = v + v \Pi D$ implies a pair of equations in real time, one of them is the Dyson equation, 
Eq.~(\ref{eq-dyson-inf}), the other is the Keldysh equation for which we can write in alternative forms,
\begin{eqnarray}
D^K &=& D^r \Pi^K_{\rm obj} D^a + (I + D^r \Pi^r_{\rm obj}) v^K (I + \Pi^a_{\rm obj} D^a) \nonumber \\
&=&  D^r \Pi^K_{\rm obj} D^a + D^r (v^r)^{-1} v^K (v^a)^{-1} D^a \\
& = & D^r\Bigl(\Pi^K_{\rm obj} + (2N_{\infty}+1)\bigl( (-v^r)^{-1} + (v^a)^{-1}\bigr) \Bigr)D^a. \nonumber
\end{eqnarray}  
Here we define the self-energy of the objects to be $\Pi_{\rm obj} = \sum_{\alpha=1}^N \Pi_\alpha$, excluding the bath at infinity. 
The first line above is a general form of the Keldysh equation \cite{Haug08} without evoking a special property of $v^K$.   For isolated systems,
the second term on the right is zero because in this case, an isolated system is nondissipative, 
satisfying $v^{-1} v^K = 0$ where $v^{-1}$ is interpreted as the differential operator.   We can use the Dyson equation to obtain the 
second line.  It is clear if the objects are absent, $\Pi_{\rm obj} = 0$, we get $D^K = v^K$, which is the Green's function for
the free field.  Assuming the free field in the absence of the objects is thermal, i.e., satisfying the fluctuation-dissipation theorem
at temperature $T_{\rm \infty}$, that is, $v^K = (2N_\infty + 1)(v^r - v^a)$, 
we obtain the last line,  from which we see that $(-v^r)^{-1}$ serves as the retarded self-energy
for the environment. 

The Meir-Wingreen formulas are the most general ones where each of the objects could be in some arbitrary nonequilibrium state.
For the energy formula, we have a more symmetric form by adding the Hermitian conjugate inside the trace and divided by 2, 
and then using the general relations, $D^r - D^a = D^> - D^<$, $D^K = D^> + D^<$, and similarly for the self-energies $\Pi$, as 
$I_\alpha = - \int_0^\infty \frac{d\omega}{2\pi} \hbar\omega {\rm Tr} \bigl(D^> \Pi_\alpha^< - D^< \Pi_\alpha^>\bigr)$.  This is the same as for the scalar photon
version, Eq.~(\ref{eq-photon-IL-WM}), except that here $D$ and $\Pi_\alpha$ are tensors with
directional indices.  

The next consequence of the Meir-Wingreen formula we discuss is a derivation of the Landauer-B\"utticker formula for the energy transport 
when local equilibrium is valid.  By local equilibrium we mean that each object has  a version of the fluctuation-dissipation theorem for the self-energies,
\begin{equation}
\Pi_\alpha^K = (2N_\alpha +1) (\Pi_\alpha^r - \Pi_\alpha^a), \quad\alpha = 1, 2, \cdots, N, N+1.
\end{equation}
Here  $N_\alpha = 1/(e^{\beta_\alpha \hbar \omega} -1)$ is the Bose function for object $\alpha$ at a local temperature of 
$T_\alpha = 1/(k_B \beta_\alpha)$.  Using this expression for $\Pi_\alpha^K$ and the Keldysh equation  $D^K = D^r \bigl(\sum_\alpha^{N+1} \Pi_\alpha^K\bigr) D^a$, we can completely eliminate the Keldysh components in favor of the retarded Green's functions and self-energies.  
Introducing $\Gamma_\alpha = i(\Pi_\alpha^r - \Pi_\alpha^a)$ as the spectrum of the bath, adding the Hermitian conjugate
of $D^r \Pi^K_\alpha + D^K \Pi_\alpha^a$ and divided by 2 to take care of the real part,  using the symmetry
$(D^K)^\dagger = - D^K$ and similarly for the self-energies, and finally the identity
$i(D^r - D^a) = D^r (\sum_\beta^{N+1} \Gamma_\beta) D^a$  (see Eq.~(\ref{eq-Datta-identity})), we obtain, after some straightforward
algebra,
\begin{equation}
I_\alpha = \int_0^\infty \frac{d\omega}{2\pi}\, \hbar \omega \sum_{\beta=1}^{N+1}(N_\alpha - N_\beta) {\rm Tr}\bigl( D^r \Gamma_\beta D^a \Gamma_\alpha\bigr).
\end{equation}
We remind the reader that here the trace means integrating over the volume and summing over the Cartesian directions, and the Green's functions
have the arguments and indices, e.g., $D^r_{\mu\nu}({\bf r}, {\bf r}', \omega)$.    We note originally the Bose function enters
as $2N_\alpha+1=\coth(\beta_\alpha \hbar \omega/2)$ where the $1$ is the contribution from zero-point motion.   However, in the final formula, the Bose function enters only as 
a difference, the 1s have been canceled out.  Thus, for energy transport, with local thermal equilibrium, the zero-point motion never contributes 
to energy transport.    The same cannot be said about force and angular momentum or if the system is not in local equilibrium.   Due to the presence of the extra differential 
operator $\hat{\bf p}$ or $\hat{\bf J}$, a similar derivation fails to go through, thus there is no equivalent Landauer formula for the 
force and angular momentum unless the operator $\hat{O}$ in front of $D$ commutes with $\Pi_\alpha$.   If there were such a formula in the sense that it 
has a factor $N_\alpha - N_\beta$, we would not have Casimir forces.
 
The Caroli-like expressions for the transmission coefficients have been derived for energy
transfer between objects modelled as dipoles \cite{ben-Abdallah11,Ekeroth17}, and also for
fluctuational-surface-current formulation in \onlinecite{Rodriguez12}.  
If we define the multiple-bath transmission coefficients as $T_{\beta\alpha} = {\rm Tr}\bigl( D^r \Gamma_\beta D^a \Gamma_\alpha\bigr)$,
in general it is not symmetric with respect to a swap of the two baths unless there are in total only two baths.  For systems that are reciprocal, 
i.e., $\Pi^T_\alpha = \Pi_\alpha$, we can show that we do have $T_{\beta\alpha} = T_{\alpha\beta}$.   Unsymmetric transmission implies an energy current known as
super-current between objects even at thermal equilibrium \cite{Zhu-prl16}, but the total $I_\alpha$
out of object $\alpha$ is still zero.  Symmetric or not, there is a sum 
rule \cite{Datta95,Latella17}, $\sum_{\beta} T_{\beta\alpha} = \sum_{\beta} T_{\alpha\beta}$, which is just a consequence of the total current 
conservation, $\sum_\alpha I_\alpha=0$.

\subsection{Prove $F=-D\Pi_\alpha$}
In proving the Meir-Wingreen formulas, we have used this relation, which means, in full form in space, contour time, and direction,
\begin{eqnarray}
F_{\mu\nu}({\bf r}\tau; {\bf r}'\tau') = \hspace{5cm} \\
 -\int d^3{\bf r}''\int d\tau'' \sum_{\lambda} D_{\mu\lambda}({\bf r}\tau; {\bf r}''\tau'')
\Pi_{\alpha,\lambda\nu}({\bf r}''\tau''; {\bf r}'\tau'), \nonumber 
\end{eqnarray}
which is a matrix multiplication in the direction indices, and convolution in space and contour time.  Note that here the self-energy in continuum form 
is for the object $\alpha$ so that $F$ function is associated with a particular object, which we have suppressed the $\alpha$ subscript
for notational simplicity.     We can argue about this relation not so rigorously with linear response.   Within the lowest order of random
phase approximation, the electrons and photons are not coupled directly so we can take the averages in each space separately.   Focusing
on the electrons first, then the response of the electrons due to the internal vector field is the current ${\bf j} = - \Pi_\alpha {\bf A}$,
here we have not yet evaluated the average on the photon space, so $\bf j$ and $\bf A$ are still quantum operators.  Again, this
equation means convolution in contour time and space, and $\Pi_\alpha$ is a $3\times 3$ tensor in directional index space.   Let us be slightly 
more precise by writing  $j(2) = - \int d(3) \Pi_\alpha(2, 3) A(3)$, here we use the abbreviation $(n) \equiv ({\bf r}_n, \tau_n, \mu_n)$,
and $\int d(3)\, \cdot\,$ means integrations over space, contour time, and summing over the directions.   Since we still treat $j(2)$ as a quantum 
operator, we can put the linear response into the definition of the Green's function $F(1,2) = \langle T_\tau A(1) j(2)\rangle/(i\hbar) = 
-\int d(3) \bigl\langle T_\tau A(1) \Pi_\alpha(2,3) A(3)\bigr\rangle /(i\hbar)$.    We can pull $\Pi_\alpha$ out from the average as it is just a number, so we have, using the
definition of $D$, $F(1,2) = - \int d(3) D(1,3) \Pi_\alpha(2,3)$.   As the (bosonic) contour ordered function is symmetric with respect to 
the permutation of their arguments by definition,  $\Pi_\alpha(2,3) = \Pi_\alpha(3,2)$, we obtain  $F = - D \Pi_\alpha$.


But there is a more rigorous proof of this relation. 
It is convenient to start with the total $F$, that is, the sum of the contributions of all the objects (not counting infinity as one) generated 
by the total current, ${\bf j} = \sum_\alpha \bf {j}_\alpha$.   As a consequence of
the Dyson equation, we can show that $F= - D\Pi$ is an exact result.   To demonstrate this, 
we note that the contour ordered Dyson equation can be written alternatively as
$D = v + v \Pi D = v + D \Pi v$.   Multiplying $v^{-1}$ from the left, we find
$D v^{-1} = I + D \Pi$.  Here the contour time differentiation
is with respect to the second argument $\tau'$.  We can compute the left-hand side explicitly and 
connect to $F$. We first note that the vector field $\bf A$ and the total current $\bf j$ are related at the Heisenberg 
operator level as $v^{-1} {\bf A} = - {\bf j}$ with $v^{-1} = - \epsilon_0 (\partial^2/\partial \tau^2 + c^2 \nabla \times \nabla \times)$
acting on $\bf A$.    Here we have generalized the real time $t$ to the Keldysh contour time $\tau$.  Acting $v^{-1}$ on $D$ from left is the same as acting on the second $\bf A$ in
$D$.  Without the contour ordering, we can immediately replace ${\bf A} v^{-1}$ by
$-\bf j$, which gives $-F$.   Note that $v^{-1}$ is symmetric in space indices so acting
from left is the same as acting from right. 
The contour order operator introduces extra terms.  We can express $D$ in terms of the step functions without the contour operator
$T_\tau$  as 
$D(\tau, \tau') = \theta(\tau, \tau') \langle A(\tau) A(\tau') \rangle/(i\hbar) + \theta(\tau',\tau) \langle A(\tau') A(\tau)\rangle/(i\hbar)$,
here we have suppressed the space and direction arguments since they are irrelevant for the reasoning.   The space differentiation
operation by $\nabla$ does not contribute extra terms as it straightly goes inside the contour order $T_\tau$.   Taking the first derivative
$\partial /\partial \tau'$ to $D$ leads to an extra term of the form $-\frac{1}{i\hbar} \delta(\tau, \tau') \bigl\langle [A(\tau), A(\tau')]\bigr\rangle$.
The delta function appears because $\partial \theta(\tau,\tau')/\partial \tau' =- \delta(\tau,\tau')$, and 
$\partial \theta(\tau',\tau)/\partial \tau' = \delta(\tau,\tau')$.  The commutator is now at equal time because of the delta-function factor, and thus is zero.
If we continue to
the second derivative $\partial^2/\partial \tau'^2$ that appears in $v^{-1}$, we find by similar steps a new term 
$\delta(\tau, \tau') \bigl\langle [\dot{A}, A] \bigr\rangle$.   
Using the canonical commutation relation, Eq.~(\ref{eqA-Pi-comm}), we find that this is precisely the identity
$I$ in space, contour time, and directional index.  Thus, we have $-F = D \Pi$.


\begin{figure}
\centering
\includegraphics[width=0.8\columnwidth]{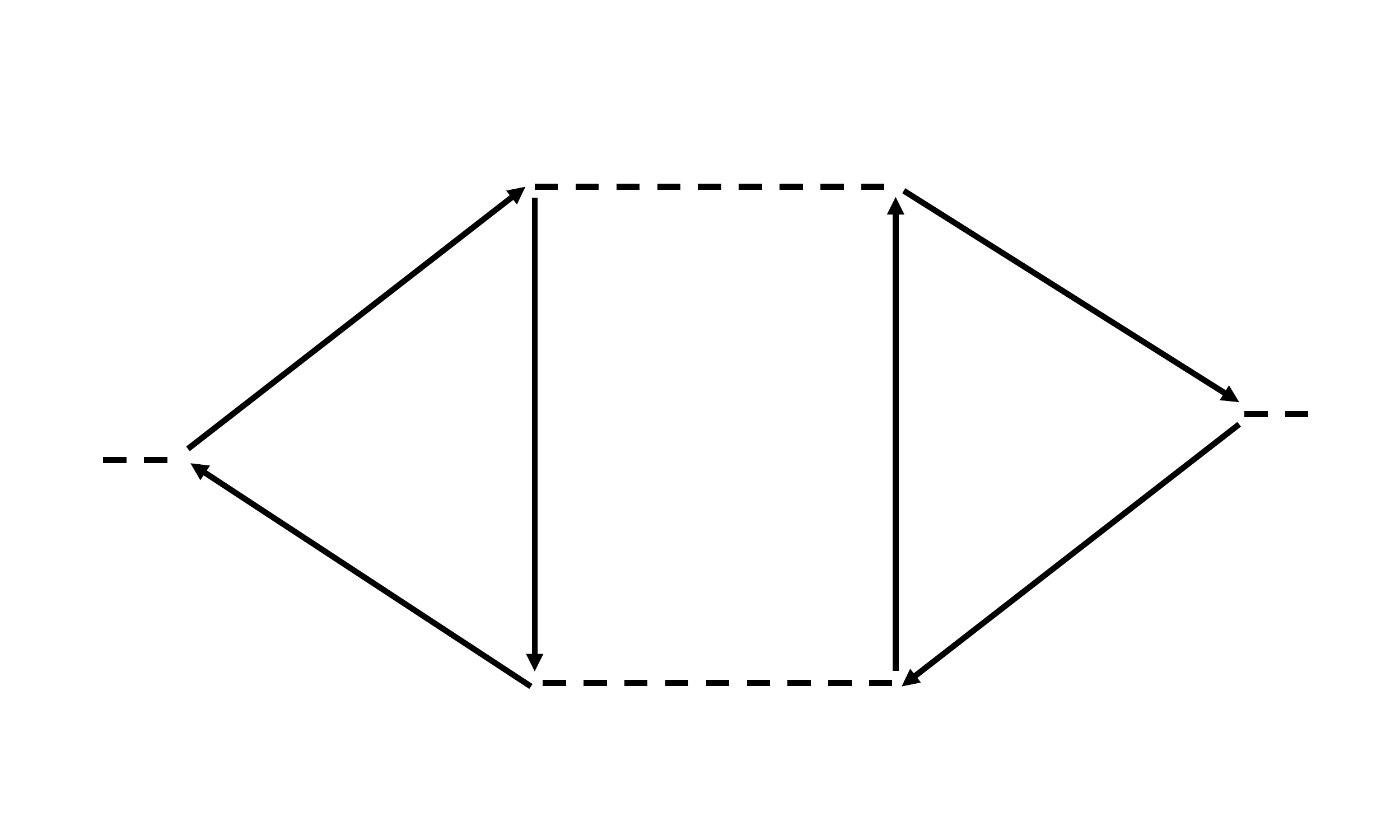}
\caption{One of the higher order contributions to the self-energy $\Pi$ in a diagrammatic expansion known as Aslamazov-Larkin diagram.  It consists of 6 electron
Green's functions (solid line with arrow) and 2 bare interactions $v$ (dotted line).  The electron line of a closed loop must belong to the same object, say, $\alpha$ and $\beta$ for the two loops, respectively.  Due to the existence of the $v$ lines which introduce mutual interaction, it is impossible 
to write the self-energy as a sum over each object individually. 
}
\label{fig:aslamazov}
\end{figure}

This is not yet the equation we are supposed to prove, which is $F_\alpha = -D \Pi_\alpha$ for object $\alpha$.   In order for $F$ to have
additivity over each object, we require that $\Pi$ is additive, i.e.,
\begin{equation}
\Pi = \sum_\alpha^N \Pi_\alpha.
\end{equation}
This additivity can be made true by selecting those Feynman diagrams of $\Pi$ with the right-most vertex associated with the second 
argument of $\Pi$ with object $\alpha$ for $\Pi_\alpha$, while the first argument has no restriction.
The additivity helps us to separate the contributions for the transported quantities unambiguously from each object. 
If we insist that $\Pi_\alpha$ means strictly from object $\alpha$, then additivity 
is certainly true at the RPA level of approximation, but it fails at higher order of approximation for $\Pi$.   A good
example is the diagram shown in Fig.~\ref{fig:aslamazov}, known as the Aslamazov-Larkin diagram \cite{Aslamazov68}.  The additivity of the self-energy breaks down at  $O(e^6)$,
where $e$ is the magnitude of electron charge. 
 We note that the additivity is false at the $e^4$ order for $\chi$, while is at $e^6$ for $\Pi$.   This is the main 
reason that we prefer to work with the irreducible diagrams from $\Pi$, instead of working with the current-current correlation $\chi$.

\section{Baths at infinity}
Since the electromagnetic field can propagate to infinity, any collection of finite objects will have some energy transmitted to infinity.
In order to account for the conservation laws, we must also count the ``object" at infinity.  An exception is the Polder and von Hove geometry of two materials with a gap; here in this problem, there is no need to consider the ``infinity''
and the space are all occupied by the objects. 
We can treat the empty space for $|{\bf r}| > R$ outside a big sphere of radius $R$ as the infinity.   It has the property that any 
energy, or momentum, or angular momentum sent to infinity is absorbed, and never reflected back.   Mathematically,
when we solve the retarded Dyson equation, we can treat the surface of the sphere at $R$ as an absorbing boundary condition.   Or
if the problem is solved in the full space including that outside the sphere, we must seek for a decaying solution that goes to zero at  $R \to \infty$.
  
Any object in our formalism is represented by the self-energy $\Pi$, including the object at infinity.  Usually, we set the 
temperature at infinity to zero, but we can also ascribe a temperature at infinity.   In this case, the finite objects are enclosed
in a black-body cavity at temperature $T_\infty$ from the environment.  
The self-energy of the environment can be expressed as a differential operator defined in the whole space, Eq.~(\ref{eq-Pi-inf-v}), $\Pi^r_\infty = - v^{-1}$. 
The question is, does it have an imaginary part?   Since an infinity domain must be dissipative, it does.  The effect of
this differential operator is realized if it acts on the actual solution of the problem.  Very often it is transformed
so that it does not appear explicitly in the end.  This is conceptually simple, but
computationally not very useful.   A more useful point of view is to think of the bath (the object) at infinity as defined precisely at the
sphere $|{\bf r}| = R$.   In this way, the degrees of freedom of the bath are solely specified by the solid angle $\Omega$.  

Here in this section, we derive an expression for the bath at infinity defined on the sphere locally as \cite{ZhangYM22prb,Peng1703arxiv}
\begin{equation}
\label{eq-Pi-inf-solid-angle}
\Pi^{r}_\infty = - i \epsilon_0 c \omega\Bigl( \stackrel{\leftrightarrow}{\bf U}  - \hat{\bf R}\hat{\bf R}\Bigr),
\end{equation}
and discuss its consequence.   To obtain this result, we consider a dust model 
\cite{Eckhardt84} of the bath at infinity, by saying that the space outside
$R$ is dissipative.   The dissipation is obtained phenomenologically by setting $\omega \to \omega + i \eta$ in the free Green's function, 
with some small positive $\eta$.   This is equivalent to give the space $|{\bf r} | > R$ as a dielectric medium with a local dielectric 
function $\epsilon \approx 1 + 2 i \eta/\omega$, or a constant,  infinitesimally small conductivity $2 \eta \epsilon_0$. 
The free Green's function is a good approximation if one of the space coordinate approaches infinity,
even if there is matter present at a finite region.  
We assume the asymptotic form of spherical wave solution $D^r \sim e^{i (\omega + i\eta) |{\bf r}|/c}$.  The effect of the dusts is to
introduce a damping over the purely oscillatory solution.  To leading order in large $r$, we have  $v^{-1} D^r \approx - 2 i \eta \epsilon_0 \omega D^r$.   Comparing with the Dyson equation,  $v^{-1} D^r = I + \Pi^r D^r$, we can identify the prefactor as the dust 
self-energy in a local form.  Note that the retarded and advanced self-energies or Green's functions are related 
by Hermitian conjugate.  In the dust picture for the bath at infinity, since the dissipative bath has been modeled explicitly, we
can put all the objects, including the dusts, in a large finite box as an overall isolated system.  In this case $-v^{-1}$ will no longer have an imaginary part
and is not interpreted as the self-energy of the bath anymore.   Thus, the meaning of the differential operator depends on
the boundary conditions choosing. 

The bath self-energy appears in the Meir-Wingreen formula as ${\rm Tr}\bigl[ \hat{O}(D^r \Pi^K_\infty + D^K \Pi_\infty^a) \bigr]$ where
$\hat{O}$ is an operator acting on the first argument of $D^r$ and $D^K$.   If we ascribe a temperature to the bath at
infinity, the Keldysh bath self-energy is related to the retarded/advanced version by the fluctuation-dissipation theorem,
$\Pi_\infty^K = (2N_\infty + 1) (\Pi^r_\infty- \Pi^a_\infty)$.   The trace here will be interpreted in two ways.  Originally, the 
trace is supposed to be the volume integral over all space with $\Pi_\infty^r$ the differential operator 
$-v^{-1}$.   Now we have an explicit formula for the self-energy over the volume outside the sphere due to the dusts.  
The field intensity decay produces a finite volume integral.   Due to the spherical symmetry,
we can work in polar coordinates.  
The integration over the volume is a solid angle integral and radial $\int_R^\infty dr r^2 \cdots$.   The exponential
decay in intensity, $D^r D^a \sim e^{-2 \eta |{\bf r}|/c}$, gives us a length $c/(2\eta)$ to multiply the volume expression of the self-energy.  
In the 
limit $\eta \to 0^+$, we obtain a finite answer for the solid angle expression independent of $\eta$, which can be interpreted as a new self-energy defined on the sphere of radius $R$ as given by Eq.~(\ref{eq-Pi-inf-solid-angle}).    The extra transverse projector which has 
angular dependence takes care that electrodynamic field in the far field is transverse.

\subsection{\label{sec-XIA}No zero-point motion contribution at infinity?}
In this subsection, we give an argument based on the Meir-Wingreen formula that there is no zero-point motion contribution to transport at
infinity.  If this were not true, we might run into problem of a divergent contribution to energy transport, for example. This is equivalent to say, that we can replace the Keldysh Green's functions and self-energies by the lesser components, 
$D^K \to 2 D^<$ and $\Pi_\infty^K \to 2 \Pi_\infty^<$ in the Meir-Wingreen formulas, Eq.~(\ref{meir-wingreen-I}) to (\ref{meir-wingreen-N}),
when applied to the quantities transmitted to infinity.   In order to show that this is true, we need an assumption that each object has local thermal equilibrium (perhaps this assumption can
be relaxed).  Applying the fluctuation-dissipation theorem, then the terms $2 N_\alpha +1$ in the Keldysh Green's functions or 
self-energies can be split as twice the lesser components plus extra terms.   We need to 
check that the extra terms are zero,
\begin{equation}
\label{eq-check-RE}
{\rm Re}\, {\rm Tr} \Bigl[\hat{O} \bigl(  D^r (\Pi^r_\infty - \Pi^a_\infty) + D^r (\Pi^r - \Pi^a) D^a \Pi^a_\infty  \bigr) \Bigr] = 0?
\end{equation}
Here $\Pi^r = (\Pi^a)^\dagger$ without a subscript is the total self-energy (including the bath at infinity).  Using the identity
$D^r - D^a = D^r(\Pi^r - \Pi^a) D^a$, the expression can be simplified as to check if the real part of 
${\rm Tr}\bigl[ \hat{O}(D^r \Pi^r_\infty - D^a \Pi^a_\infty) \bigr]$ is zero.  For the case of $\hat{O} = - \hbar\omega$, the trace is 
imaginary since the two terms are related by complex conjugate.    But the power $I_\infty$ must be real, so the result must be zero.
For the general situations when $\hat{O} = -\hbar \omega$, $\hat{\bf p}$, or $\hat{\bf J}$, we note
that the three factors $\hat O$, $D^r$, and $\Pi^r_\infty = (-v)^{-1}$ are operators in the 
space $\bf r$ and direction $\mu$.   The operator $\hat O$ is Hermitian.  Importantly,
$\hat O$ commutes with $\Pi^{r}_\infty$,
$[\hat{O}, \Pi^r_\infty] = 0$; this is a consequence of time translation, space displacement, and rotation symmetries of the free field, which we can check explicitly using the operator
representation of $\Pi^r_\infty$.   Using these properties of the operators and
the cyclic permutation property of trace, we find again that 
the two terms are related by complex conjugate and 
the expression (\ref{eq-check-RE}) has no real part. 
With these arguments, we can compute the physical observables at infinity as
\begin{equation}
I_\infty (\hat{O}) = {\rm Re} \int_0^{\infty} \frac{d\omega}{\pi} {\rm Tr}\bigl[ \hat{O}(D^r \Pi^<_\infty + D^< \Pi_\infty^a) \bigr].
\end{equation}
Here $\hat{O} = -\hbar\omega$, $\hat{\bf p}$, and $\hat{\bf J}$ for energy, momentum, and angular momentum transfer, respectively.
Due to our sign convention, $I_\infty = I_\infty(-\hbar\omega)$ is the energy out of infinity, i.e., $-I_\infty$ is the energy
absorbed at infinity, ${\bf F}_\infty$ is the force applied to infinity  (momentum transferred to infinity), and similarly ${\bf N}_\infty$ is the torque applied to infinity.   If the temperature at infinity is set to 0, $\Pi_\infty^<=0$, only the second $D^<\Pi_\infty^a$ term contributes.  
We obtain for the energy transferred to infinity per second (power) by the objects from a finite region as,
\begin{equation}
\label{eq-I-inf}
I_\infty = \int_0^{\infty} \frac{d\omega}{\pi} {\rm Re}\, {\rm Tr} \Bigl[ (-\hbar\omega) D^r \sum_{\alpha=1}^N\Pi^<_\alpha D^a \Pi_\infty^a\Bigr]. 
\end{equation}
We have already derived the same formula using the Poynting vector expectation value and normal order on a sphere of radius $R \to \infty$,
see Eq.~(\ref{eq-URRDls}).  
Here ${\rm Tr}[ \cdots ]= \int d\Omega R^2 \sum_{\mu} \cdots$, i.e., integrating over the surface of the sphere and summing over the direction
index.  We obtain the formulas for force and torque similarly, just by replacing the operator $(-\hbar \omega)$ by the momentum 
or angular momentum operator, which can be shown to agree with a direct calculation using Maxwell's stress tensor on
the surface of a large sphere.

\section{A consistency check with {Kr\"uger} {et al.} theory}

In the literature, the susceptibility $\chi$ is used as a primary material property instead of the self-energy (or polarizability) $\Pi$.
Here $\chi$ is the charge-charge correlation for the scalar field theory, and is the current-current correlation for the full theory.    The 
self-energy $\Pi$  is obtained from $\chi$ by keeping only the irreducible diagrams with respect to the interaction $v$ in a diagrammatic
expansion. The two are
related by $\chi = \Pi + \Pi v \Pi + \cdots = \Pi + v \Pi \chi$.  This gives rise to alternative expressions for the same physical quantities.    For
the scalar field result of the Landauer formula, Eq.~(\ref{eq-landauer-IL}), if we assume that the system is reciprocal \cite{Asadchy20}, i.e.,
both $\Pi$ and $\chi$ are symmetric, we can transform the Caroli formula into a form given by \onlinecite{Yu17}.   
The reciprocity allows us to use $\Pi^r - \Pi^a = 2 i\,{\rm Im}\, \Pi^r$ needed in the transformation \cite{Wang18pre}.
Note that the assumed reciprocal relation, $\Pi = \Pi^T$, does not hold when the electrons are in a magnetic field. 

Similarly, in pioneering works, \onlinecite{Krueger12prb} \cite{Krueger11prl,Bimonte17} give a series of formulas for various cases of energy transfer and (Casimir) forces for 
arbitrary objects.   Again, if we assume
reciprocity, the Meir-Wingreen formulas can also be transformed into the Kr\"uger {\it et al.}\ form.  It seems the assumed reciprocity,
$\Pi^T = \Pi$, is important for this equivalence to hold.   The symmetry of $\Pi$ implies the symmetry of $\chi$ as the free Green's
function is symmetric.  For the case of current-current correlation, the transpose here is in the combined space of coordinate and
directional index, i.e., $\Pi = \Pi^T$ means  
\begin{equation}
\Pi_{\mu\nu}({\bf r},{\bf r}',\omega) = \Pi_{\nu\mu}({\bf r}',{\bf r},\omega).
\end{equation}
Since the retarded $\Pi$ is related to the dielectric function by $\Pi^r = -\epsilon_0 \omega^2(\epsilon - 1)$, in a local theory,
reciprocity means $\epsilon^T  = \epsilon$ as a 3 by 3 matrix, which is the original meaning of ``reciprocity'' used by Lorentz \cite{Lorentz1896}. 

As an illustration, let us consider a simple case of one object at a temperature $T$ emitting energy to infinity which is at zero temperature.
Our result is given by Eq.~(\ref{eq-I-inf}).  To transform into Kr\"uger {\it et al.}\ form of Eq.~(37),  the key term in our formula is 
${\rm Tr} \bigl( D^< \Pi^a_\infty\bigr)$.   The trace here is interpreted as a volume integral over the whole space and sum over the
directional index.  The retarded Green's function is  $D = v + v \chi v$, here
all of them are retarded version.   We omitted the superscript $r$ for simplicity.   The lesser component of $D$ can be expressed
in terms of the retarded Green's function by the Keldysh equation 
$D^< = D^r \Pi^< D^a = N D (\Pi - \Pi^\dagger) D^\dagger$.  We have used the fluctuation-dissipation theorem for the object, where
$N$ is the Bose function at temperature $T$;  bath at infinity is at zero temperature. 
 We now need a relation between the imaginary part of
$\Pi$ and the imaginary part of $\chi$.  Using the relation  $\chi = \Pi + \Pi v \Pi + \cdots$, we can check that 
\begin{equation}
(1+\chi v) (\Pi - \Pi^\dagger)(1+v^\dagger \chi^\dagger) = \chi - \chi^\dagger + \chi ( v^\dagger  - v) \chi^\dagger.
\end{equation}
With this identity, we have $D^< \Pi_\infty^a = N v (1+\chi v)(\Pi - \Pi^\dagger) (1+ v^\dagger \chi^\dagger) v^\dagger (-v^{-1})^\dagger
= -N v \bigl((\chi - \chi^\dagger) + \chi (v^\dagger - v) \chi^\dagger\bigr)$.   We have used the representation for the bath at infinity
by $\Pi_\infty^a =  (-v^{-1})^\dagger$ which gets canceled by a factor $v^\dagger$ from the $D^\dagger$ term on the right side
of the Keldysh equation.  Finally, putting the integral over frequency back, using the reciprocity  so that $\chi - \chi^\dagger = 
2 i\, {\rm Im}\, \chi$, and similarly for $v$, we get 
\begin{equation}
-I_\infty  = \int_0^{\infty} \frac{d\omega}{\pi} 2 \hbar\omega\, N {\rm Im}\, {\rm Tr} 
\Bigl[ v {\rm Im}\, \chi - v \chi ({\rm Im}\, v )\chi^*  \Bigr].
\end{equation}
This is Eq.~(37) in Kr\"uger {\it et al}.\ if we note a slight notational change, $\chi = - {\cal T}/\mu_0$, where
$\cal T$ is the scattering operator, and $v = - \mu_0 {\cal G}_0$ the free Green's function, with the product $v\chi = {\cal G}_0 {\cal T}$ remaining the same. Here $\mu_0$ is the vacuum permeability. 

To show the equivalence of the Landauer/Caroli formula with Kr\"uger, {\it et al}.\ formulation with heat transfer or force 
between two objects, 1 and 2, we need to solve the Dyson equation in a block matrix form as we need only part of the matrix elements 
of $D^r$ connecting the two objects, as $\Pi^r$ is block diagonal.  By doing this, we can express $D^r_{12}$ in  terms of 
$\chi_1$ and $\chi_2$ of individual objects together with the free field $v_{12}$ connecting the two objects.  Similar steps show 
indeed they are equivalent, e.g., Eq.~(57) in \onlinecite{Krueger12prb} and also is equivalent to Yu {\it et al}.\ form when ${\rm Im}\, v = 0$.   A recent result for torque by
\onlinecite{Strekha22} is believed to agree with our formula also. 

\section{Breaking reciprocity by current drive -- far field results} 

\begin{figure}
\centering
\includegraphics[width=0.9\columnwidth]{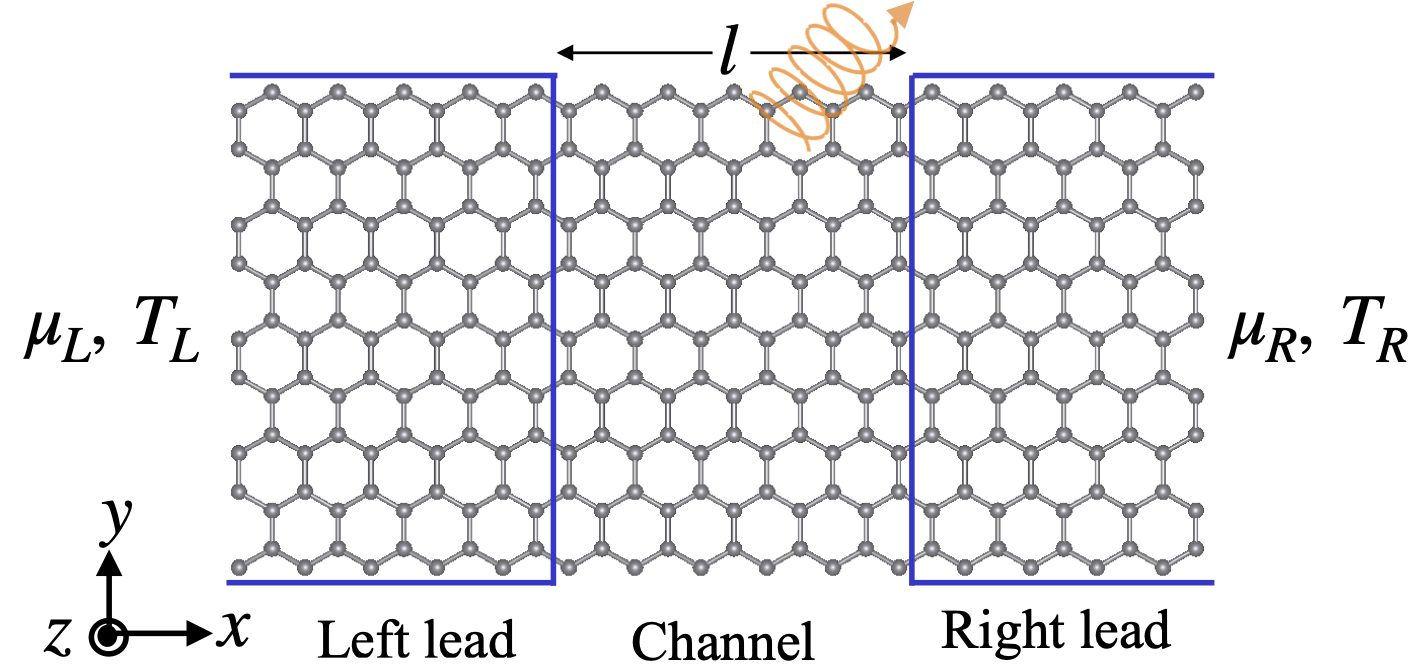}
\caption{Illustration of bias-induced emissions of energy, momentum, and angular momentum from the conducting channel of a two-terminal transport device of a graphene strip. The length of the central channel is $l$. The semi-infinite left and right leads are extensions of the pristine graphene strip. 
}
\label{fig:figA}
\end{figure}

In this last section, we apply the above developed general theory for the transport of energy, momentum, and angular momentum due to
a driven current in a nanoscale piece of metal or semiconductor, more specifically, a segment of nanoribbon of graphene as an example, 
see Fig.~\ref{fig:figA}.  The driven current is realized by applying a bias voltage to the two leads connecting the piece, typical
of mesoscopic ballistic transport setup. 

In applying the radiation-to-infinity formulas,  Eq.~(\ref{eq-I-inf}), for practical calculations, we need to make two more approximations.
First, we replace the full Green's function, $D$, by the free field one, $v$.   This means that we ignore the multiple reflections inside matter,
and take only the first term in the Dyson expansion,  $D = v + v \Pi v + \cdots$.   The correction to $v$ is a kind of screening effect
given by a factor $(1 + v\chi ) = \epsilon^{-1}$, so that $D = \epsilon^{-1}v$.  For the far field, the longitudinal component (corresponding to the scalar field $\phi$ in a transverse gauge)
decays to zero quickly, thus does not contribute to far-field radiation.  The screening effect for the transverse components is much smaller.  
We can make an order of magnitude estimate of the omitted terms based on a dimensional analysis.   The self-energy is 
$\Pi = - \omega^2 \epsilon_0 a^3 (\epsilon - 1)$ on a lattice in a discrete representation, where $a$ is lattice constant
and $\epsilon$ is a dimensionless dielectric function.   The
free Green's function is $v \sim  1/(4\pi\epsilon_0 c^2 R)$; we take the distance $R \sim a$.  Multiplying the two, we find 
$v\Pi  \approx v\chi  \sim (\omega a/c)^2$, which is a ratio of two velocities.   If we take the velocity $\omega a$ to be the electron speed
$at/\hbar$ where $t$ is a hopping parameter of order eV, we find 
$\Pi v \sim 10^{-4}$.  If we take $\hbar \omega$ to be the thermal energy of order $k_BT$, the result is even smaller. 
Using the approximation $D \approx v$ works for small molecules or graphene sheet where most of the atoms are exposed to space,
the emission and then re-absorption are negligible.   But inside the bulk of a material, even though the omitted single scattering term is small, 
the cumulative effect is large, so $D \approx v$ is not a good approximation.   In such a case, we need to solve the Dyson equation
$D = v + v \Pi D$ by some means. 

The second approximation is to use multipole expansion for the Green's function.   Since the object is finite, and the observation is at 
$R \to \infty$, multipole expansion is a good approximation when the wavelength of the field is much larger than the relevant size of the object. 
Multipole expansion simply means we Taylor expand the second argument in
\begin{equation}
D_{\mu\nu}({\bf R},{\bf r}) = D_{\mu\nu}({\bf R}, {\bf 0}) +
 {\bf r} \cdot \bigl(\nabla_{{\bf r}'} D_{\mu\nu}({\bf R},{\bf r}')\bigr)\Big|_{{\bf r}'={\bf 0}} + \cdots. 
\end{equation}    
We have omitted the frequency $\omega$ argument for simplicity.  After replacing $D$ by $v$, the function is translationally invariant, 
which depends only on the difference $|{\bf R} - {\bf r}|$.  So, we can also write as 
$v({\bf R}, {\bf r}) = v({\bf R}) - {\bf r} \cdot \nabla_{\bf R} v({\bf R}) + \cdots$.  It is sufficient to keep to linear order in ${\bf r}$ for
our purpose. 

We now discuss the evaluation of the Keldysh term $D^r \Pi^< D^a$ at far field and the solid angle integration.  For the energy, force, and the spin part of the torque contribution, the second $(\stackrel{\leftrightarrow}{\bf U} - 3 \hat{\bf R}\hat{\bf R})$ term in Eq.~(\ref{eq-v-real-space}) can be omitted as it decays to zero faster than $1/R$.  Only the first $1/R$ transverse term 
contributes to the far field.  The $1/R^2$ factor from the product of $D^r$ and $D^a$ is canceled
by the surface area element $R^2 d\Omega$, picking up a finite result in the limit $R \to \infty$.  In this limit, we can replace the gradient
operation $\nabla_{\bf R} v$ by $i (\omega \hat{\bf R}/c) v$.  We have already given the monopole
term after the solid angle integration by Eq.~(\ref{eq-radiation-farpiless}).   The linear terms in ${\bf r}$ do not contribute to 
the total energy radiation, because it contains an odd number of $\hat {\bf R}$.  The next non-vanishing terms appear in
quadruple form, ${\bf r} {\bf r} \Pi^<$.   We can also give an order of magnitude estimate of the extra correction terms.   Since
each time we take a gradient, we obtain a dimensionless factor ${\bf r} \cdot \hat{\bf R}\,\omega/c$.   So, the correction terms 
are smaller by a factor  $(a \omega/c)^2$  than the monopole term.  Here we agree that the typical distance is the lattice constant.    
This is at a similar order of magnitude as due to the screening corrections.  Since $\omega/c \sim 1/\lambda$ is the inverse wavelength, 
the higher order terms are smaller by a factor $(a/\lambda)^2$.

For the force, if we keep only the monopole term, the result is zero as it contains an average of an odd number of the unit vectors ${\hat{\bf R}}$.  
In this case, the dipole term is essential in order to have a nonzero value.  We need to compute the average of four $\hat{\bf R}$'s for
the solid angle integration.   We can check explicitly
\begin{equation}
\overline{{\hat R}_\alpha {\hat R}_\beta {\hat R}_\gamma {\hat R_\mu}} = 
\frac{1}{15} \bigl(\delta_{\alpha\beta}\delta_{\gamma\mu} + \delta_{\alpha\gamma} \delta_{\beta\mu} + \delta_{\alpha\mu} \delta_{\beta\gamma}  \bigr).
\end{equation}
Here the overline means average over the solid angles, $\frac{1}{4\pi} \int d\Omega \cdots$.  The Greek subscripts indicate the Cartesian
components of the unit vectors.  We also recall  $\overline{{\hat R}_\alpha \hat{R}_\beta} = \frac{1}{3} \delta_{\alpha\beta}$. 
After some algebra, the force is \cite{ZhangYM22prb}
\begin{eqnarray}
F^\mu_\infty &= & \int_0^\infty\!\!\! d\omega { \hbar \omega^3 \over  60 \pi^2 \epsilon_0 c^5 }  
\sum_{\alpha,l,l'}\Bigg[ 4\,  \Pi^{<}_{l\alpha,l'\alpha}(r_\mu^l -  r_\mu^{l'}) \nonumber\\
&& -  (r_\alpha^l - r_\alpha^{l'}) \Pi^{<}_{l\alpha,l'\mu} 
 -  \Pi^{<}_{l\mu,l'\alpha} (r_\alpha^l - r_\alpha^{l'}) \Bigg].\qquad 
\end{eqnarray}
For the monopole expressions, like the total energy emission, we only need the sum total of $\Pi_{l\mu,l'\nu}$ over the sites $l$ and $l'$.  For the force,
it is the first moment respect to the distance ${\bf r}^l - {\bf r}^{l'}$ that is needed. 
For systems with reciprocity, $(\Pi^<)^T = \Pi^<$, this expression is zero.  Thus, 
we need to break reciprocity in order to have a nonvanishing force.

For the orbital angular momentum contribution to the torque, since ${\bf r} \times \hat{\bf p}$ term is proportional to $R$ at large $\bf r$,
the leading $1/R^2$ term in the product of the Green's functions is canceled to zero.  We need to work harder to pick up the $1/R^3$ terms from
$\nabla_{\bf R} D^r D^a$.   In this case, we need to keep track of the next order term in the gradient operation.  It turns out that 
the contribution from the spin part is equal to the contribution from the orbital angular momentum part (different from the conclusion in \onlinecite{Khrapko19}), given a total that is in agreement with
a direct calculation with Maxwell's stress tensor result \cite{Zhang20prb},
\begin{equation}
N_\infty^\mu = \int_0^\infty\!\!\!d\omega { \hbar \omega \over 6 \pi^2 \epsilon_0 c^3} \sum_{\alpha\beta} \epsilon_{\mu\alpha\beta} \Pi^<_{\beta\alpha}.
\end{equation}
Here $\epsilon_{\mu\alpha\beta}$ is the anti-symmetric Levi-Civita symbol,  $\Pi^<_{\alpha\beta}$ without the site indices $(l,l')$ means 
the sum total.  Since $(\Pi^<)^\dagger=- \Pi^<$, there is no need to take the real part; the expression is real.

\subsection{Graphene strip calculation}

\begin{figure}
\centering
\includegraphics[width=0.85\columnwidth]{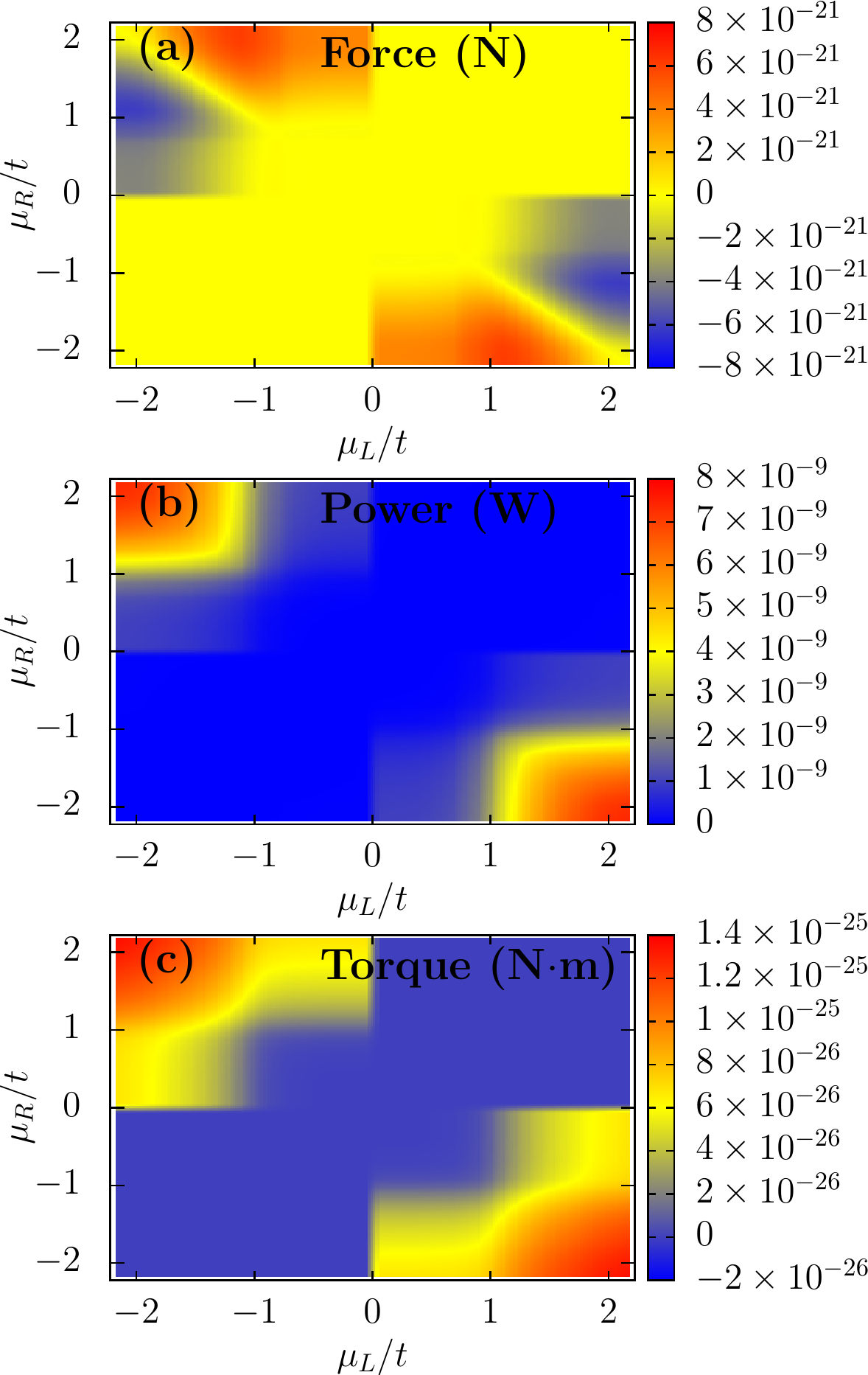}
\caption{Results of (a) force component in the $x$ direction, (b) power, and (c) torque component in the $z$ direction as a function of the chemical potentials $\mu_{L}/t$ and $\mu_{R}/t$ for the two-terminal graphene strip system.
}
\label{fig:figB}
\end{figure}

We consider the radiation of energy, as well as force and torque experienced by
a finite piece of a zigzag graphene nano ribbon, as shown in Fig.~\ref{fig:figA}.  The
nonequilibrium state is modeled in the usual way of mesoscopic transport by connecting the piece with baths from the left and right. The two leads are in their respective equilibrium states, where the electrons follow the Fermi distribution $f_{L(R)} = 1 / \textrm{exp}\big[(E - \mu_{L(R)})/\big( k_B T_{L(R)}\big) + 1\big]$. Here, $\mu_{L(R)}$ and $T_{L(R)}$ are chemical potential and temperature of the left (right) lead, respectively. The {C-C} bond length is $a = 1.4\,$\AA. In the numerical calculation, we choose the hopping parameter $t = 2.7\,$eV, the length of the central channel $l = 5 \sqrt{3}\, a$, and the with $W=9$ (number of zigzag lines). We set the temperatures of the two leads to be equal, with $T_{L} = T_{R} =300\,$K. 

We show in Fig.~\ref{fig:figB} the results of the force, power, and torque as a function of the chemical potentials of the two leads.  Both the power and torque are symmetric with respect to $\mu_{L} = \pm \mu_{R}$, while the force is antisymmetric with $\mu_{L} = \pm \mu_{R}$. Figure~\ref{fig:figB}(a) only shows the $x$ component of the force, as the $y$ component is very small, and the $z$ component is zero due to symmetry. The force on the strip due to the light emission is a nonequilibrium effect induced by the driving current in the $x$ direction,
the direction of electron transport. The torque perpendicular to the surface in Fig.~\ref{fig:figB} (c) is due to the asymmetric structure in the central channel with width of $W=9$, while we have checked that the driving current cannot induce nonzero torque for mirror symmetric strip when the width is even, for example, $W=8$.    As we can see from the figures that the force and torque are rather small even
with large drive.   It is a challenge to find realistic systems that give signals that experimentalist can detect.

Radiation of angular momentum from a benzene molecule driven out of equilibrium by two
leads unsymmetrically attached to the molecule is presented in \onlinecite{Zhang20prb}, while the emission of angular momentum from a two-dimensional Haldane model is 
calculated in \onlinecite{ZhangYM21JP}. 
In \onlinecite{ZhangYM22prb}, the emissions of energy, momentum, and angular momentum
from a semi-infinite graphene edge are calculated using the above formalism, taking into 
account translational invariance in the driven direction by going into wave-vector space.    We 
note here, for the angular momentum it is truly an edge effect as the total is not proportional to the area but only length of the graphene.  Also, it is truly a nonequilibrium
effect, as in thermal equilibrium, both the momentum and angular momentum emissions (or
the negative of force and torque applied to graphene) is zero.   The nonequilibrium electron
states are set up in a cheaper way by setting the Fermi functions based on the sign of
the group velocity due to the left and right baths far away, which is valid in the ballistic regime.  Such problem is beyond the ability of the usual fluctuational electrodynamics.
 
We setup a periodic boundary condition in the electron transport direction 
(open boundary condition in the width direction) \cite{ZhangYM22prb}.   With periodic 
boundary condition, the self-energy $\Pi^<$ can be calculated in an eigenmode representation of the electrons.   
In the limit $\eta \to 0^+$, the lesser Green's function can be represented in the eigenmode as
\begin{equation}
G^<(E) = 2\pi i \sum_n f_n |n\rangle \langle n| \, \delta(E - E_n),
\end{equation}
where $H|n\rangle = E_n |n\rangle$ solves the single electron eigenvalue problem, and $f_n$ is the Fermi function in state $n$ with
energy $E_n$.   $G^>$ is obtained by a replacement of $f_n$ to $f_n-1$.   Then $\Pi^<(\omega)$ can be computed according to Eq.~(\ref{eq-Pils}).
This gives a more efficient
method as the self-energy is a sum of delta functions, from which, the frequency integration can be performed analytically.    In such a 
setup, the different baths are given implicitly based on the sign of the group velocity of electrons to the Fermi distribution.     

The power emitted in this mode-space approximation is a Fermi-golden rule result 
\cite{Kibis07},
\begin{eqnarray}
-I_\infty &=& \frac{4\alpha}{3\hbar c^2} \sum_{\mu,n,n'} (E_n - E_{n'})^2 
\theta(E_n - E_{n'}) \nonumber \\
&&  \times \Bigl|\langle n| V^{\mu} |n'\rangle\Bigr|^2 f_n (1-f_{n'}), 
\end{eqnarray}
where $\alpha \approx 1/137$ is the fine-structure constant, $\theta$ is the step function,
and $V^\mu$ is the velocity matrix in $\mu$ direction.  We observe energy emission
when an electron jumps from an occupied state to an empty state.  
Similar formulas are obtained for the torque and force with the help of the velocity matrices \cite{ZhangYM22prb}.

\vfill

\section{Conclusion}
In this review, we presented the NEGF formalism for photon transport in the framework of
Coulomb interactions among electrons in the non-retardation limit, as well as scalar and
transverse vector potential formulations.   In the near field, Coulomb interaction
is the most important contribution, equivalent to keeping only $p$ polarization (TM mode)
in a parallel plate geometry.   For far-field radiation, we must also compute the
vector potential contribution due to a finite speed of light, since only this term contributes to thermal radiation at
infinity.    To keep track of both the scalar and vector potentials in transverse gauge is
a bit clumsy, thus, we also offer another choice of gauge which is the more
economical temporal gauge.   Not only the transmission of energy, but also of momentum and angular momentum, can be computed in a unified fashion with a Meir-Wingreen-like formula. 

The solution to electrodynamics is formulated to solve the retarded Dyson equation;
this is completely equivalent to solving the Lippmann-Schwinger equation in a scattering approach.  The distribution or correlation function of the field is given by the 
Keldysh equation.   This is equivalent to applying the fluctuation-dissipation theorem
in fluctuational electrodynamics (FE).   In FE, it is applied in the last step,
but here we in some sense incorporate it upfront.   The NEGF approach offers the
option of not applying the local thermal equilibrium used in FE, thus opening the door for
more general settings of nonequilibrium steady states.  We have given several examples
of this sort.  The steady states are established
through the applications of multiple electron baths at different temperatures or chemical
potentials.    

The Meir-Wingreen formula, or if local equilibrium is used for energy, the Landauer formula
with the Caroli form for the transmission, is simple.  But the complexity hides in the 
solution to the Dyson equation.   In the Kr\"uger {\sl et al.}\ approach, one focuses on a
single object at a time, and then combine the results to get the full solution,
incorporating multiple reflections.  This gives computational efficiency, but with a more
involved formula for the transport quantity.   Another advantage of treating each object
separately as a scattering problem is that one can handle moving objects, by Lorentz boost.  It is
not clear how such problems can be handled in the NEGF framework.   An individual 
treatment of each object by computing its scattering operator will
fail to work if the property of one object is strongly influenced by nearby objects
due to extreme proximity. 

Our point of view here is to couple the calculation of
the materials properties with the electrodynamic calculation closely so that the materials
properties are calculated {\it ab initio} through the random phase approximation (RPA) for
the electrons.   One can also use the machinery of many-body physics to go beyond  RPA,
such as the self-consistent GW calculation or the contribution to the self-energy $\Pi$ from the Aslamazov-Larkin diagram.  
Nonlinear effects in the field can be handled by including high order diagrams.
These mutual correlations, not in the standard FE, may be important at extreme near field.  Unlike the frequency-dependent local dielectric function,
the photon self-energy $\Pi$ is nonlocal to start with, which is a better physical 
quantity when the structure is described atomistically.  In the examples shown, we only considered
electrons as our materials, but phononic contributions can be, in principle, 
incorporated.    In fact, there is a formula for the longitudinal inverse dielectric function due to phonons in a crystal given in \onlinecite{Dolgov89} 
as
\begin{eqnarray}
\epsilon^{-1}({\bf q},\omega) &=& \frac{1}{\epsilon_\infty} + \\
&&\!\!\!\!\!\frac{e^2}{\Omega\epsilon_0\epsilon_\infty} \sum_{n,\kappa,\kappa'} 
\frac{Z_\kappa Z_{\kappa'}}{\sqrt{M_{\kappa}M_{\kappa'}}}
{ \hat{\bf q} \cdot {\bf e}_\kappa({\bf q}n)\,
 \hat{\bf q} \cdot {\bf e}^{*}_{\kappa'}({\bf q}n)  \over (\omega + i \eta)^2 - \omega^2_n({\bf q}) },  \nonumber
\end{eqnarray}
where $\epsilon_\infty$ is the high-frequency dielectric constant from electrons, $Z_\kappa$ is the Born effective charge for the ion $\kappa$ in a unit cell,
$\Omega$ is unit cell volume, ${\bf e}_\kappa({\bf q}n)$ is the phonon polarization vector of
wavevector $\bf q$ and phonon branch $n$, and $\omega_n({\bf q})$ is the phonon 
dispersion relation (note the 
relation $\epsilon^{-1} =1 + v \chi$).   But the
real challenge is to treat electrons, phonons,
photons, and their mutual interactions together consistently.   

\section*{Acknowledgements}

We thank Gaomin Tang, Jingtao L\"u, and Mauro Antezza for collaborations.  J.-S. W. thanks
Mehran Kardar for hosting a visit at MIT.  He also thanks Shanhui Fan, Philippe Ben Abdallah, Matthias Kr\"uger, and Zhuomin Zhang for discussion.
This work is supported by an MOE tier 2 grant R-144-000-411-112.   Parts of the manuscript
were written while visiting Kavli Institute for Theoretical Physics, University of California Santa Barbara, supported in part by the
National Science Foundation under Grant No. NSF PHY-1748958. 

\clearpage
\bibliography{references}

\end{document}